\documentclass[iop]{emulateapj}

\newcommand{\myemail}{catherine.walsh@qub.ac.uk}

\usepackage{graphics}
\usepackage{url}
\usepackage{subfigure}
\usepackage{captcont}
\usepackage{wasysym}
\usepackage{savesym}
\savesymbol{iint}
\savesymbol{iiint}
\usepackage{amsmath}
\usepackage{url}

\shorttitle{Chemical Processes in Protoplanetary Disks}
\shortauthors{Walsh et al.}

\begin{document}

\title{Chemical Processes in Protoplanetary Disks}

\author{Catherine Walsh and T. J. Millar} 
\affil{Astrophysics Research Centre, School of Mathematics and Physics, Queen's University
Belfast, University Road, Belfast, UK, BT7 1NN}
\and 
\author{Hideko Nomura}
\affil{Department of Astronomy, Graduate School of Science, Kyoto University, Kyoto
606-8502, Japan}
 
\email{\myemail} 

\begin{abstract}

We have developed a high resolution combined physical and chemical model 
of a protoplanetary disk surrounding a typical T Tauri star. 
Our aims were to use our model to calculate the chemical structure 
of disks on small scales (sub-milli-arcsecond in the inner disk 
for objects at the distance of Taurus, $\sim$~140~pc)   
 to investigate the various 
chemical processes thought to be important in disks and to determine potential molecular tracers of 
each process. 
Our gas-phase network was extracted from the UMIST Database for Astrochemistry to which we 
added gas-grain interactions including freeze out and thermal 
and non-thermal desorption (cosmic-ray induced desorption, 
photodesorption and X-ray desorption) and a grain-surface network.  
We find that cosmic-ray induced desorption has the least effect on our disk chemical structure 
while photodesorption has a significant effect, enhancing the abundances of most 
gas-phase molecules throughout the disk and affecting the abundances and distribution 
of HCN, CN and CS, in particular.  
In the outer disk, we also see enhancements in the abundances of H$_2$O and CO$_2$.   
X-ray desorption is a potentially powerful mechanism in disks, acting to homogenise the 
fractional abundances of gas-phase species across the depth and increasing 
the column densities of most molecules although there remain significant uncertainties 
in the rates adopted for this process. 
The addition of grain-surface chemistry enhances the fractional abundances of several 
small complex organic molecules including CH$_3$OH, HCOOCH$_3$ and CH$_3$OCH$_3$ to potentially 
observable values (i.e.\ a fractional abundance of $\gtrsim$~10$^{-11}$). 
\end{abstract}

\keywords{astrochemistry --- planetary systems: protoplanetary disks --- 
stars: formation --- solar system: formation --- ISM: molecules}

\section{INTRODUCTION}
\label{introduction}

Protoplanetary disks are 
crucial objects in low-mass star formation, possessing three vital functions: they  
(i) aid the dissipation of angular momentum away from the  young stellar system, 
(ii) allow the efficient accretion of matter onto the young star and 
(iii) contain all material, dust and gas, 
which may end up in a planetary system orbiting the main-sequence star.  

In this work, we investigate the chemistry and molecular composition of a
protoplanetary disk surrounding a young star which 
will evolve into a main-sequence star resembling our Sun.    
At the low temperatures encountered in many astrophysical regions 
($\sim$~10~K to $\sim$~100~K), 
molecules are readily excited into higher rotational energy states and 
subsequently emit radiation at (sub)millimeter wavelengths. 
Early observations of T Tauri stars at these 
wavelengths revealed the presence of molecular material in a flattened disk-like 
structure (e.g.\ \citet{dutrey94}) and in Keplerian rotation about the parent star 
(e.g.\ \citet{guilloteau94}).  
Since then, molecular rotational line emission originating from a disk 
has been observed in several 
T Tauri systems including 
TW Hydrae \citep{kastner97,vanzadelhoff01,vandishoeck03,ceccarelli04,thi04,qi04,qi06,qi08}, 
DM Tauri \citep{dutrey97,ceccarelli04,ceccarelli05,guilloteau06,dutrey07,pietu07} and 
LkCa 15 \citep{vanzadelhoff01,aikawa03,qi03,thi04,dutrey07,pietu07}.  
Most species detected are small simple molecules, molecular ions and radicals such as 
CO, HCO$^+$, CN, HCN, CS, C$_2$H and N$_2$H$^+$, along with several associated  
isotopologues (e.g.\ $^{13}$CO, C$^{18}$O, DCO$^+$, H$^{13}$CN, H$_2$D$^+$ and C$^{34}$S).  
The most complex species observed to date is the small organic molecule, 
formaldehyde, H$_2$CO \citep{dutrey97,aikawa03, dutrey07} 
with methanol, CH$_3$OH, thus far eluding detection (e.g.\ \citet{thi04}).  
\citet{ceccarelli05} report a detection of 
deuterated water, HDO, in the disk of DM Tau, although this result 
has since been disputed by \citet{guilloteau06}.  

Infra-red emission has also been observed originating from disks  
embedded in young stellar objects and arising from 
vibrational transitions in gas-phase molecules capable of survival 
in the warmest regions ($>$~350~K).  
Thus, infra-red emission probes not only a different physical region of 
the disk to that probed by (sub)mm emission, but also uses
different molecules as tracers, hence providing complimentary chemical 
information.  
The molecules detected thus far at infra-red wavelengths are CO, HCN, OH, H$_2$O, 
CO$_2$ and C$_2$H$_2$ \citep{carr04,lahuis06,carr08,salyk08,pascucci09} 
with an upper limit determined for CH$_4$ \citep{gibb07}. 
   
Observations of molecular line emission from disks, to date, have been 
hampered by the small angular size of these objects on the sky and 
the limitations of existing facilities, explaining why the species 
detected are those which are abundant and possess
relatively simple rotational energy spectra (e.g.\ CO).   
Single-dish facilities which operate at (sub)mm wavelengths 
such as the 15~m James Clerk Maxwell Telescope (JCMT) in Hawaii and 
the IRAM~30~m telescope in Spain, have been predominantly 
used in the detections of the molecular species in the T Tauri systems  
listed.  
With beam-sizes much larger than the typical source size, 
usually a single molecular line profile is generated characterising 
emission from the entire disk.  
In order to spatially resolve the emission and hence trace the 
radial and vertical physical and chemical structure, interferometry 
must be employed and indeed, \citet{qi08} report spatially 
resolved emission arising from molecular rotational transitions 
in the disk of TW Hya using the Sub-Millimeter Array (SMA).   

The discipline of (sub)mm astronomy is scheduled for a revolutionary transformation  
with the first light of the Atacama Large Millimeter Array (ALMA) in Chile 
expected in 2012 (see \url{http://www.almaobservatory.org}).  
ALMA, with its 50 12~m telescopes and fully variable configuration, will have the 
spatial resolution necessary to observe molecular line emission from protoplanetary 
disks on sub-milli-arcsecond scales and enable the 
tracing of the molecular content of disks to within $\approx$~0.1~AU of 
the parent star at its highest operational frequencies.  
It is anticipated that the sensitivity and high spectral resolution of ALMA 
will lead to the potentially overwhelming detection of many further 
molecular species, including complex organic molecules considered the 
building blocks of life, in many astrophysical sources including protoplanetary disks.    

Motivated by the impending completion of ALMA, we 
have constructed a high resolution combined chemical and physical 
model of a protoplanetary disk surrounding a typical T Tauri star using 
as comprehensive a chemical network as computationally possible.  
In the work presented here, our objectives were (i) to calculate 
the chemical structure of protoplanetary disks on small 
(sub-milli-arcsecond in the inner disk) scales, 
(ii) to investigate the influence of various chemical 
processes, such as non-thermal desorption and grain-surface chemistry, 
thought to be important in disks, and  
(iii) to subsequently determine potential molecular tracers of each process.  
We also used our model to (i) compute molecular line emission profiles 
for rotational transitions which have been observed in disks using 
existing facilities, (ii) compare our modelled line profiles and intensities with existing 
observations and (iii) produce molecular line emission maps at the expected 
spatial resolution of ALMA for disks at various distances and inclinations.  
This second study and corresponding set of results will be covered in a 
subsequent paper (Walsh et al. in preparation).  
Our study also aims to help answer some 
fundamental questions concerning the evolution of stars, planets and ultimately, life.  
Is it possible for primordial (possibly organic) material created 
in a young star's protoplanetary 
disk to survive the assimilation into planets and other planetary system objects?   
Is our solar system's chemical and thus, planetary composition unique?  
How intrinsically linked 
are star formation, planet formation and the origin of life in the universe? 
These questions are evermore important as we move into the era of exoplanet research 
and the hunt for planets and the signatures of life in external stellar systems.  

In Section~\ref{diskmodel} we 
describe the theoretical foundation and generation 
of the physical model used to characterise our protoplanetary disk (Section~\ref{physicalmodel}) and
the chemical network we have 
collated and used in our calculation of the disk chemical evolution (Section~\ref{chemical model})
including gas-phase chemistry (Section~\ref{gasphasechemistry}), 
 photochemistry (Section~\ref{photochemistry}), gas-grain interactions 
(Section~\ref{gasgraininteractions}) and 
grain-surface chemistry (Section~\ref{grainsurfacechemistry}).  
The results of our chemical evolution calculations are 
covered in Section~\ref{results} where we discuss the 
chemical structure and stratification in the disk 
(Section~\ref{chemicalstructure}),  
the effects of our included chemical processes  
(Sections~\ref{nonthermaldesorptioneffects} and \ref{grainsurfacechemistryeffects}), 
the disk ionisation fraction (Section~\ref{diskionisationfraction}) and the radial 
molecular column densities (Section~\ref{columndensities}).  
We briefly discuss our work in relation to similar projects by other research groups 
in Section~\ref{comparison} and finally, 
in Section~\ref{summary}, we summarise our work and outline our main conclusions 
and further work we intend to undertake.  

\section{PROTOPLANETARY DISK MODEL}
\label{diskmodel}

\subsection{Physical Model}
\label{physicalmodel}

The physical model of a protoplanetary disk we use in this work is from 
\citet{nomura05} with the addition of X-ray heating as described in 
\citet{nomura07}.  
They self-consistently modelled the density and temperature 
profiles of gas and dust in a protoplanetary disk accounting for  
UV and X-ray irradiation by the central star and 
subsequently computed molecular hydrogen line 
emission at ultraviolet and infrared wavelengths.  
Here, we have used this model to compute 
the chemical structure of a protoplanetary disk with the 
ultimate aim to expand on their work by 
calculating molecular line emission from disks at  (sub)mm wavelengths.  
In the remainder of this section, we give a brief overview of our physical model and 
we refer readers to the original papers for the mathematical and 
computational details.  

We consider an axisymmetric disk surrounding 
a typical T Tauri star with mass, $M_\ast$~=~0.5~$M_\odot$, radius, $R_\ast$~=~2~$R_\odot$ 
and temperature, $T_\ast$~=~4000~K \citep{kenyon95}.  
The density and temperature distributions are determined through iteratively solving 
the equations for hydrostatic equilibrium in the vertical direction and the 
local thermal balance between the heating and cooling of the gas.  
The theoretical foundation of this model comes from the 
\emph{standard accretion disk model} of \citet{lynden74} and \citet{pringle81} which 
defines a surface density distribution for the disk given the parent star's 
mass and radius and a disk mass accretion rate, $\dot{M}$.  
The kinematic viscosity in the disk is parameterised according to the 
work of \citet{shakura73}, the so-called \emph{$\alpha$-prescription}.  
We adopt a viscous parameter, 
$\alpha$~=~0.01 and a mass accretion rate, $\dot{M}$~=~10$^{-8}$~$M_\odot$~yr$^{-1}$.  

The heating mechanisms included are grain photo-electric heating by 
far-ultraviolet photons and X-ray heating due to hydrogen ionisation by 
X-ray photons with cooling via gas-grain collisions and line transitions.  
We use a model 
spectrum created by fitting the observed XMM-Newton X-ray spectrum of the classical 
T Tauri star, TW Hya (e.g.\ \citet{kastner02}) with a two-temperature thin thermal plasma model 
(MEKAL model see e.g.\ \citet{liedahl95}).  
The X-ray luminosity is $L_{X}$~$\sim$~$10^{30}$~erg~s$^{-1}$ 
 and the resulting X-ray spectrum is given in Figure~1 of \citet{nomura07}.
  
The UV radiation field in disks has two sources, the star and the interstellar medium.  
In this disk model, the radiation field due 
to the T Tauri star has three components: black-body emission at the star's effective temperature, 
optically thin hydrogenic bremsstrahlung emission and strong Lyman-$\alpha$ line emission.  
All components are necessary to accurately model the excess UV emission observed towards 
classical T Tauri stars thought to arise from an accretion shock as 
disk material impinges upon the stellar surface 
(e.g.\ \citet{valenti00, johnskrull00}).  
The total FUV luminosity  in our model is $L_{UV}$~$\sim$~$10^{31}$~erg~s$^{-1}$ 
 with the calculation of the radiation field in the disk described in detail in 
Appendix~C of \citet{nomura05} and the resulting spectrum shown in Figure~C.1 in that paper.

We assume the dust and gas in the disk is well-mixed and adopt 
the dust-size distribution model which reproduces the observational extinction 
curve of dense clouds \citep{weingartner01}.  
 The calculation of the dust opacity in the disk is as described in 
Appendix~D of \citet{nomura05} with the 
resulting monochromatic absorption coefficient shown in Figure~D.1. 
We note here that this is an over-simplification of the treatment of the dust-size distribution in 
protoplanetary disks and we are currently working on improving our model by adding in the effects 
of dust-grain settling and coagulation.

\begin{figure*}
\subfigure{\includegraphics[width=0.5\textwidth]{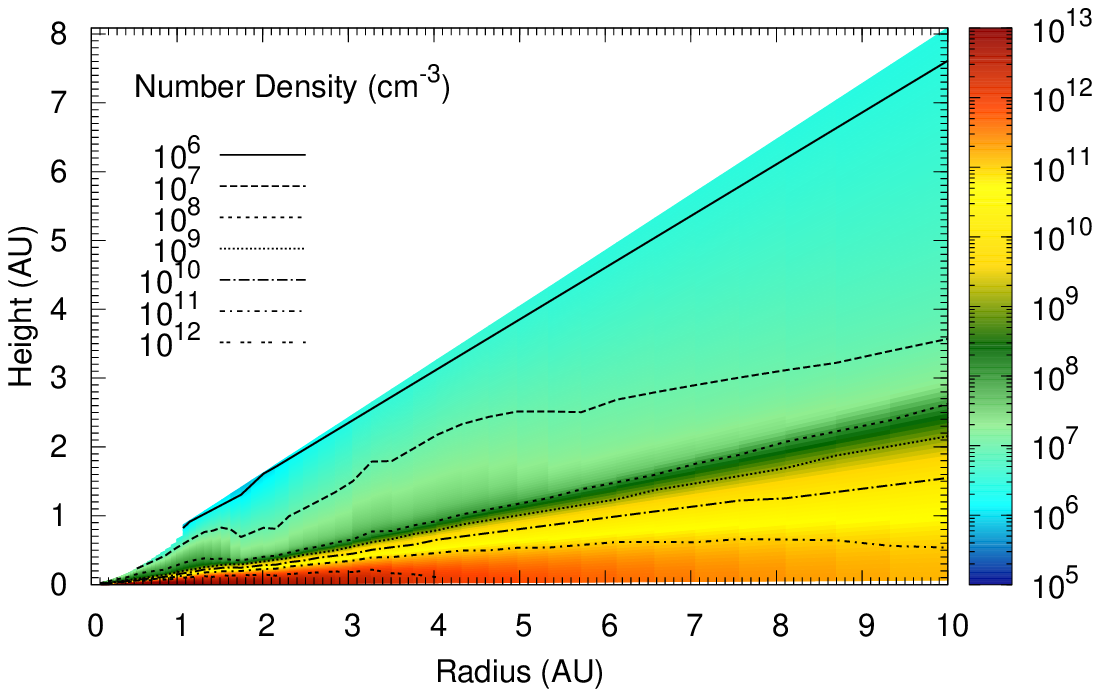}}
\subfigure{\includegraphics[width=0.5\textwidth]{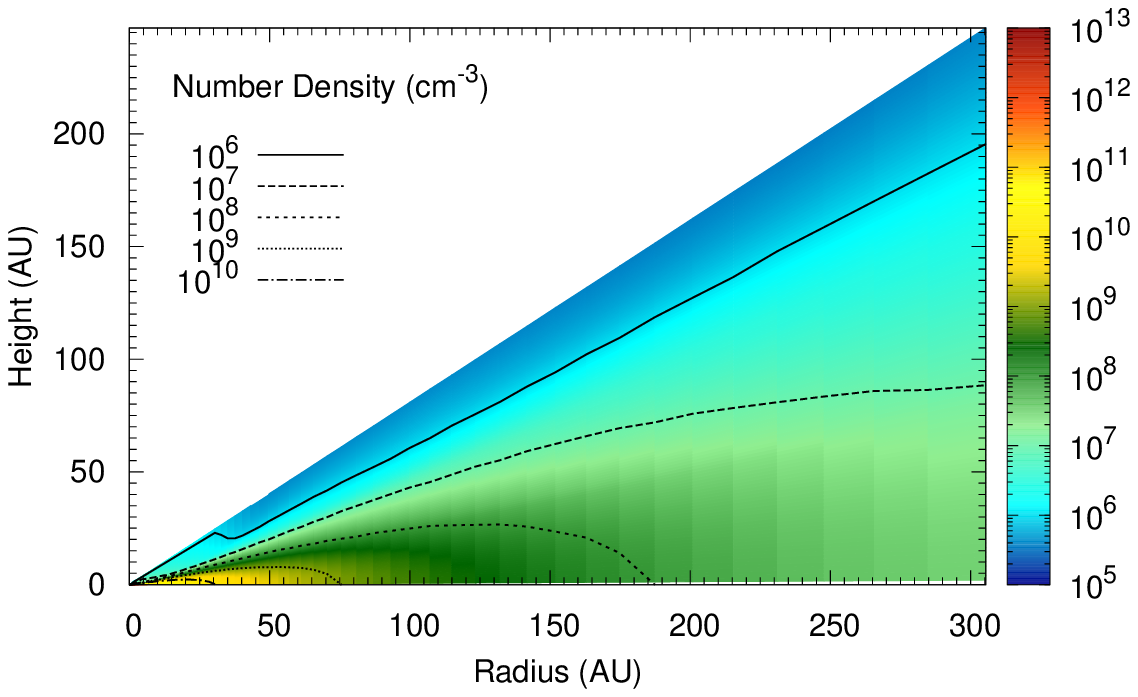}}
\subfigure{\includegraphics[width=0.5\textwidth]{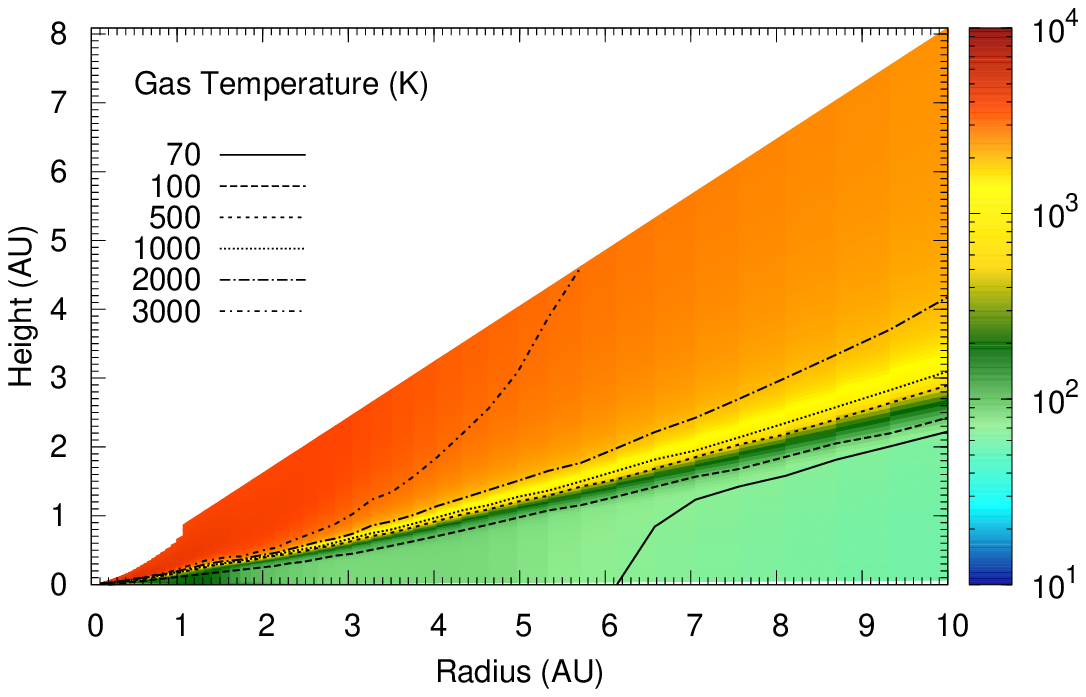}}
\subfigure{\includegraphics[width=0.5\textwidth]{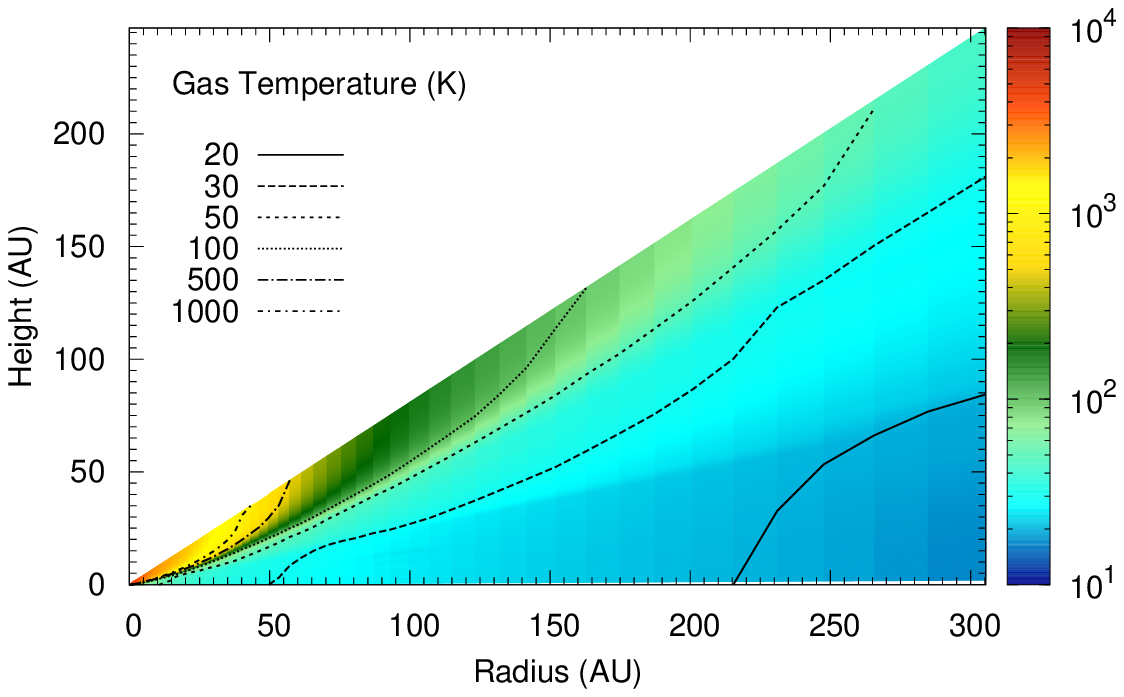}}
\subfigure{\includegraphics[width=0.5\textwidth]{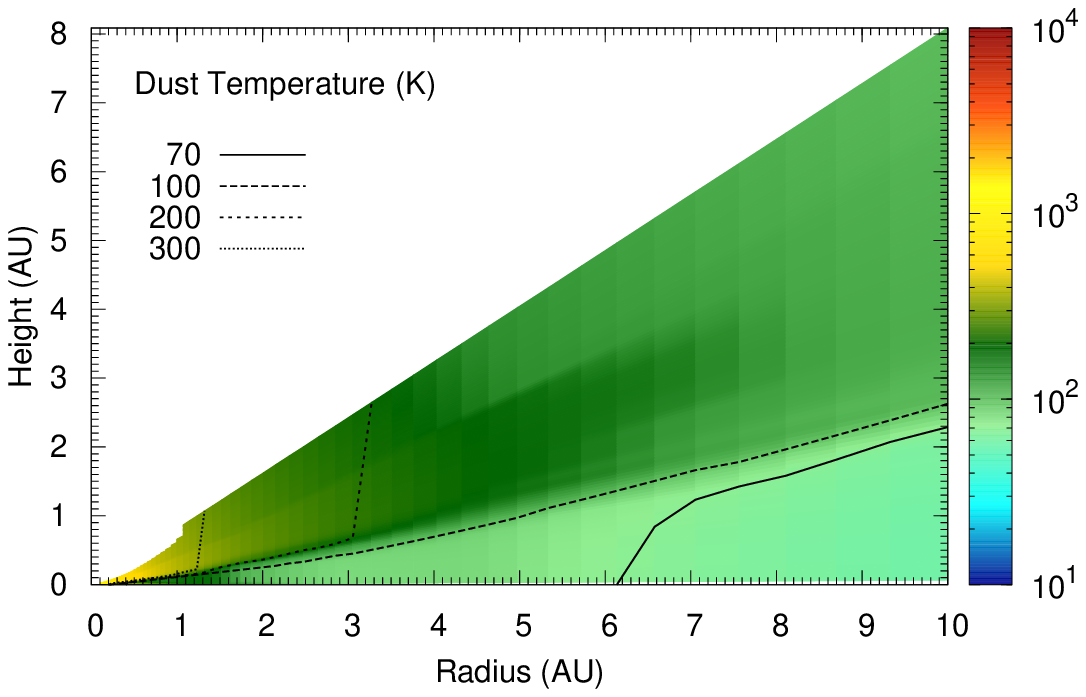}}
\subfigure{\includegraphics[width=0.5\textwidth]{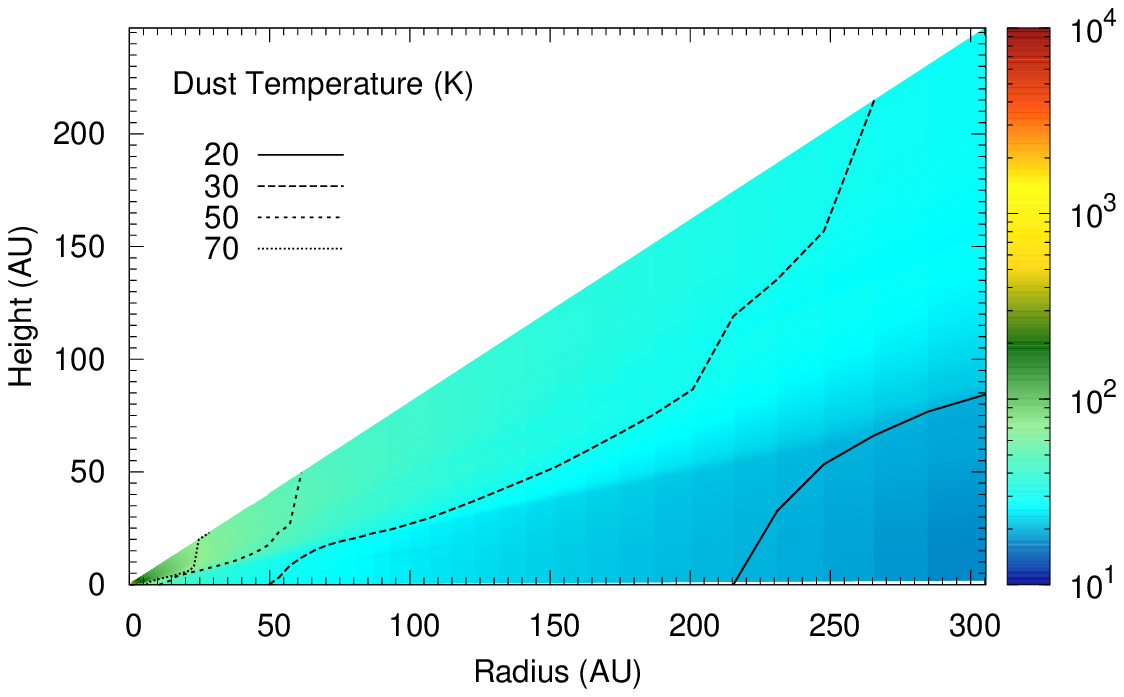}}
\caption{Number density (top),  gas temperature (middle) and  dust temperature (bottom) 
as a function of disk radius and height up to maximum radii of $r$~=~10~AU (left) and 305~AU (right).}
\label{figure1}
\end{figure*}

In Figure~\ref{figure1} we display the resulting number density (cm$^{-3}$),  gas temperature (K) 
and dust temperature (K) 
as a function of disk radius and height (top, middle and bottom rows, respectively).  
To illustrate the extreme vertical gradients in the physical conditions 
at small radii, we display the density and temperature both within 10~AU (left panels) and 
305~AU (right panels).  
We describe the physical structure of our disk in the Appendix.  

\subsection{Chemical Model}
\label{chemical model}

The structure of the disk described in the preceding section leads to 
a multitude of different physical regimes and as such, we need 
to account for every chemical process which may occur.  
The axisymmetric structure results in a cold, dense midplane where even the 
most volatile molecules are expected to freeze out onto dust grains creating 
an icy mantle and depleting the gas of molecules.  
Moving in the vertical direction, the density decreases and the temperature
increases driving the evaporation of molecules from grain surfaces 
and stimulating a rich gas-phase chemistry resulting in further molecular 
synthesis.  
Further towards the surface, the radiation fields increase in strength 
dissociating and ionising molecules into constituent radicals, atoms and ions.  
A similar stratification is expected in the radial direction as the temperature 
and density in the disk midplane both increase with decreasing distance from the 
star.  
When the midplane dust temperature reaches a value higher than the desorption temperature 
of a particular molecule, it is returned to the  gas phase.  
This point is known as the \emph{snow line} and can occur at a unique radius 
for each molecule.  
At small radii, due to the high densities found in the midplane, 
there is a significant column density of material shielding this region from 
the intense UV and X-ray fields of the star such that molecules are expected to 
survive in the midplane at radii within $\sim$~0.1~AU.  

In order to investigate the chemical structure thoroughly, we used 
a large gas-phase network supplemented with 
gas-grain interactions, including freeze out and thermal desorption.   
We considered various non-thermal desorption mechanisms, namely, cosmic-ray induced desorption, 
 photodesorption and X-ray desorption.  
To probe the efficacy of molecular synthesis on grain-surfaces 
we also added a large grain-surface reaction network.  
 
\subsubsection{Gas-Phase Chemistry}
\label{gasphasechemistry}

Our gas-phase chemistry is extracted from the latest release of the `dipole-enhanced' version 
of the UMIST Database for Astrochemistry (\url{http://www.udfa.net}), henceforth referred to as `Rate06' 
\citep{woodall07}.  
We include almost the entire Rate06 gas-phase network removing only those species 
(and thus reactions) which contain 
fluorine, F, and phosphorus, P, in order to reduce computation time.    
We deemed the loss of F- and P-containing species to have a minimal impact on the remaining chemistry.  
Our gas-phase network thus consists of 4336 reactions involving 378 species composed of the 
elements H, He, C, N, O, Na, Mg, Si, S, Cl and Fe.  
The initial elemental fractional abundances (relative to total H nuclei density) 
we use are the set of oxygen-rich low-metallicity abundances from 
\citet{graedel82}, listed in Table~8 of \citet{woodall07}.  
 We find by $10^{6}$ years, the typical age of protoplanetary disks, the chemistry has 
forgotten its origins, justifying our use of initial elemental abundances. 
We intend in future models to calculate the chemical evolution of a parcel of gas as it 
follows a streamline in the accretion flow in which case the input abundances should reflect 
the molecular make-up of the ambient cloud material.
Our model grid has over 12,000 grid points in 129 
logarithmically spaced radial steps from 0.04~AU to 305~AU.  

\subsubsection{Photochemistry}
\label{photochemistry}

 In the models presented here, we have approximated our photoreaction rates at each point in the 
disk, $k^{ph}(r,z)$, by scaling the rates from Rate06 (which assume the interstellar UV field) using 
the wavelength integrated UV flux calculated at each point, $G_{FUV}(r,z) = \int_{912\AA}^{2000\AA} G_{FUV}(\lambda,r,z) \;
\mathrm{d}\lambda$.  
Hence, the rate for a particular photoreaction at each $(r,z)$ is given by
\begin{equation}
k^{ph} = \frac{G_{FUV}}{G_0}k_0 \quad \mathrm{s}^{-1},
\end{equation}
where $G_0$ is the interstellar UV flux and $k_0$ is the rate expected in the interstellar medium.

\subsubsection{Gas-Grain Interactions}
\label{gasgraininteractions}

Gas-grain interactions are important in large areas of protoplanetary disks 
as the dust temperature can reach values lower than the freeze-out temperatures 
of molecules.  
If the  freeze out of gas-phase species is allowed, then the evaporation of molecules from 
dust grains must also be included.  
In this work, we consider both the thermal and non-thermal desorption of molecules from dust grains.  
For the thermal desorption of a particular molecule to occur, the dust-grain temperature must exceed the 
freeze-out temperature of that molecule.  
Non-thermal desorption requires an input of energy from an external source and is thus independent 
of dust-grain temperature.  
As protoplanetary disks are irradiated by UV and X-ray photons from the central star 
as well as UV photons and cosmic-rays originating from the interstellar medium,
the non-thermal desorption mechanisms we investigate are cosmic-ray induced desorption, 
 photodesorption, and X-ray desorption.  
Our gas-phase chemical network has thus been supplemented with an additional 
1154 gas-grain interactions involving 149 surface species. 

 The accretion rate (or freeze-out rate), $k_i^a$, of species $i$ onto dust-grain surfaces is treated 
using the standard prescription \citep{hasegawa92},
\begin{equation}
k_i^a = S_i \sigma_d \left< v_i \right> n_d \quad \mathrm{s}^{-1},
\label{accretionrate}
\end{equation}
where $S_i$ is the sticking coefficient, here assumed to equal unity for all species, 
$\sigma_d = \pi a^2$ is the geometrical cross section of a dust grain with radius, $a$, 
$<v_i> = (k_BT/m_i)^{1/2}$ is the thermal velocity of species $i$ at a temperature, 
$T$ and with mass, $m_i$, $k_B$ is Boltzmann's constant, 
and $n_d$ is the number density of dust grains.  

The thermal desorption rate, $k_i^d$, of species $i$ is dependent on dust-grain temperature, 
$T_d$ \citep{hasegawa92}, and is given by
\begin{equation}
k_i^d = \nu_0(i) \exp \left( \frac{-E_d (i)}{T_d}\right) \quad \mathrm{s}^{-1},
\label{thermaldesorption}
\end{equation}
where $E_d(i)$ is the binding energy of species $i$ to the dust grain in units of K.  
The characteristic vibrational frequency of each adsorbed species in its potential well, 
$\nu_0 (i)$, is represented by a harmonic oscillator relation \citep{hasegawa92}, 
\begin{equation}
\nu_0(i) = \sqrt{\frac{2n_s E_d(i)}{\pi^2 m_i}} \quad \mathrm{s}^{-1},
\label{vibrationalfrequency}
\end{equation}
where, here, $E_d(i)$ is in units of erg and $n_s = 1.5 \times 10^{15}$~cm$^{-2}$ is the number
density of surface sites on each dust grain.   
The binding energies, $E_d$, for several important molecules (mainly following those collated by
\citet{hasegawa92} and \citet{willacy98}) are listed in Table~\ref{table1}.  
We intend to conduct a review of our set of desorption energies in light of 
more recent experimental results  e.g.\ a recent investigation into the desorption of 
methanol by \citet{brown07} determined a binding energy of $\approx$~5000~K, as opposed to  
the theoretical value of 2140 K used here (see Table~\ref{table1}).  
This binding energy was determined for pure methanol ice as opposed to methanol adsorbed 
onto, or mixed with, water ice.  
We find throughout our model that the ratio of methanol to water ice is less than 1\%.  
Similar experiments for methanol adsorbed onto water ice (Brown, private communication) 
suggest that the binding energy of methanol in this complex is comparable with that determined for 
pure methanol but due to overlapping desorption features the results are difficult to analyse.  
Recent work by \citet{bottinelli10} comparing laboratory data with observations of methanol in young stellar 
objects (YSOs) 
suggests that methanol ice in these environments likely exists as pure ice or mixed with CO and/or CO$_2$ ice 
which is consistent with its formation via hydrogenation of CO on dust grains.  
Considering the latter molecules are non-polar, it is possible that the binding energy of methanol in
astrophysical ices is lower than that determined in the laboratory experiments.  
Increasing the binding energy of methanol to a value of $\approx$~5000~K 
will increase the desorption temperature from $\approx$~30~K to 40~K to $\sim$~100~K.  
This will push the `snow line' for methanol closer to the star but should have little effect on the 
outer disk methanol abundances where the dust temperature is $<$~30~K.  
We expect non-thermal desorption to dominate over thermal desorption in the upper layers of the disk.  

\begin{deluxetable}{lcc}
\tablecaption{Molecular Binding Energies \label{table1}}
\tablewidth{0pt}
\tablehead{\colhead{Molecule} & \colhead{Binding Energy (K)} & \colhead{Reference}}
\startdata
CO         & 960  & 1\\
N$_2$      & 710  & 2\\
HCN        & 4170 & 2\\
CO$_2$     & 2690 & 3\\
H$_2$O     & 4820 & 4\\
NH$_3$     & 3080 & 4\\
CH$_4$     & 1080 & 2\\
C$_2$H$_2$ & 2400 & 2\\
H$_2$CO    & 1760 & 5\\
CH$_3$OH   & 2140 & 5 
\enddata
\tablerefs{(1) \citet{sandford88}, (2) \citet{yamamoto83}, (3) \citet{sandford90}, 
(4) \citet{sandford93}, (5) \citet{hasegawa93}}
\end{deluxetable}

To calculate the cosmic-ray induced desorption rate for each species, $k_i^{crd}$, 
we use the method of \citet{leger85} and \citet{hasegawa93}.  
 They assume that dust grains with a radius of 0.1~$\mu$m are impulsively heated by the impact of 
relativistic Fe nuclei with energies of 20 to 70 MeV nucleon$^{-1}$ which deposit, on average, 
an energy of 0.4~MeV into each dust grain.  
Assuming that the majority of molecules desorb around 70~K, the cosmic-ray induced desorption 
rate can be approximated by
\begin{equation}
k_i^{crd} \approx f(70\;\mathrm{K})k_i^d(70\;\mathrm{K}) \quad \mathrm{s}^{-1},
\label{cosmicraydesorption}
\end{equation}
where $k_i^d(70\;\mathrm{K})$ is the the thermal desorption energy of species $i$ at a temperature 
of 70~K, calculated using Equation~(\ref{thermaldesorption}).  
The parameter, $f(70\;\mathrm{K})$, is the fraction of time spent by grains in the vicinity of 70~K 
and can loosely defined as the ratio of the desorption cooling time ($\approx 10^{-5}$~s$^{-1}$) to 
the time interval between successive heatings to 70~K (3.16~$\times 10^{13}$~s) so that 
$f(70\;\mathrm{K})\approx 3.16 \times 10^{-19}$ (for further details see \citet{hasegawa93}).  

Note that the method of calculating the cosmic-ray induced desorption rates is species  
dependent and a function of surface binding energy.  
In contrast, the  photodesorption rates are indiscriminate, based on the 
experimental results of \citet{westley95} and \citet{oberg07}.  
Their results suggest each photon absorbed by the grain mantle returns a particular 
number of molecules, independent of binding energy, to the  gas phase
so that the desorption rate of each species varies according to its fractional abundance on 
dust-grain surfaces.  
The overall  photodesorption rate is calculated, similar to the work of \citet{willacy00} and 
\citet{willacy07}, using
\begin{equation}
k^{pd} = F_{UV}Y_{UV}\sigma_{d}x_{d} \quad \mathrm{s}^{-1},
\label{photodesorption1}
\end{equation}
where, $F_{UV}$ is the UV radiative flux in units of photons~cm$^{-2}$~s$^{-1}$, $Y_{UV}$ 
is the experimentally determined  photodesorption yield in units of molecules photon$^{-1}$, 
$\sigma_{d}$ is the geometrical dust-grain cross section in cm$^2$ and $x_{d}$ is the 
fractional abundance of dust grains.  
 Note that the attenuation of UV radiation is accounted for in our calculation of $F_{UV}$. 
Hence, the  photodesorption rate for a specific species, $k_i^{pd}$, is calculated using
\begin{equation}
k_i^{pd} = k^{pd} \frac{n_i^s}{n_{tot}^s} \quad \mathrm{s}^{-1},
\label{photodesorption2}
\end{equation}
where, $k^{pd}$ is given by Equation~(\ref{photodesorption1}), $n_i^s$ is the number density 
of species $i$ frozen out onto grain surfaces and $n_{tot}^s$ is the total 
number density of grain-surface species.  
More recent experiments by \citet{oberg09a,oberg09b} suggest that 
 photodesorption rates are also dependent on the depth of the ice layer on grain 
surfaces with the molecular yield also dependent on ice composition.    
We intend to explore these experimental results in future models.  

For the X-ray desorption rates, we follow the same formulation as for  photodesorption, 
covered in the theory of \citet{leger85} and \citet{najita01}.  
At this point, it is worth noting that X-ray desorption is the least 
theoretically or experimentally 
constrained of all the non-thermal desorption mechanisms considered here.  
The overall X-ray desorption rate, $k^{xr}$, is given by 
\begin{equation}
k^{xr} = F_{XR}Y_{XR}P_{abs}\sigma_d x_d \quad \mathrm{s}^{-1},
\label{xraydesorption}
\end{equation} 
where, $F_{XR}$ is the X-ray photon flux in units of photons~cm$^{-2}$~s$^{-1}$, 
$Y_{XR}$ is the desorption yield in units of molecules~photon$^{-1}$ and the 
product, $P_{abs}\sigma_{d}$, is the effective  cross section with 
$P_{abs}$, the probability of X-ray absorption by the dust grain. 
The X-ray desorption rate for each 
individual species, $k_i^{xr}$, is calculated according to 
the fractional abundance of species $i$ on the dust grains  
following Equation~(\ref{photodesorption2}).
Here, we adopt a value $Y_{XR} = 200$ from the investigations of 
\citet{najita01} and for the effective grain  cross section we 
use values from the work of \citet{dwek96} regarding energy deposition into grains 
by energetic photons in the energy range 10~eV to 1~MeV.  
\citet{najita01} consider X-ray desorption from grains of 
various compositions and morphologies and conclude that both have a significant influence
on the X-ray desorption yields calculating values for 
$Y_{XR}$ ranging between 10 and $\approx$~4000 molecules photon$^{-1}$.  
In this work, we adopt a conservative estimate of the yield 
of 200 molecules photon$^{-1}$ as this is the value for the dust morphology 
which most closely matches our simple dust-grain model.  
Given the large X-ray luminosities of T Tauri stars (see e.g.\ \citet{kastner97,kastner02}), we plan 
a more thorough study on 
the effects of X-ray desorption in protoplanetary disks taking into consideration the 
X-ray energy spectrum as a function of disk radius and height and  
investigating the full parameter space considered in the work of \citet{najita01}.  
 
\subsubsection{Grain-Surface Chemistry}
\label{grainsurfacechemistry}

We use the grain-surface network from \citet{hasegawa92} and \citet{hasegawa93} 
which has 221 reactions involving an additional 9 surface species which do not have 
a gas-phase equivalent (e.g.\ CH$_3$O).  
To calculate the reaction rate coefficients, we use the theory outlined
in detail in \citet{hasegawa92}.  
The rate coefficient for a grain-surface reaction between species $i$ and $j$ can be defined as
\begin{equation}
k_{ij} = \kappa_{ij}\left( R_{diff}(i) + R_{diff}(j)\right) \left( 1/n_d \right) \quad \mathrm{cm}^{3}\;\mathrm{s}^{-1}.
\label{grainsurface1}
\end{equation}
Here, $\kappa_{ij}$ is the probability that the reaction happens upon encounter and is equal to unity 
for an exothermic reaction without an energy barrier.  
For reactions with an activation energy, $E_A$, and at least one light reactant 
i.e.\ H or H$_2$, $\kappa_{ij} = \exp(-2b/\hbar\sqrt{2\mu E_A})$ where 
$b$ is the barrier thickness and $\mu = m_im_j/(m_i+m_j)$ is the reduced mass 
of the reaction system.  
This expression is the exponential part of the quantum mechanical probability for 
tunneling through a rectangular barrier of thickness, $b$.  
The term, $R_{diff}$, is the diffusion rate of an adsorbed species and 
is the inverse of the diffusion time, $t_{diff}$, defined as $t_{diff} = N_s t_{hop}$ s, 
where $N_s$ is the total number of surface sites per dust grain and $t_{hop}$ is the timescale 
for an adsorbed species to `hop' from one surface site to another.  
The expression for $t_{hop}$ depends on the mass of the species and is given by 
\begin{equation}
t_{hop} = 
	\begin{cases}
        {\nu_0(i)}^{-1} \exp\left( \frac{2b}{\hbar}\sqrt{2m_iE_b(i)}\right) \; \mathrm{s} \;  
        & \; \mbox{H/H$_2$} \\       
        {\nu_0(i)}^{-1} \exp\left( \frac{E_{b}(i)}{kT_d} \right) \; \mathrm{s} 
        & \; \mbox{other species}  
        \end{cases}
\label{grainsurface2}
\end{equation}
where $E_b(i)\approx 0.3 E_d(i)$ is the energy barrier between surface sites.  
All other parameters have been defined previously.

\section{RESULTS}
\label{results}

We calculate the chemical abundances in the disk as a function of disk radius, 
height and time.  
The results displayed here are extracted at a time of 10$^6$~yr, the 
typical age of visible T Tauri stars with accompanying protoplanetary disks.  
Throughout this section, fractional abundance refers to the abundance of each species 
with respect to total particle number density.  
In Section~\ref{chemicalstructure}, we display and discuss results from 
model PH+CRH only, to illustrate the global chemical structure and 
stratification in the disk.  
 Table~\ref{table2} lists the names and ingredients of each model for which we present results.  
Our `fiducial' model is model PH+CRH since most current chemical models 
of protoplanetary disks include photodesorption and cosmic-ray induced desorption by default.  
In model CRH we remove photodesorption to investigate the influence of cosmic-ray induced desorption,
in model XD we look at the effects of X-ray desorption only, and in model GR, we investigate the 
addition of grain-surface chemistry to our fiducial model.  
Of course, there are many more permutations of the ingredients 
which are worthwhile considering in the future e.g. X-ray desorption plus grain-surface
chemistry.

\begin{deluxetable*}{lccccc}
\tablecaption{Chemical Models \label{table2}}
\tablewidth{0pt}
\tablehead{&\colhead{0}&\colhead{CRH}&\colhead{PH+CRH}&\colhead{XD}&\colhead{GR}}
\startdata
Thermal desorption & \checkmark  & \checkmark & \checkmark & \checkmark & \checkmark  \\
Cosmic-ray induced desorption  & & \checkmark & \checkmark &            & \checkmark  \\
 Photodesorption         & &            & \checkmark &            & \checkmark  \\
X-ray desorption               & &            &            & \checkmark &             \\
Grain-surface chemistry        & &            &            &            &  \checkmark
\enddata
\end{deluxetable*}

\subsection{Chemical Structure}
\label{chemicalstructure}

Figure~\ref{figure2} displays the fractional abundances of those molecules observed in disks 
(CO, HCO$^+$, HCN, CN, CS, C$_2$H, H$_2$CO and N$_2$H$^+$)  
as a function of disk radius and height, up to maximum radii of 10~AU (left column) and 305~AU 
(right column).

The global abundance distribution of molecules is governed by the binding energy of each 
molecule to dust grains and the UV radiation field strength.  
We see most molecules existing predominantly in a molecular layer of varying thicknesses 
at a height, $z/r$~$\approx$ 0.3 to 0.5 in the outer disk and $\approx$~0.2 to 0.3 in the inner disk 
with  freeze out causing depletion 
in the  midplane and photolysis causing destruction in the upper layers.  
CO is an exception to this and is abundant 
(x(CO)~$\approx$~$10^{-4}$) throughout the majority of the 
depth of the outer disk ($>$~50~AU) with depletion due to  freeze out 
in the disk  midplane only occurring beyond a radius of $\approx$~250~AU.   
In the inner disk ($r$~$<$~50~AU), gas-phase CO is abundant in the disk  midplane due to its 
low binding energy to the dust grains.  
In this region, however, we see most molecules confined to the `molecular layer'.    
In the outer disk HCO$^+$ has a peak fractional abundance of 
$\sim$~10$^{-10}$ to $\sim$~$10^{-9}$ throughout most of the disk, mirroring the 
distribution of CO.
Within $\approx$~50~AU, it is confined to a thin layer at a 
height $z/r$~$\approx$~0.3 with x(HCO$^{+}$)~$\sim$~10$^{-6}$ which coincides 
with the transition zone where the gas composition changes from molecular 
to atomic hydrogen.    

Gas-phase HCN has a peak fractional abundance of x(HCN)~$\sim$~$10^{-7}$ 
existing in the molecular layer throughout the disk.   
HCN can remain frozen out onto dust grains within radii $\approx$~1~AU 
of the parent star demonstrating the effects of the vastly different desorption energies 
of CO and HCN (960~K and 4170~K, respectively). 
The distribution of CN is complementary to that of HCN as it is 
predominantly formed via the  photodissociation of the latter molecule.  
Hence, throughout the disk, CN exists in a layer above that of 
HCN with a fractional abundance $\sim$~10$^{-6}$.  
In the outer disk, CN can survive in the surface region, however, the 
increasing UV field strength in the inner disk means that 
CN is also destroyed by  photodissociation in the disk surface.  

The distribution of the radicals, CS and C$_2$H are similar to that of CN 
since both are formed predominantly via the UV photolysis of larger precursor 
molecules (e.g.\ H$_2$CS and C$_2$H$_2$).  
H$_2$CO reaches its maximum fractional abundance (x(H$_2$CO)~$\sim$~$10^{-8}$) in the outer disk,    
although, within 10~AU, this value is reduced to $\sim$~$10^{-10}$ 
and H$_2$CO is confined to the molecular layer.    
H$_2$CO is returned to the  gas phase in the disk  midplane within $\approx$~1~AU.  

The fractional abundance distribution for N$_2$H$^+$ is different to 
any of the molecules considered thus far.  
N$_2$H$^+$ reaches its maximum fractional abundance of $\approx$~$10^{-10}$ in the outer 
disk only and is present where gas-phase CO is depleted e.g.\ 
in the disk  midplane beyond a radius of 250~AU.  
In dense regions, the main destruction mechanism of N$_2$H$^+$ is via reaction 
with CO.  
In the upper layers, N$_2$H$^+$ increases in abundance due to the increased 
abundance of both N$_2$ and cations, such as, H$_3$$^+$.  
Within 10~AU, x(N$_2$H$^+$) remains less than $\approx$~10$^{-13}$.   

\begin{figure*}
\subfigure{\includegraphics[width=0.5\textwidth]{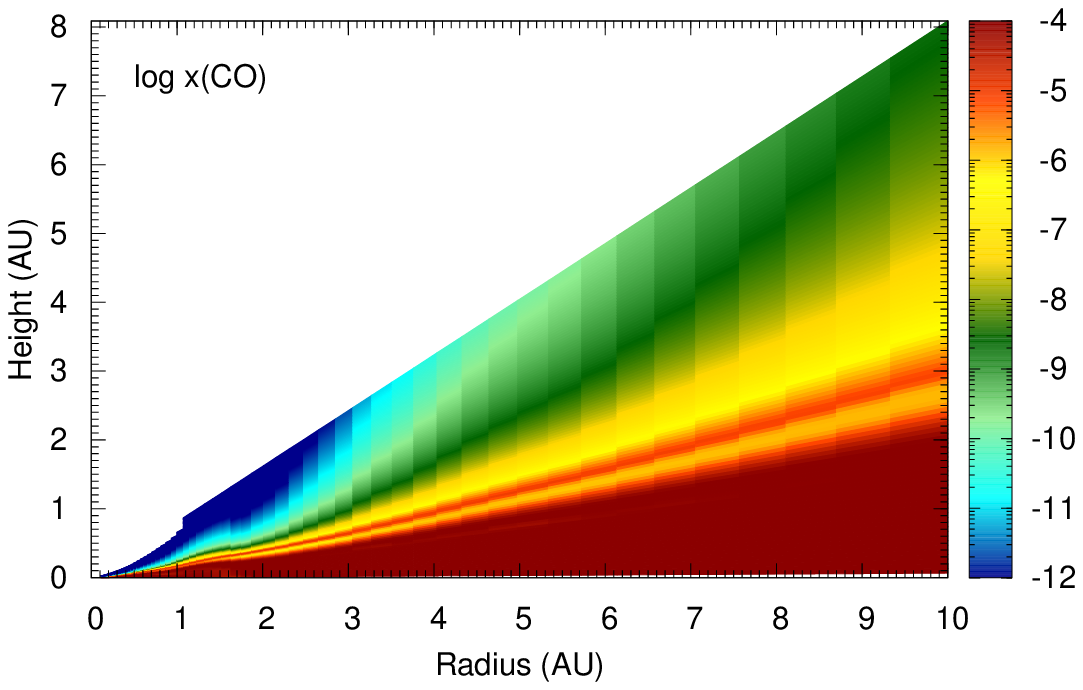}}
\subfigure{\includegraphics[width=0.5\textwidth]{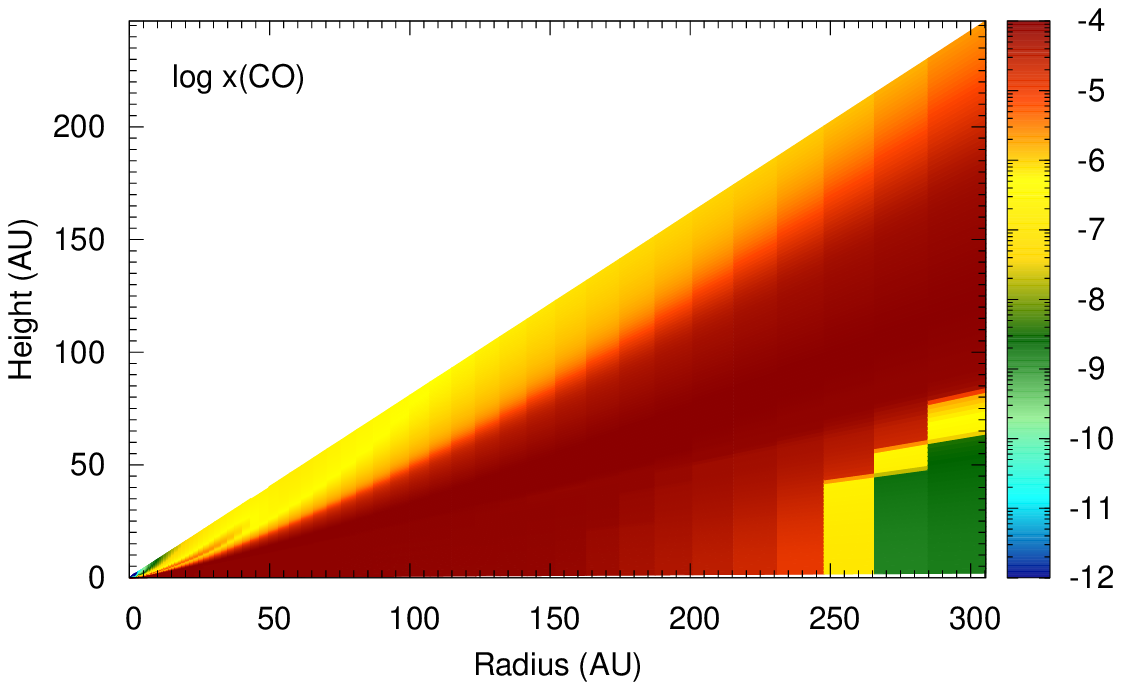}}
\subfigure{\includegraphics[width=0.5\textwidth]{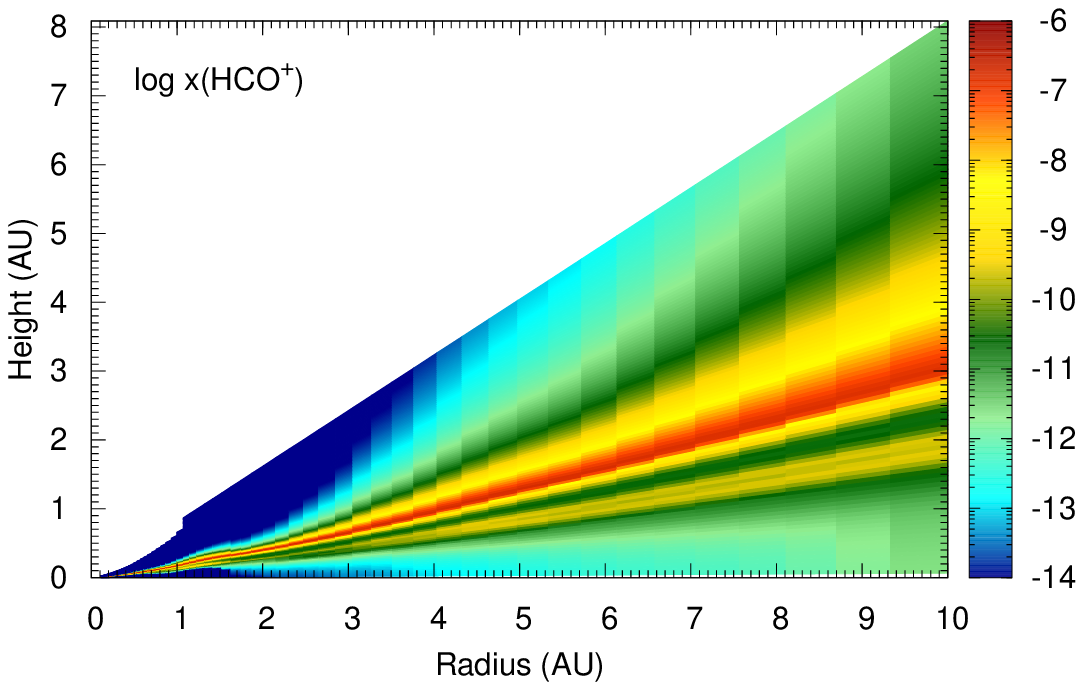}}
\subfigure{\includegraphics[width=0.5\textwidth]{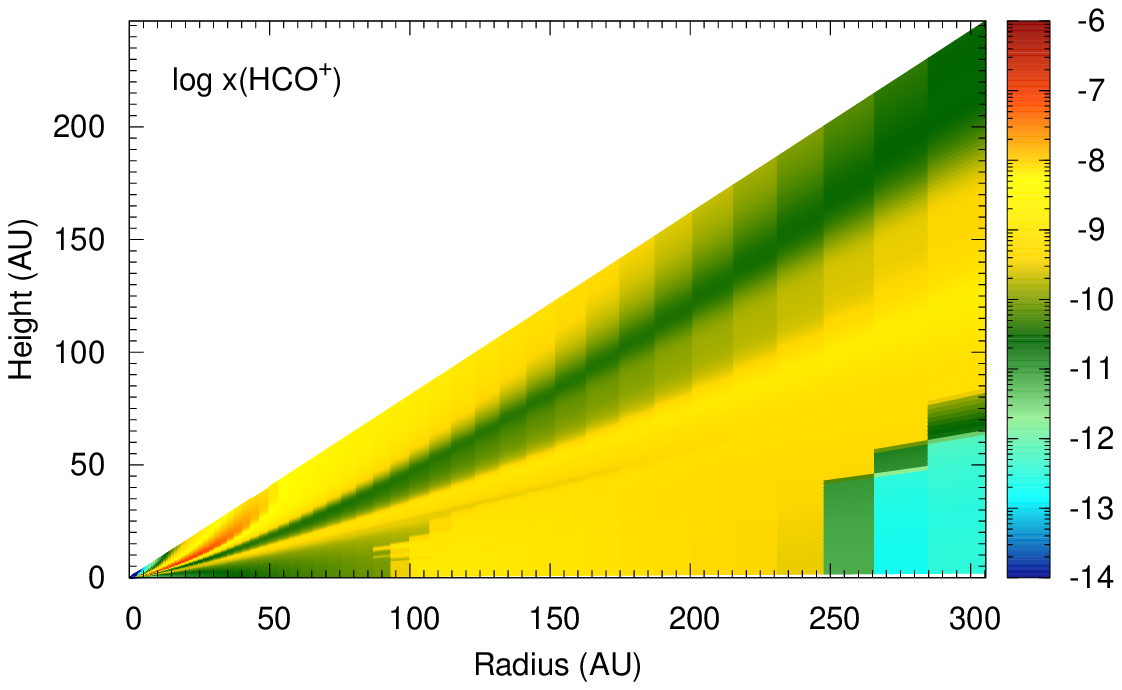}}
\subfigure{\includegraphics[width=0.5\textwidth]{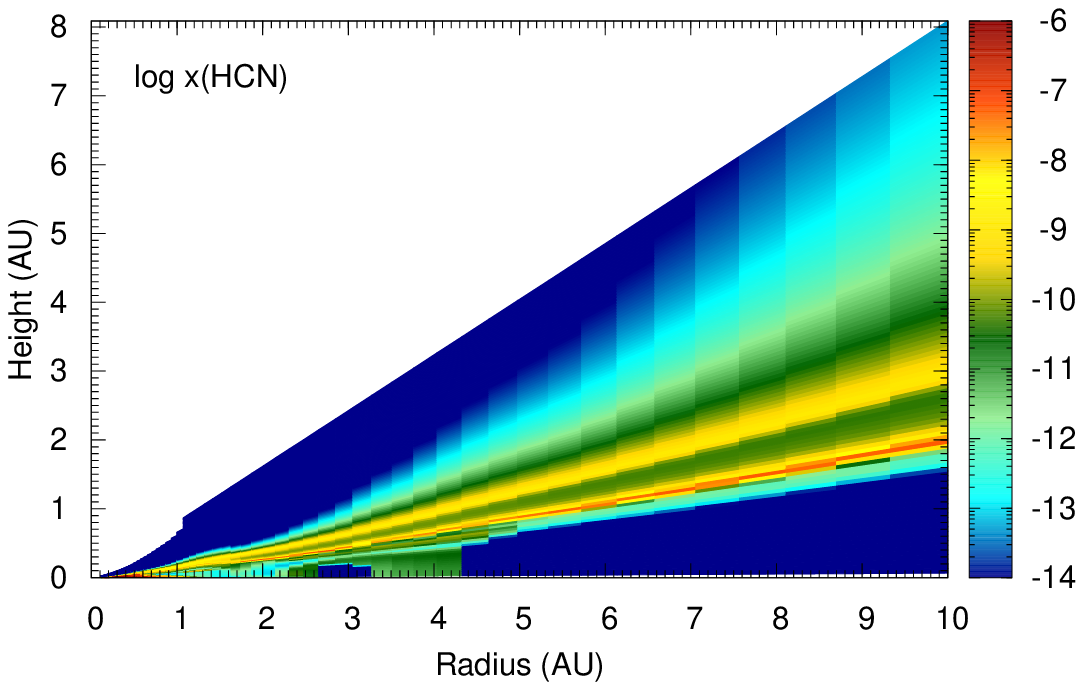}}
\subfigure{\includegraphics[width=0.5\textwidth]{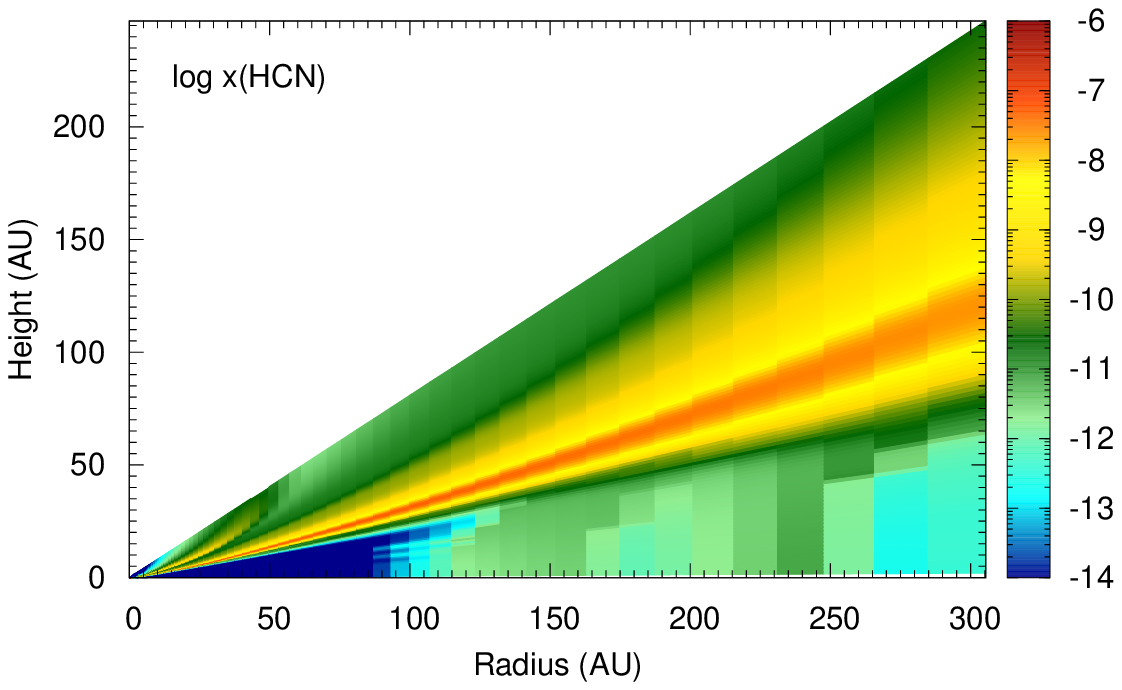}}
\subfigure{\includegraphics[width=0.5\textwidth]{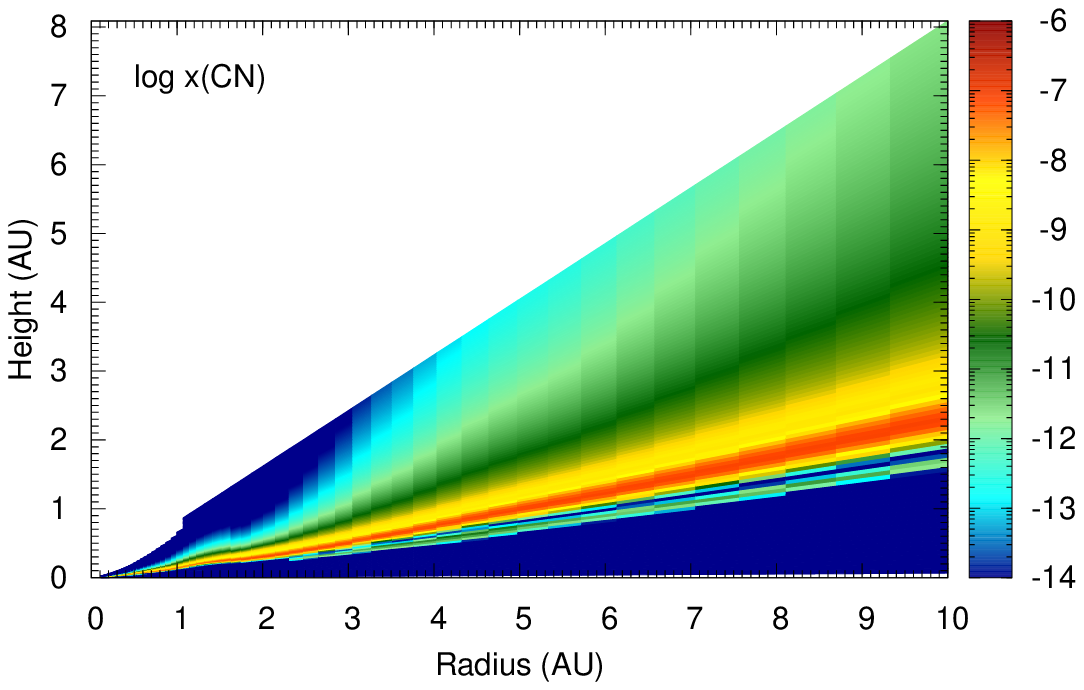}}
\subfigure{\includegraphics[width=0.5\textwidth]{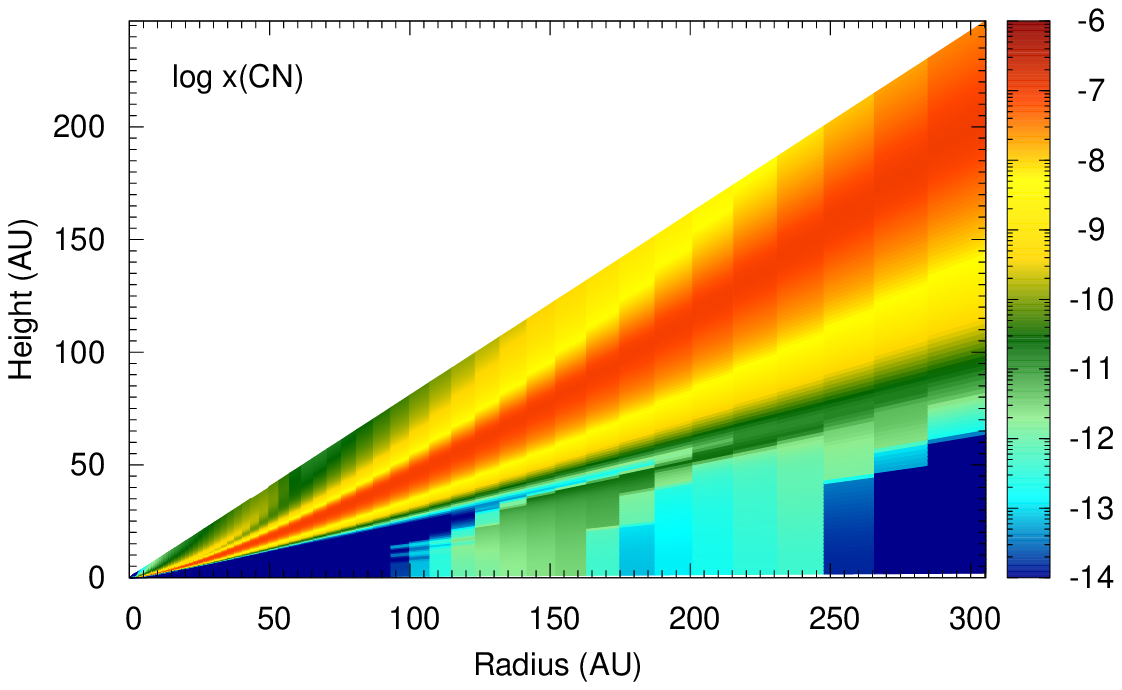}}
\captcont{Fractional abundances of  several molecules observed in disks as a function of 
disk radius and height up to maximum radii of 10~AU (left) and 305~AU (right).  }
\label{figure2}
\end{figure*}
\begin{figure*}
\subfigure{\includegraphics[width=0.5\textwidth]{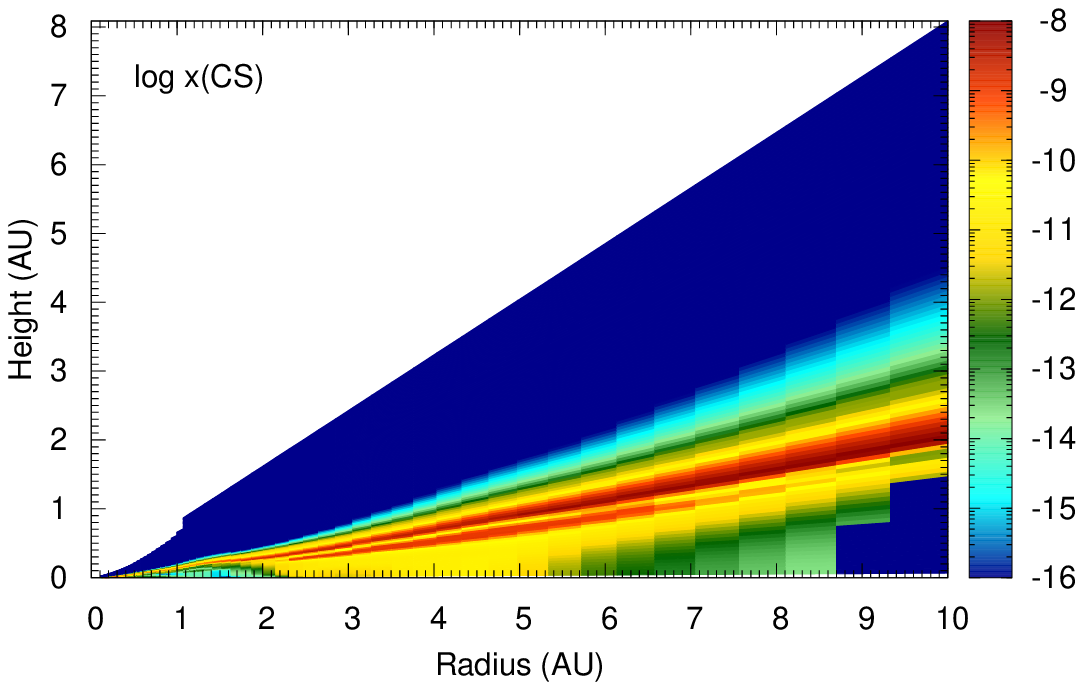}}
\subfigure{\includegraphics[width=0.5\textwidth]{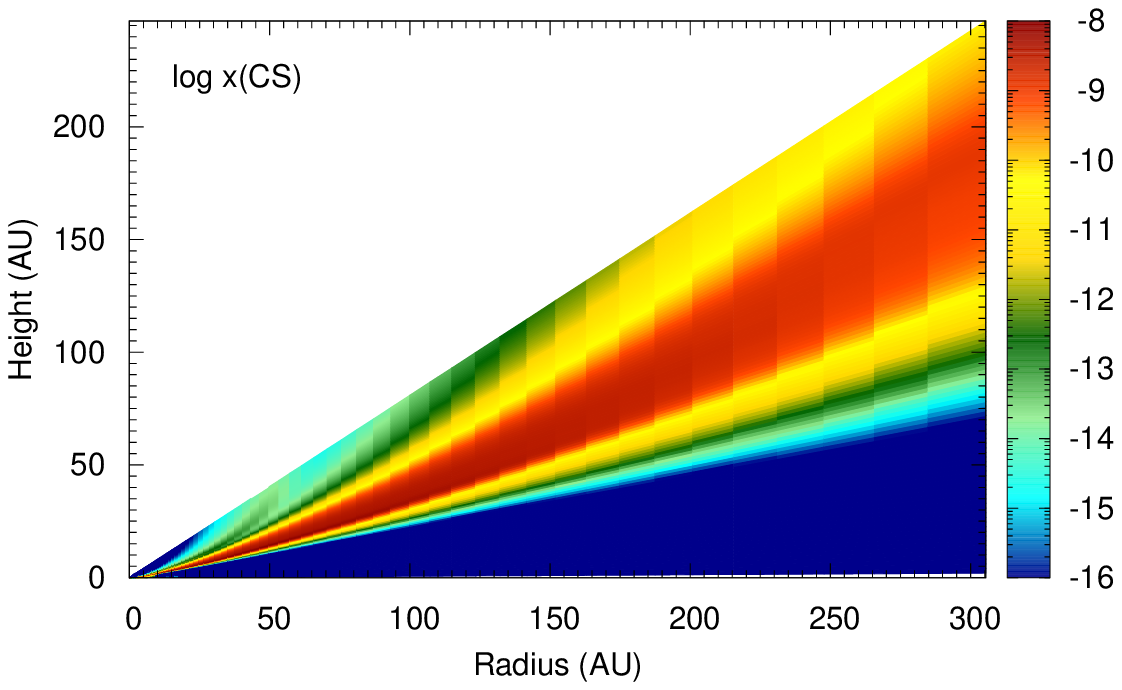}}
\subfigure{\includegraphics[width=0.5\textwidth]{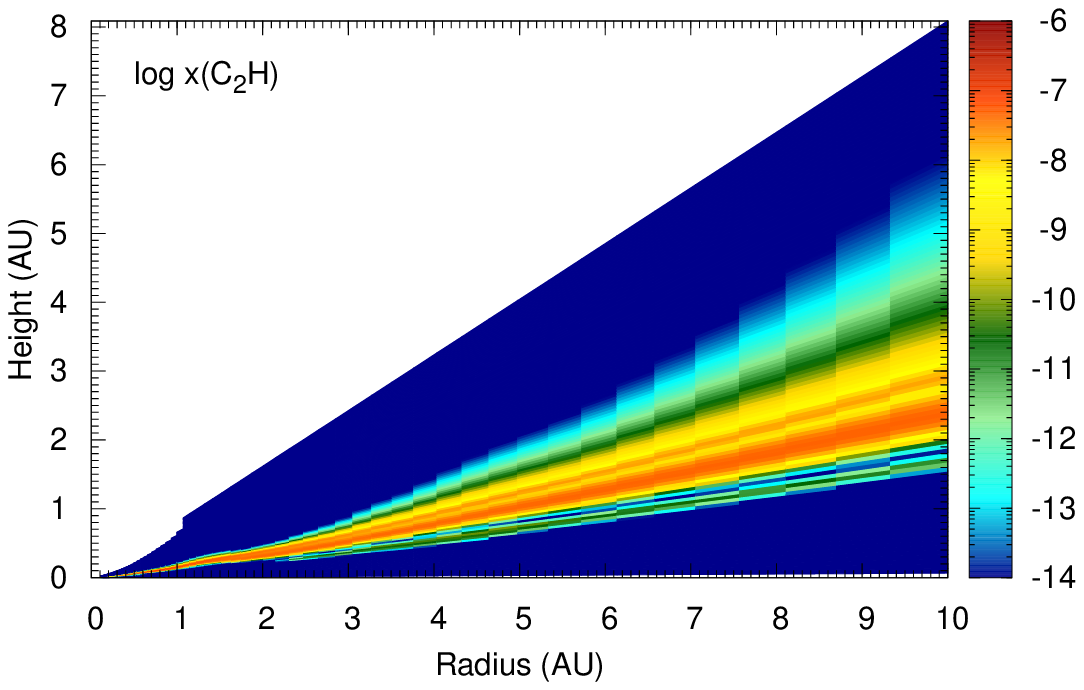}}
\subfigure{\includegraphics[width=0.5\textwidth]{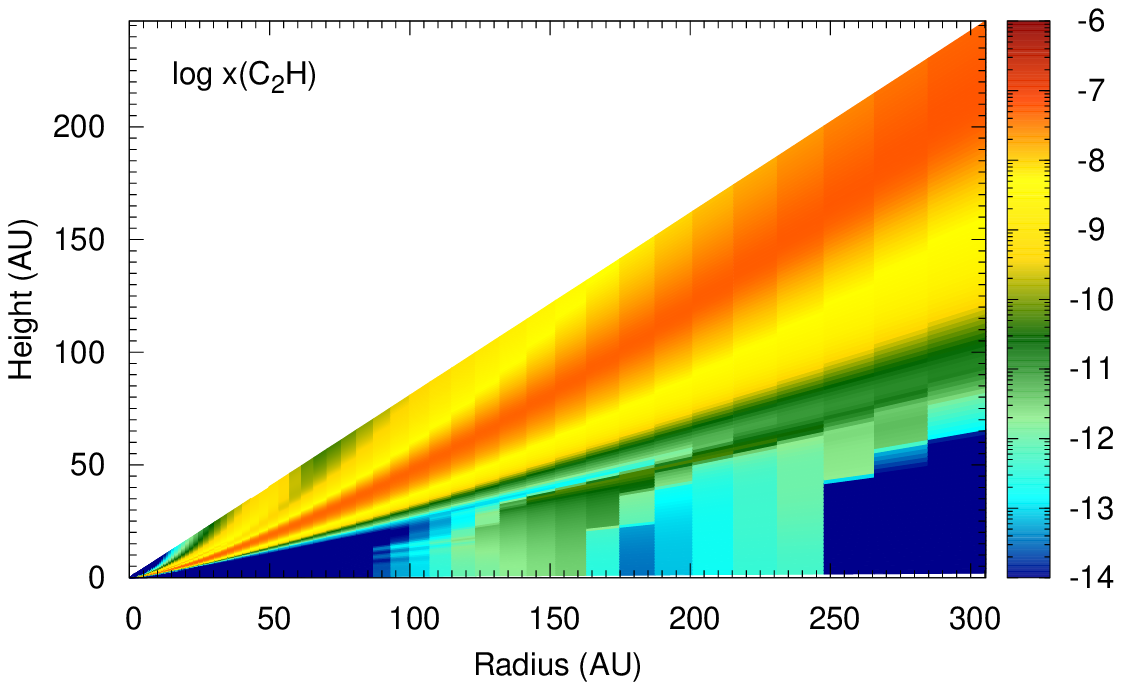}}
\subfigure{\includegraphics[width=0.5\textwidth]{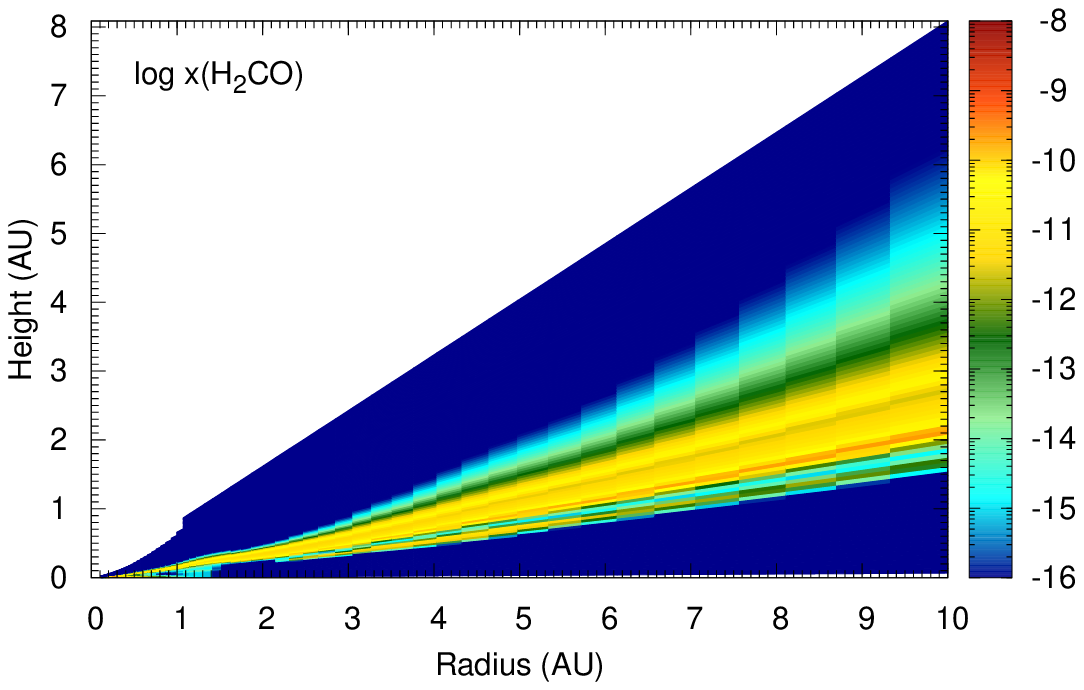}}
\subfigure{\includegraphics[width=0.5\textwidth]{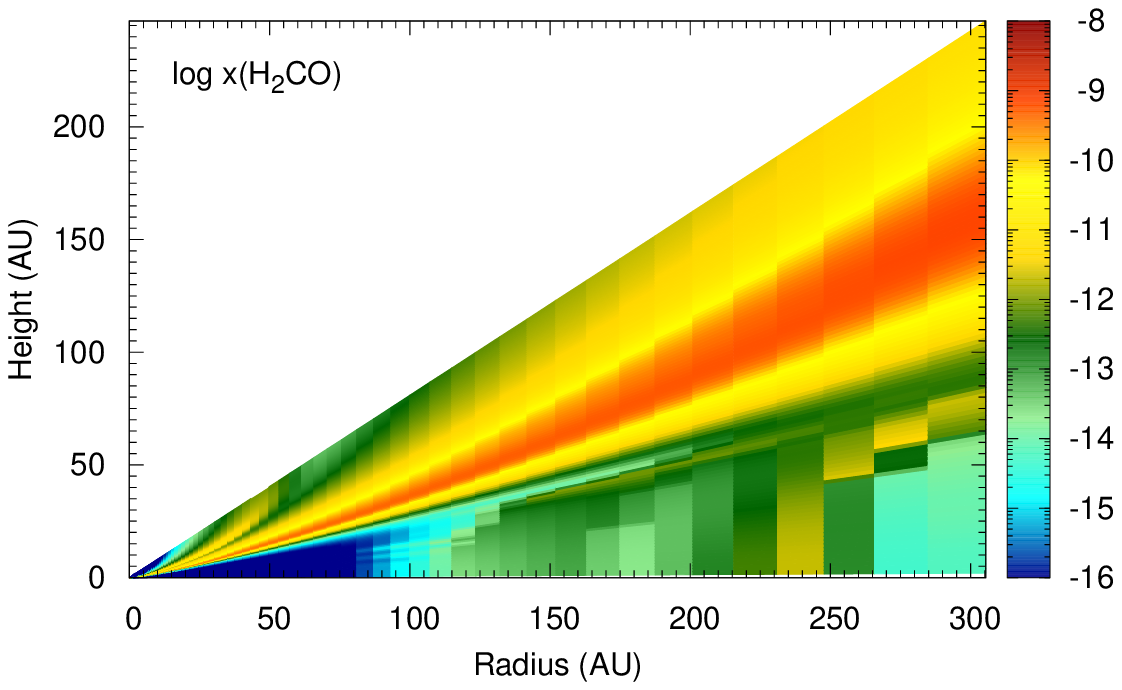}}
\subfigure{\includegraphics[width=0.5\textwidth]{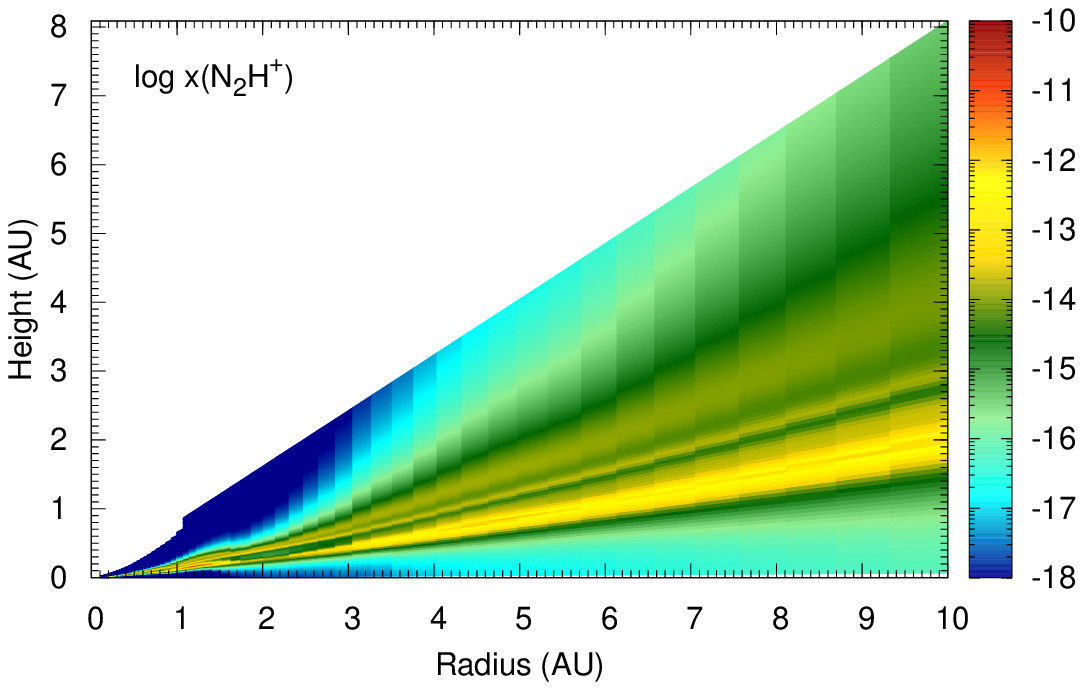}}
\subfigure{\includegraphics[width=0.5\textwidth]{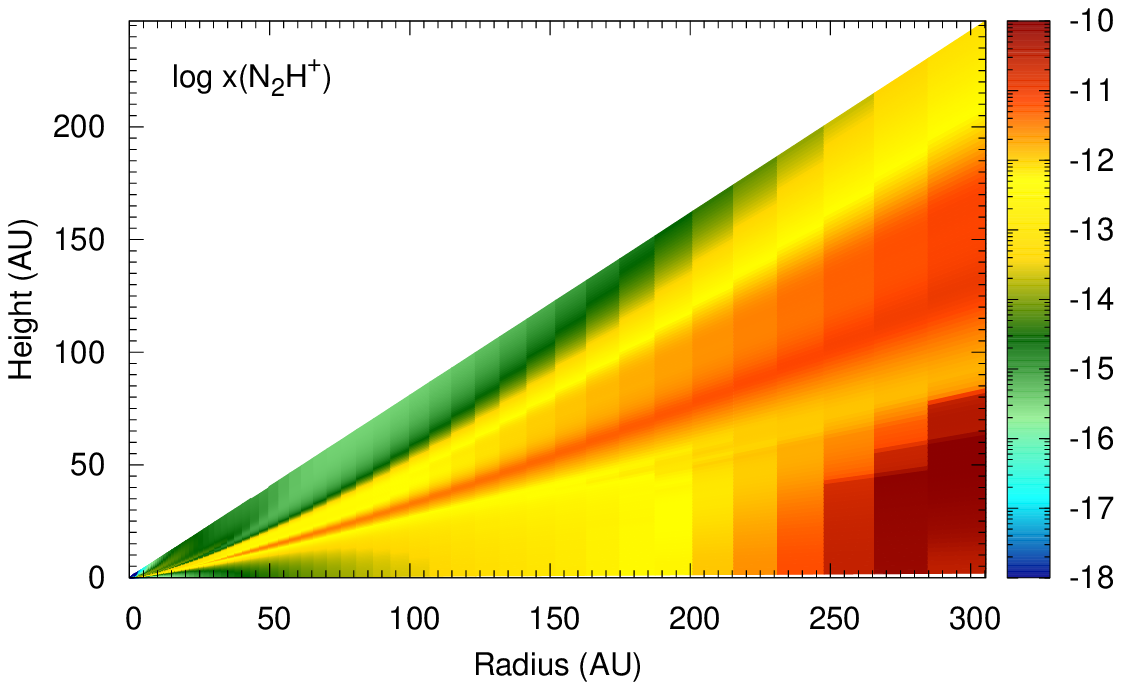}}
\caption{(Continued.)}
\end{figure*}

\begin{figure*}
\subfigure{\includegraphics[width=0.5\textwidth]{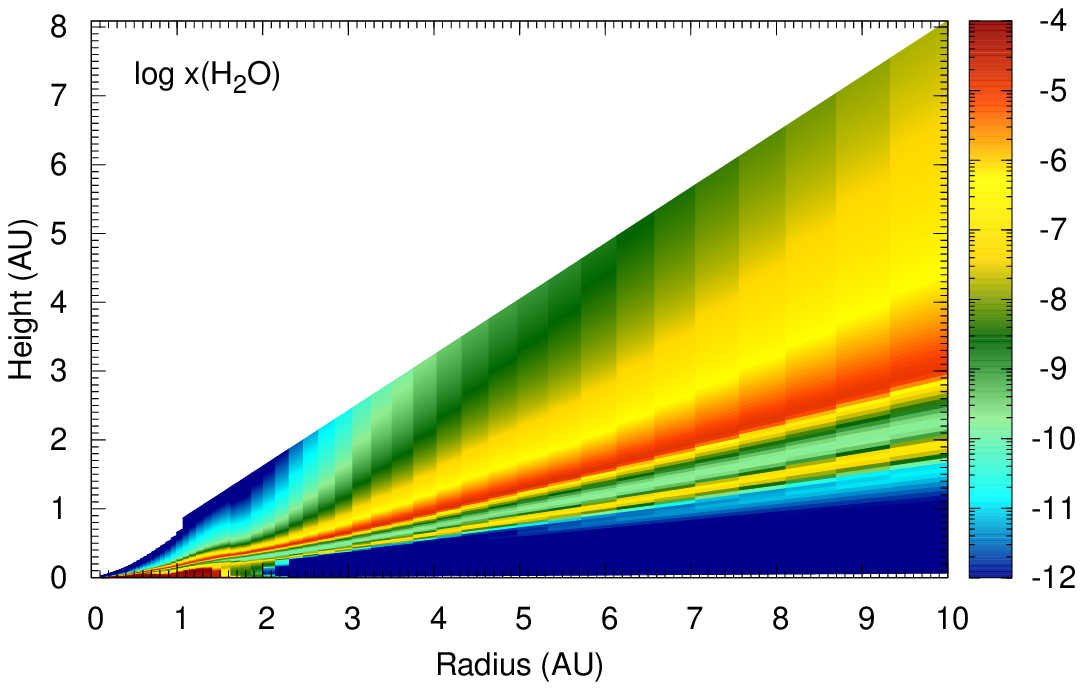}}
\subfigure{\includegraphics[width=0.5\textwidth]{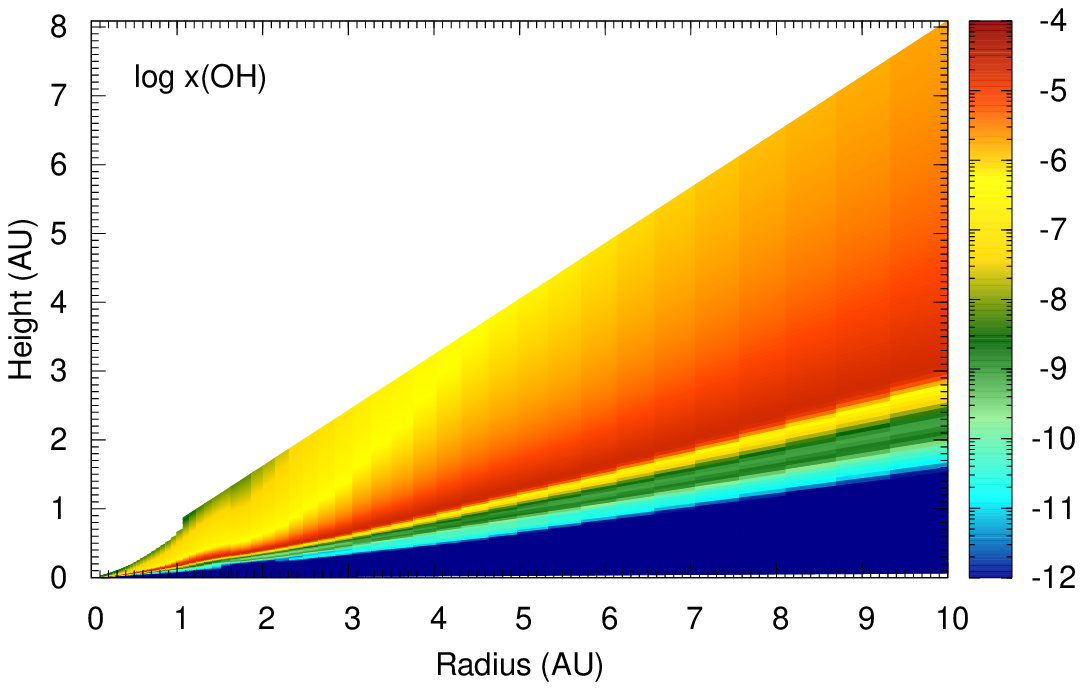}}
\subfigure{\includegraphics[width=0.5\textwidth]{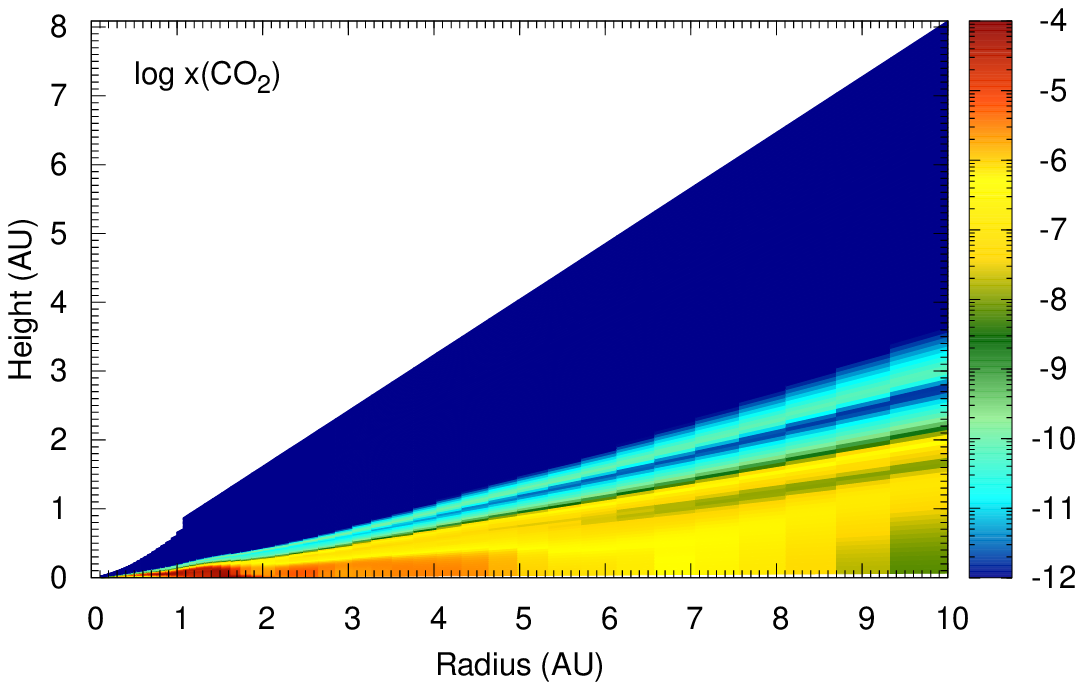}}
\subfigure{\includegraphics[width=0.5\textwidth]{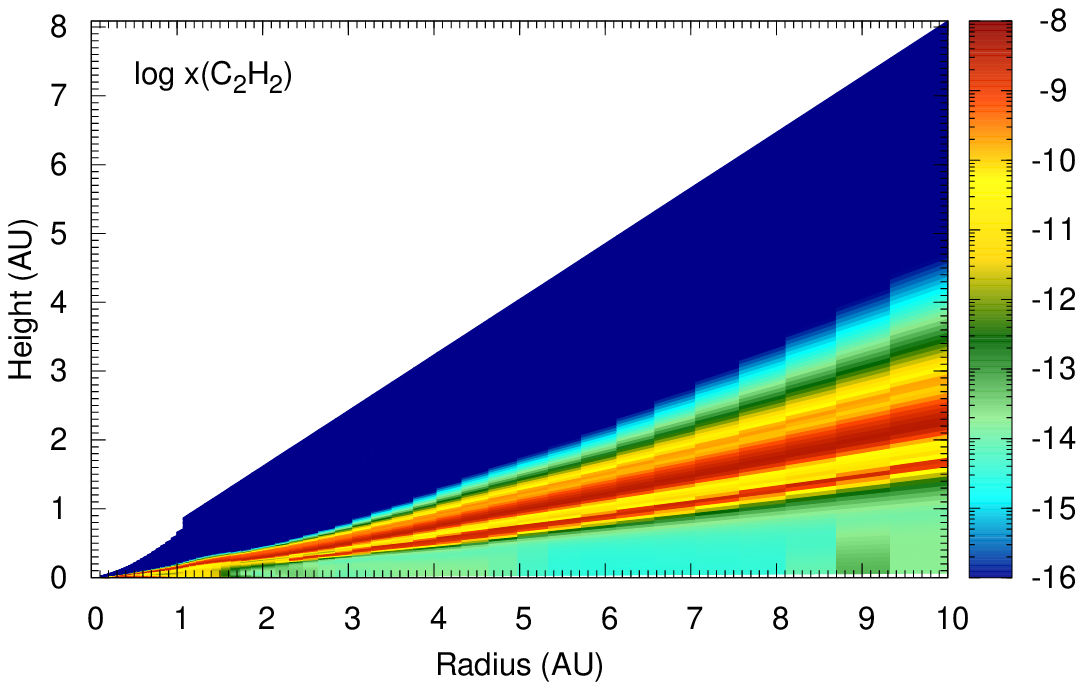}}
\caption{Fractional abundances of H$_2$O (top left), OH (top right), 
CO$_2$ (bottom left) and C$_2$H$_2$ (bottom right) 
as a function of disk radius and height up to a radius of 10~AU.}
\label{figure3}
\end{figure*}

Figure~\ref{figure3} displays the fractional abundances of those  additional molecules observed 
at infrared wavelengths,  H$_2$O (top), OH (second), 
CO$_2$ (third) and C$_2$H$_2$ (bottom), as a function of disk radius and height up to a maximum 
radius of 10~AU.  
We display results from within 10~AU only as infrared emission 
originates from the inner hot, dense disk material.    

Gas-phase H$_2$O is confined to the molecular layer with a fractional 
abundance x(H$_2$O)~$\sim$~$10^{-4}$.  As above,  freeze out is responsible for 
depletion in the  midplane and photolysis for depletion in the upper layers.  
H$_2$O is returned to the  gas phase in the  midplane at a radius $\approx$~1 to 2~AU.
The distribution of OH  is complimentary to that of H$_2$O residing throughout the disk in a 
layer above that of 
gas-phase H$_2$O, reaching a peak fractional abundance of $\sim$~$10^{-4}$.  
The distribution of CO$_2$ is similar to that of CO in the inner disk, existing only 
in the midplane with a maximum value of $10^{-4}$ within a few AU of the star.  
The snow-line for CO$_2$, however, is at the much smaller radius of $\sim$~10~AU 
(as opposed to $\approx$~250~AU).
Acetylene, C$_2$H$_2$, reaches a peak fractional abundance of $\sim$~$10^{-8}$, 
in the molecular layer.  
C$_2$H$_2$ and similar molecules are formed in hotter regions where oxygen is 
depleted from the gas phase 
via freeze out of oxygen containing molecules onto dust grains driving a 
carbon chemistry and hydrocarbon synthesis.  

In Figure~\ref{figure10} (online only) we display the fractional abundances of  the gas-phase 
molecules discussed  above, along with constituent atoms and 
grain-surface analogues where applicable, as a 
function of disk height at radii $r$~=~0.1~AU, 1~AU, 10~AU 
and 100~AU.   

\subsection{Effects of Non-thermal Desorption}
\label{nonthermaldesorptioneffects}

\begin{figure}
\centering
\includegraphics[width=0.5\textwidth]{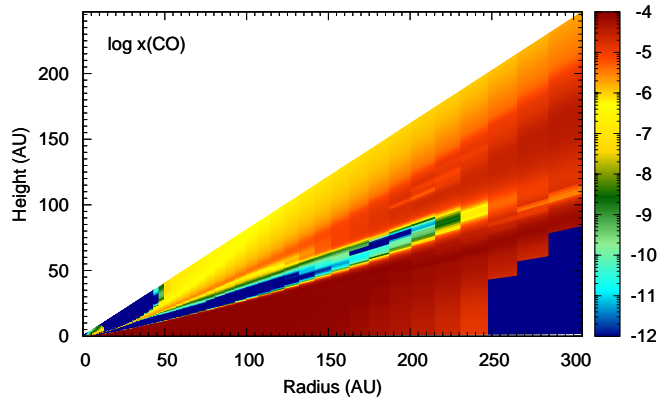}
\caption{Fractional abundance of gas-phase CO as a function of disk radius and height using 
results from model 0.  }
\label{figure4}
\end{figure}

In this section, we discuss the effects of each of our non-thermal desorption mechanisms, 
cosmic-ray induced desorption,  photodesorption and X-ray desorption, on the disk chemical 
structure (models CRH, PH+CRH and XD in Table~\ref{table2}, respectively).  
We call our control model, which includes thermal desorption only, model 0.  
First, we show in Figure~\ref{figure4} those regions of the disk in which molecules 
are depleted if we take into account thermal desorption only   
presenting, as an example, the gas-phase CO fractional abundance as a function of disk radius and 
height from model 0.  
We discuss how efficiently each non-thermal desorption mechanism works against 
depletion throughout the disk in Sections~\ref{cosmicraydesorptioneffects} 
to \ref{xraydesorptioneffects}.  
We display the fractional abundances of several gas-phase species as a function of disk radius and 
height comparing results from each of our non-thermal desorption models in 
Figure~\ref{figure11} (online only).  

In Figure~\ref{figure4} there are three notable areas where CO is depleted from the  gas phase,
(i) in the  midplane beyond a radius of $\approx$~250~AU, (ii) in a layer at a
height of $z/r$~$\approx$~0.3 and (iii) in the disk surface between a radius 
of a few AU to $\approx$~50~AU.  
In region (i), the depletion is due to freeze out of CO onto dust grains as the dust temperature 
here is below the desorption temperature of CO.  The absence
of any non-thermal desorption means that CO is completely removed from the  gas phase.  
For region (ii), the depletion is due to the destruction of CO by UV radiation.  
At this point, the 
UV field is strong enough to dissociate CO into its constituent atoms (atomic carbon and oxygen),  
however, the dust temperature is also low enough for the  freeze out of molecules which have 
binding energies larger than that of CO e.g.\ H$_2$O.  
By a time of $10^6$~years, the time at which we extract our abundances, atomic carbon and oxygen 
are trapped in molecules contained in the icy mantle.  
Above this height, the dust temperature becomes high enough for many molecules to desorb thermally 
thus replenishing the stock of C and O to reform CO.  
In region (iii), a similar effect to that in region (ii) occurs 
due to the decoupling of the dust and gas temperatures in 
the disk surface with the dust temperature up to two orders of magnitude lower than the 
gas temperature.  
The thermal desorption rate is dependent only on the dust temperature ($\propto$~$\exp(-E_d/T_d)$) 
whereas the accretion rate depends both on the gas temperature and the number density of dust grains 
(see Equations~\ref{accretionrate} and \ref{thermaldesorption}).  
Since the gas temperature can be much higher than the dust temperature in this region, 
the accretion rate can supersede that of thermal desorption so that by $10^6$ years, 
molecules which can survive the intense UV radiation field are able to  freeze out onto dust grains.  
Specifically, H$_2$O molecules are able to survive as ice on dust grains thereby depleting 
the  gas phase of oxygen-bearing molecules such as CO.  

\subsubsection{Cosmic-ray Induced Desorption}
\label{cosmicraydesorptioneffects}

The left columns of Figures~\ref{figure5} and \ref{figure6} display the fractional abundances 
of several molecules and molecular 
ions as a function of disk height at radii $r$~=~10~AU and 305~AU, respectively, 
comparing the results from model 0 (solid lines) with model CRH (dotted lines).  
Cosmic-ray induced desorption has the smallest effect on the gas-phase abundances and 
only in the outer disk  midplane.  
The fractional abundances of gas-phase CO and H$_2$O are enhanced in the disk  midplane 
at 305~AU in the results for model CRH, compared with those for model 0, although 
the values reached remain orders of magnitude smaller than those in the upper disk layers.  
The fractional abundance of N$_2$H$^+$ is also enhanced, due in part to the release of 
N$_2$ from dust grains in this region by cosmic-rays and, in fact, reaches 
its maximum fractional abundance in the disk  midplane.  
As our model disk is truncated at 305~AU, we would expect the effects of cosmic-ray induced 
desorption to continue beyond this radius.

\subsubsection{Photodesorption}
\label{photodesorptioneffects}

 Photodesorption is the most experimentally constrained non-thermal desorption mechanism 
we have included in our model.  
 Photodesorption has an effect in the upper disk layers where the UV radiation 
field has reached an appreciable strength yet the temperature remains low 
enough for  freeze out to occur.  
We compare the results from model 0 (solid lines) with those from model 
PH+CRH (dotted lines) in the middle plots of Figures~\ref{figure5} and \ref{figure6}.  
 Photodesorption enhances the abundance of molecules in the molecular region 
of the disk, counteracting the depletion effects discussed at 
the beginning of Section~\ref{nonthermaldesorptioneffects}.  
Clearly seen at a radius of 10~AU is the smoothing of molecular 
abundances throughout the middle region of the disk with both model results 
converging higher in the surface.  
At a radius of 305~AU, there is a smoothing out of abundances counteracting depletion 
but also a slight difference in the distribution of molecules 
in the surface regions.  
The fractional abundances of CO and CS are enhanced in model PH+CRH 
relative to model 0, whereas those of HCO$^+$, H$_2$O, HCN and N$_2$H$^+$ are reduced.  
This is due to the alteration of gas-phase chemistry when  photodesorption is included as 
a significant amount of all molecules can remain in the  gas phase in the mid and upper 
layers of the disk.  Thus, those molecules which ordinarily would be frozen out in the absence 
of photodesorption are available to take part in gas-phase reactions e.g.\ N$_2$H$^+$ is 
destroyed via reaction with gas-phase CO so that an enhancement in the abundance of the latter leads to a 
corresponding drop in that of the former.    

\subsubsection{X-ray Desorption}
\label{xraydesorptioneffects}

\begin{figure*}
\centering
\subfigure{\includegraphics[width=0.32\textwidth]{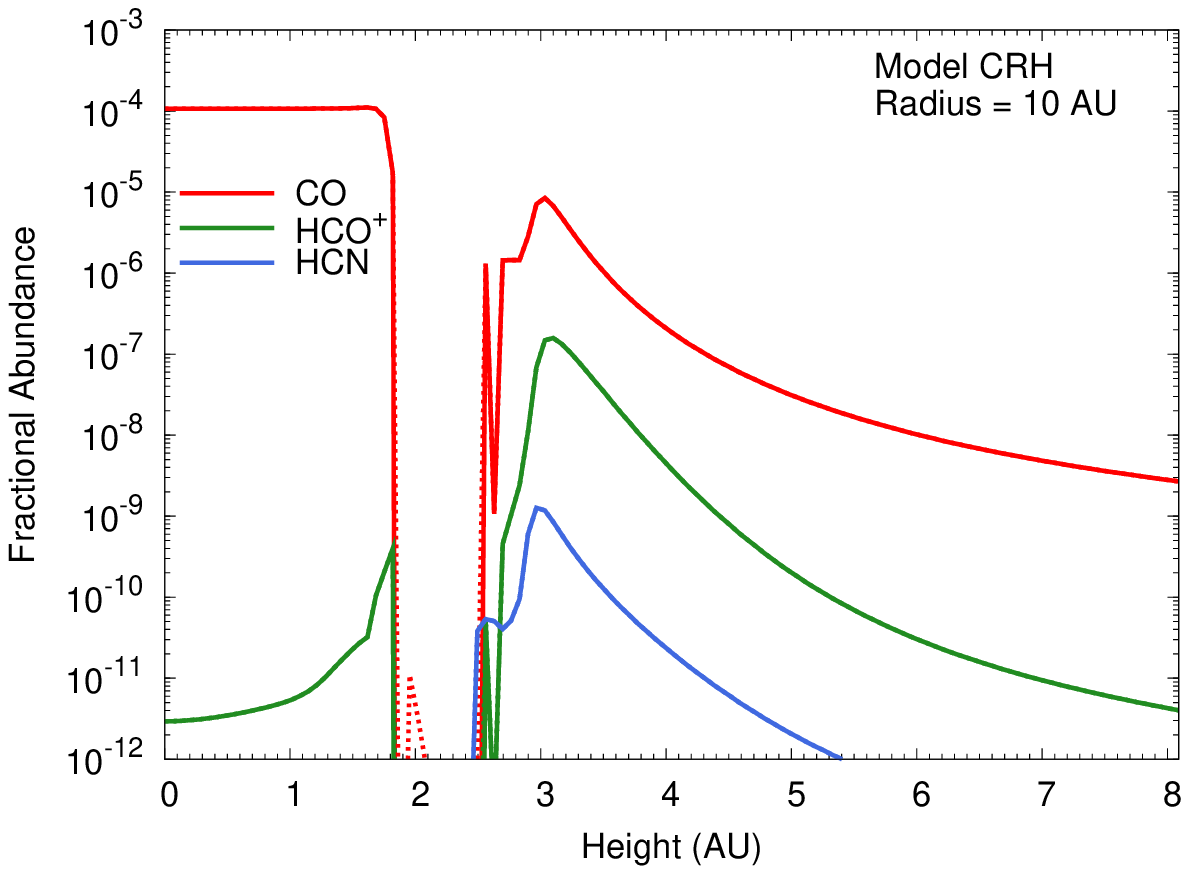}}
\subfigure{\includegraphics[width=0.32\textwidth]{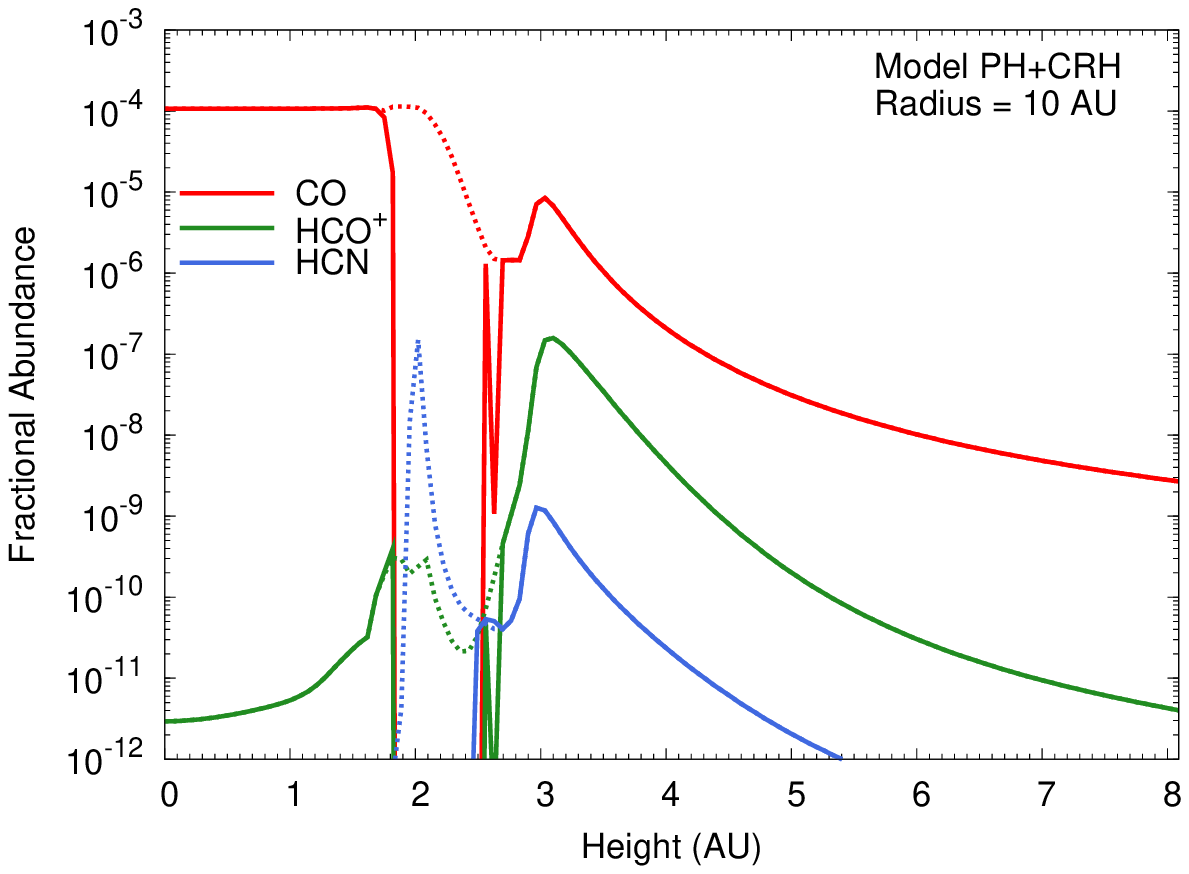}}
\subfigure{\includegraphics[width=0.32\textwidth]{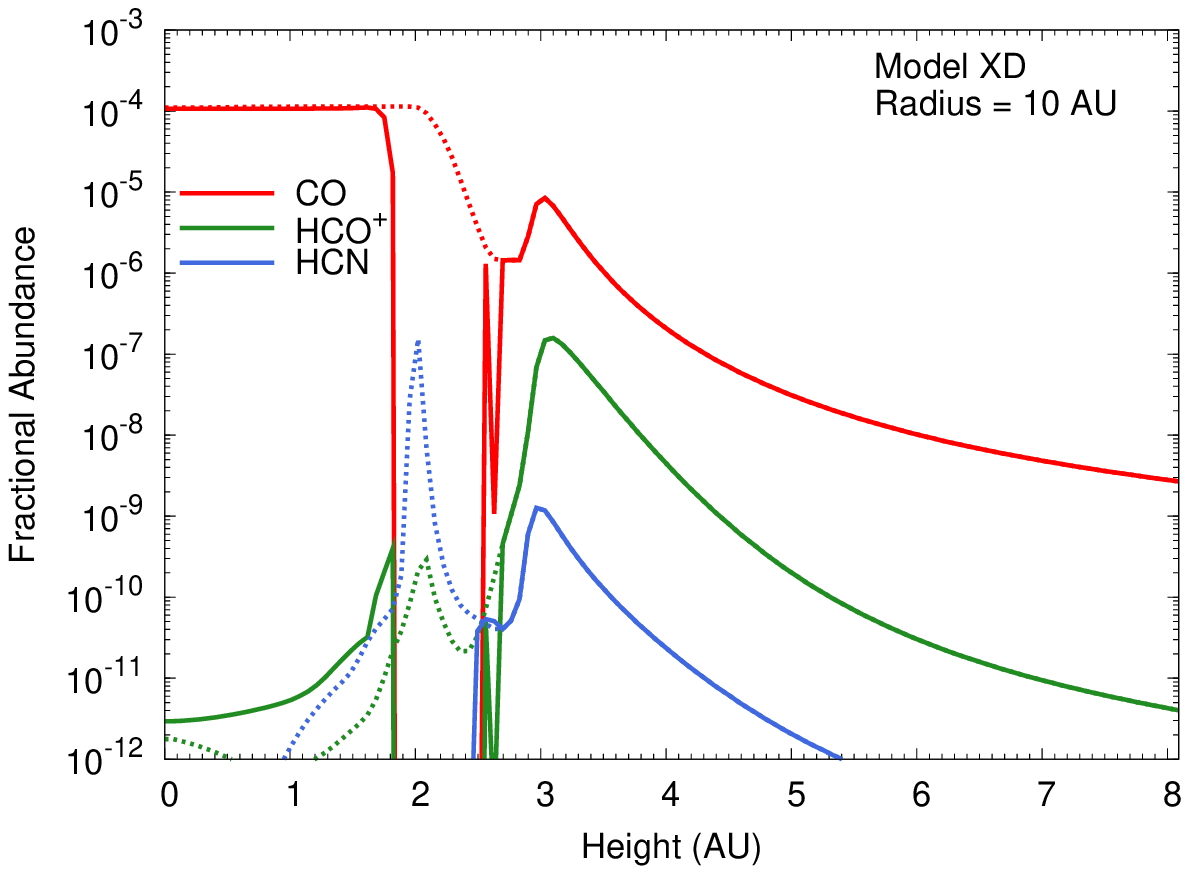}}
\subfigure{\includegraphics[width=0.32\textwidth]{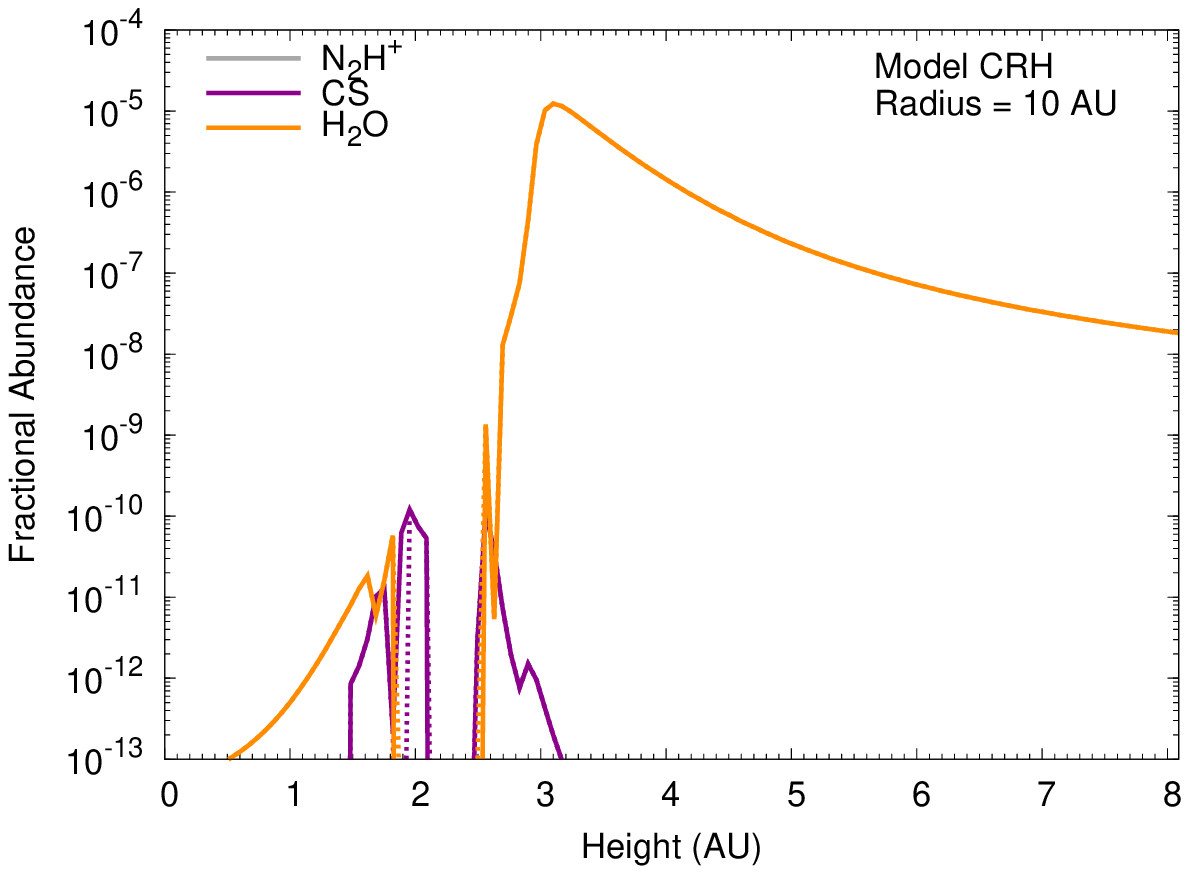}}
\subfigure{\includegraphics[width=0.32\textwidth]{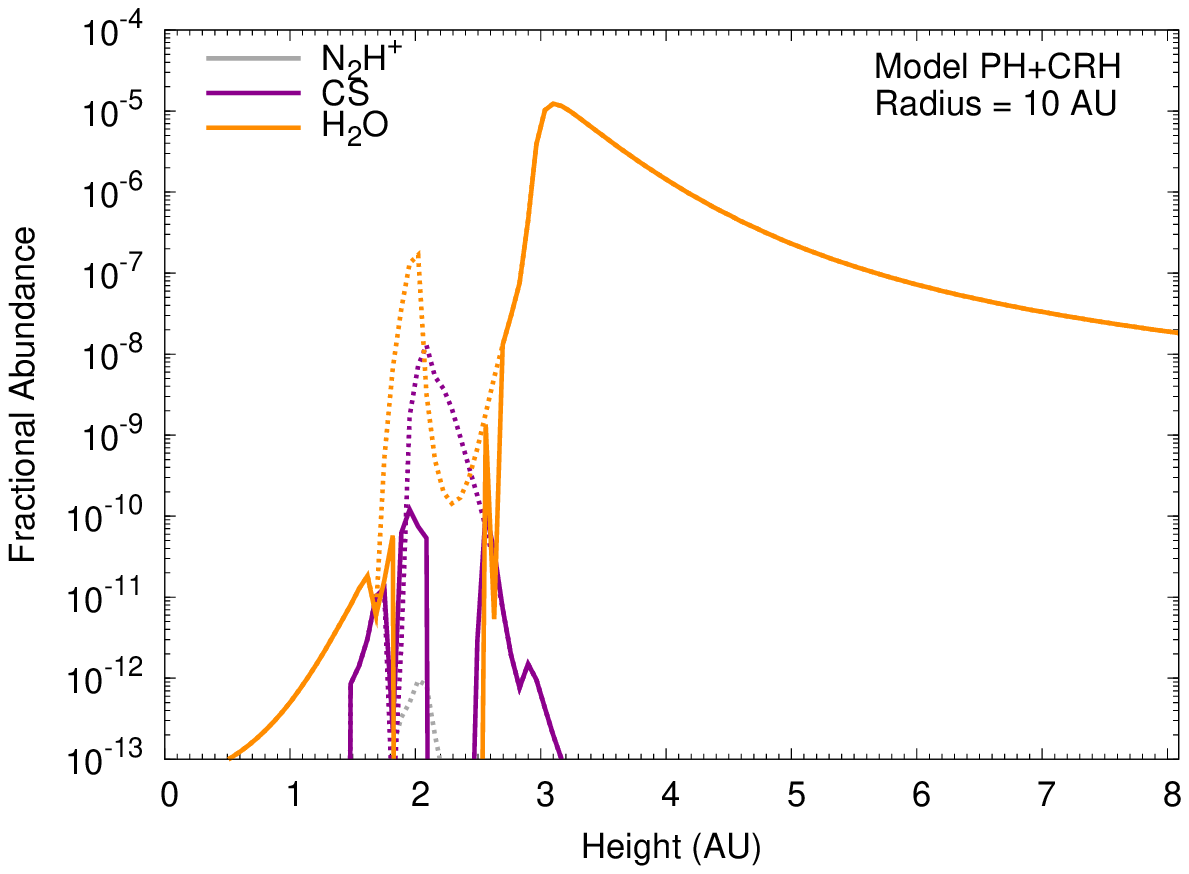}}
\subfigure{\includegraphics[width=0.32\textwidth]{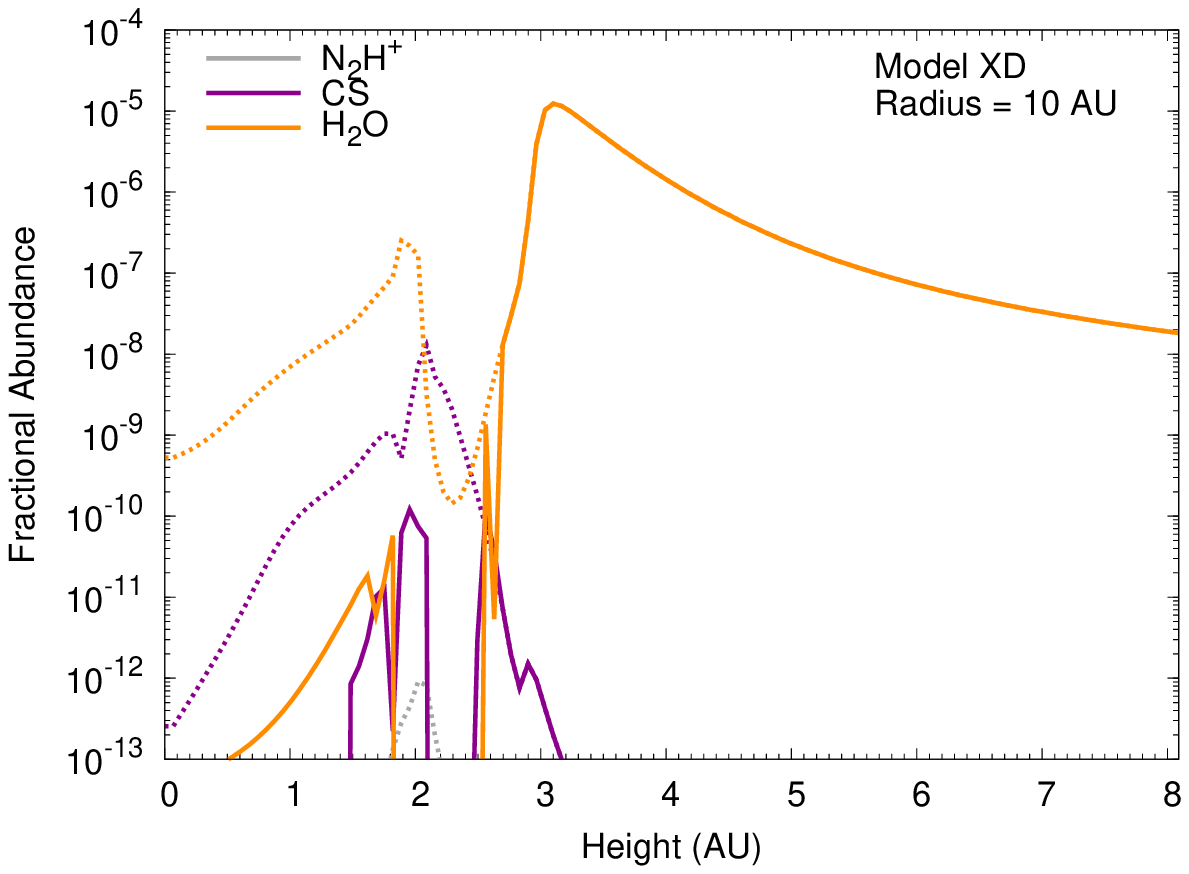}}
\caption{Fractional abundances of several gas-phase molecules and molecular ions 
as a function of disk height at a radius, $r$~=~10~AU comparing 
results from model 0 (solid lines) with each non-thermal desorption model (dotted lines), 
CRH (left), PH+CRH (middle) and XD (right).  
 Note that the results for N$_2$H$^+$ from model 0 are too small to appear on our plot.}
\label{figure5}
\end{figure*}

\begin{figure*}
\centering
\subfigure{\includegraphics[width=0.32\textwidth]{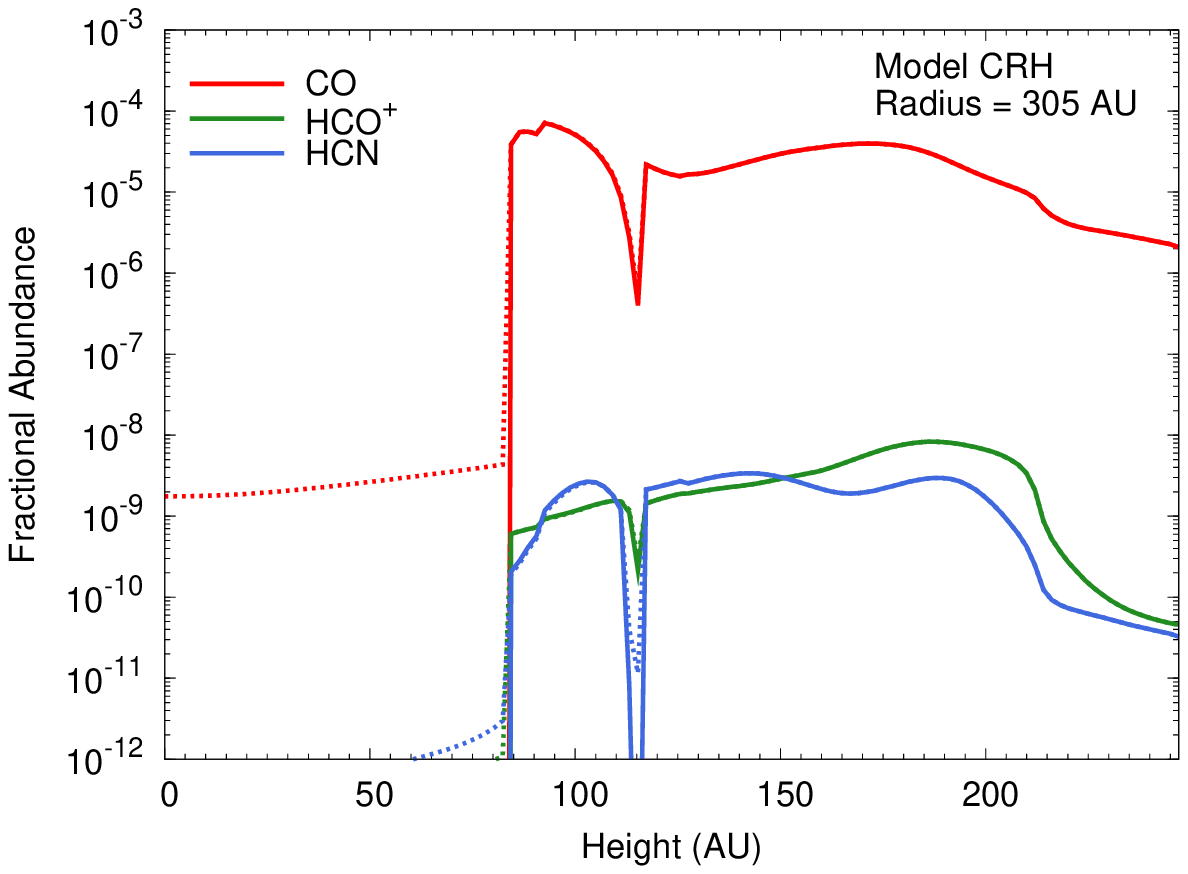}}
\subfigure{\includegraphics[width=0.32\textwidth]{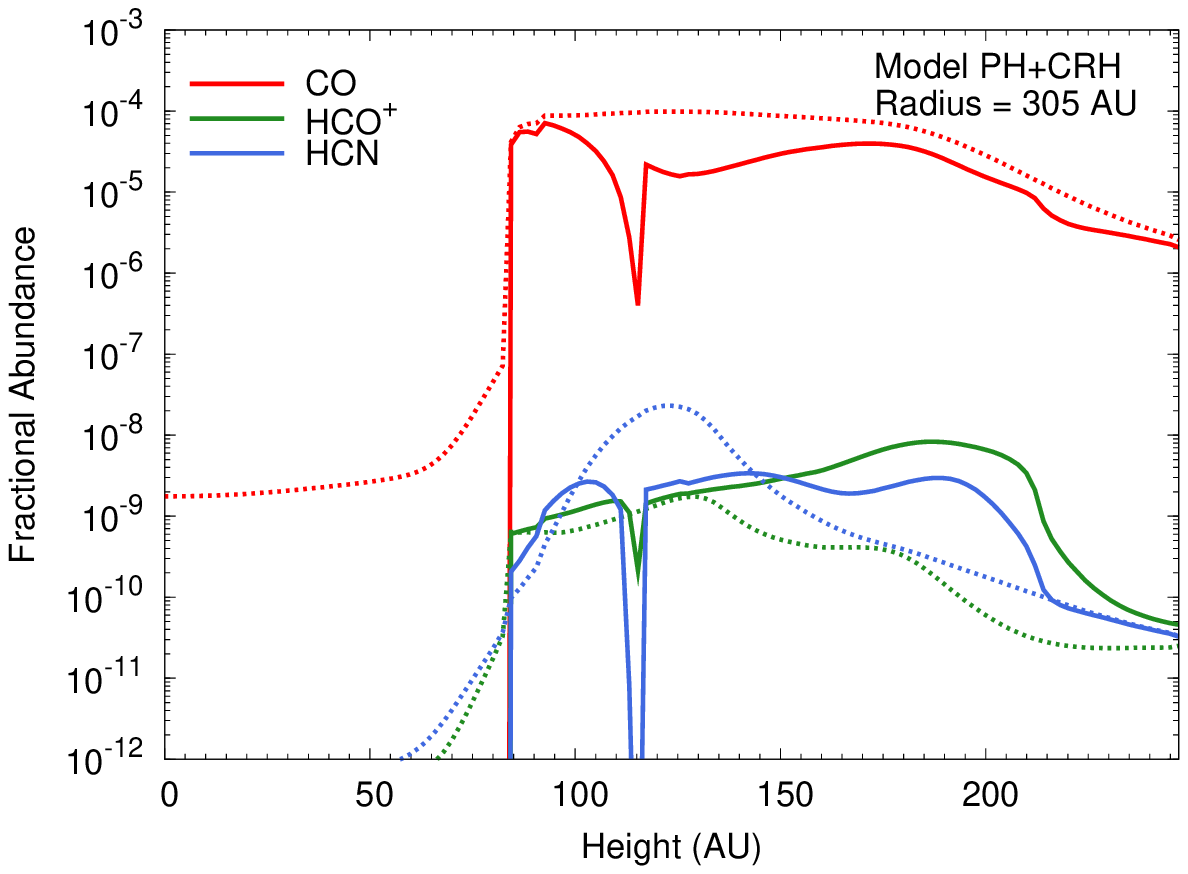}}
\subfigure{\includegraphics[width=0.32\textwidth]{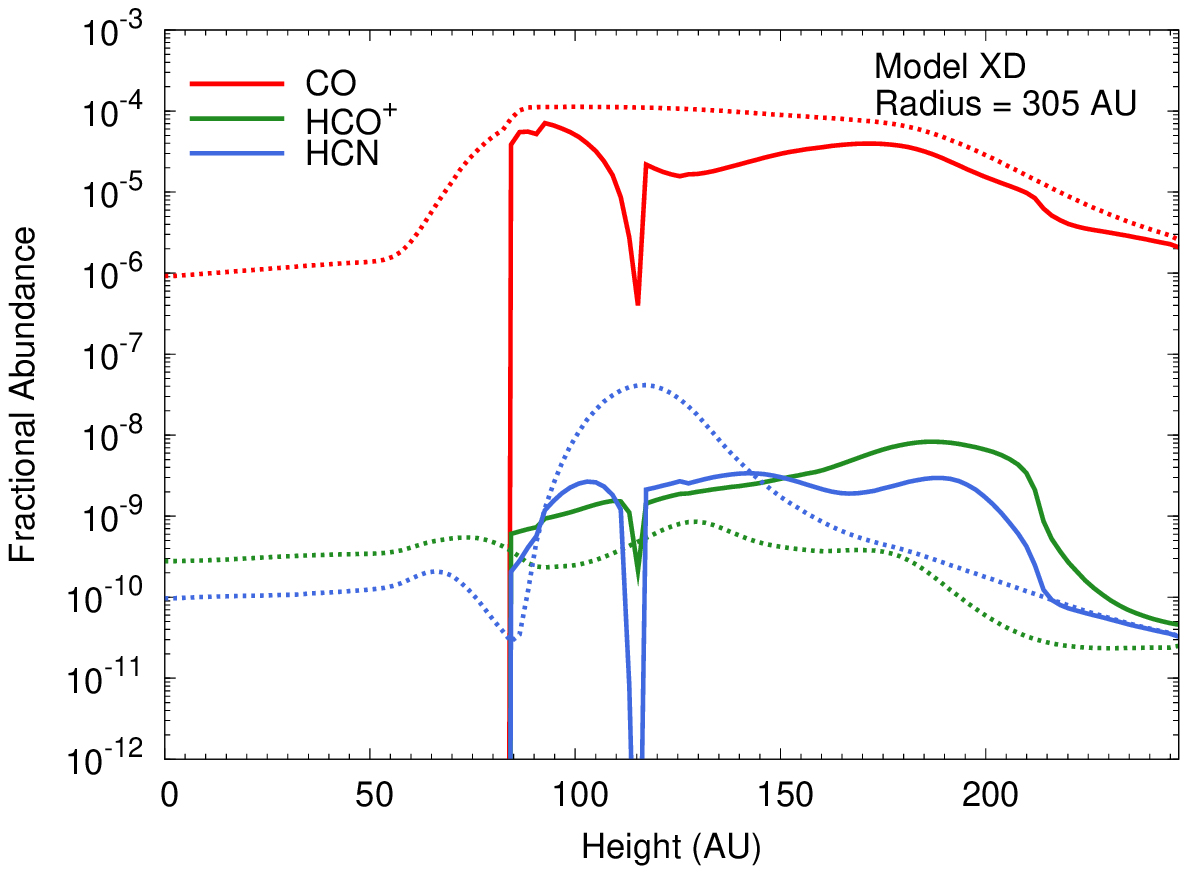}}
\subfigure{\includegraphics[width=0.32\textwidth]{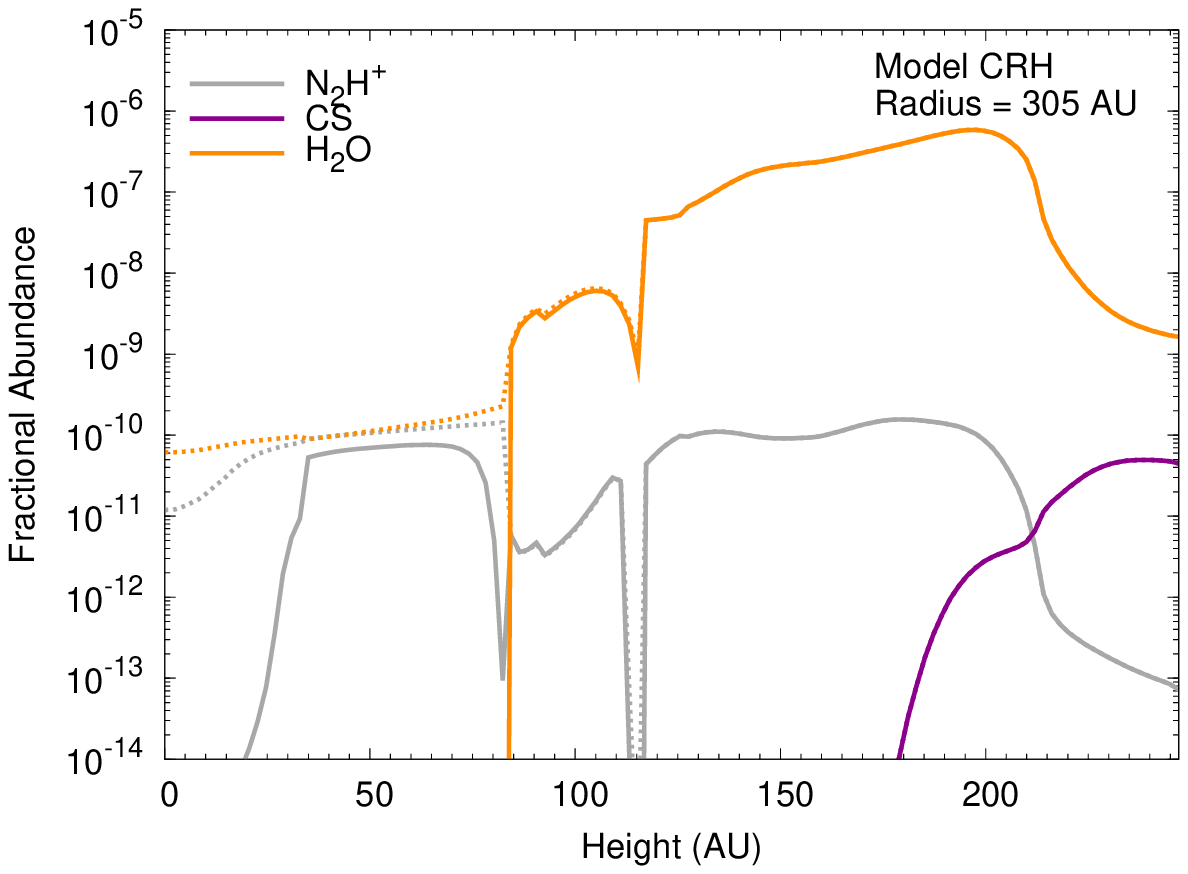}}
\subfigure{\includegraphics[width=0.32\textwidth]{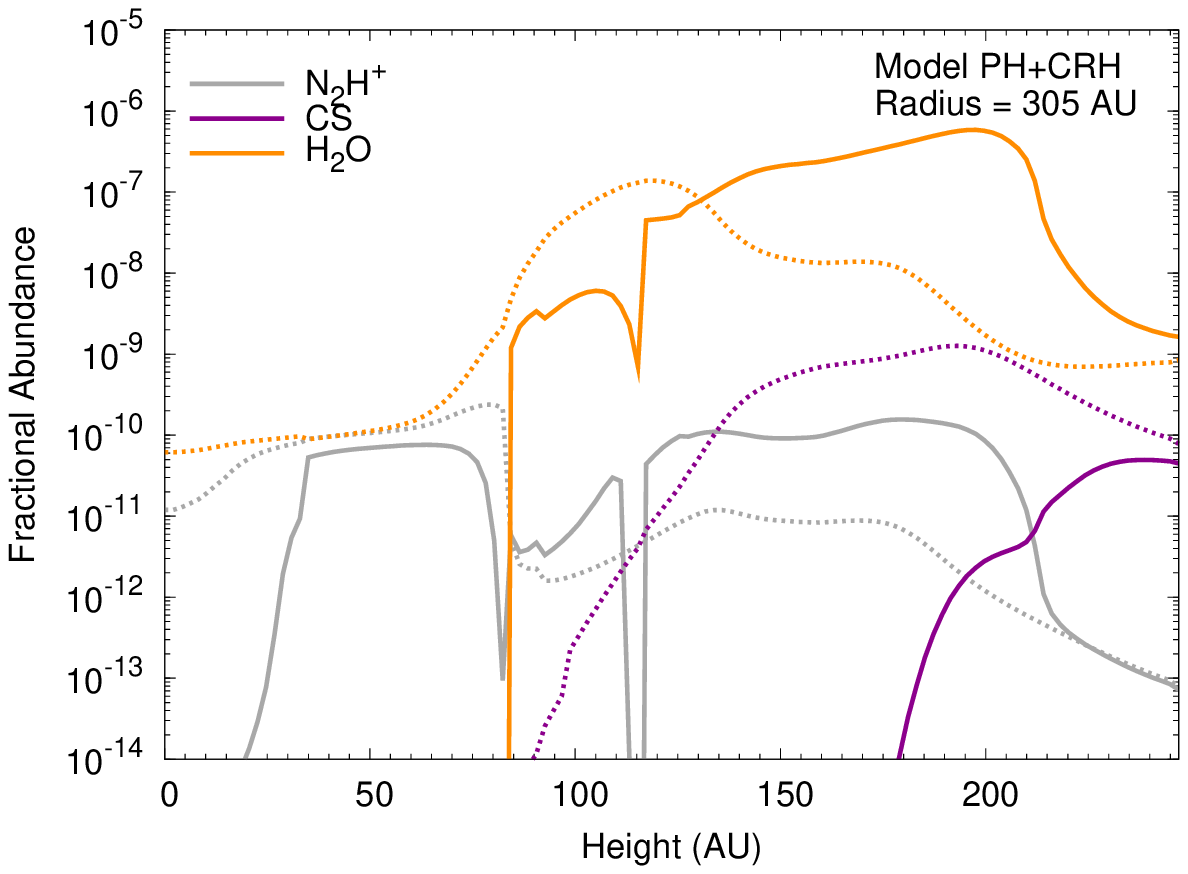}}
\subfigure{\includegraphics[width=0.32\textwidth]{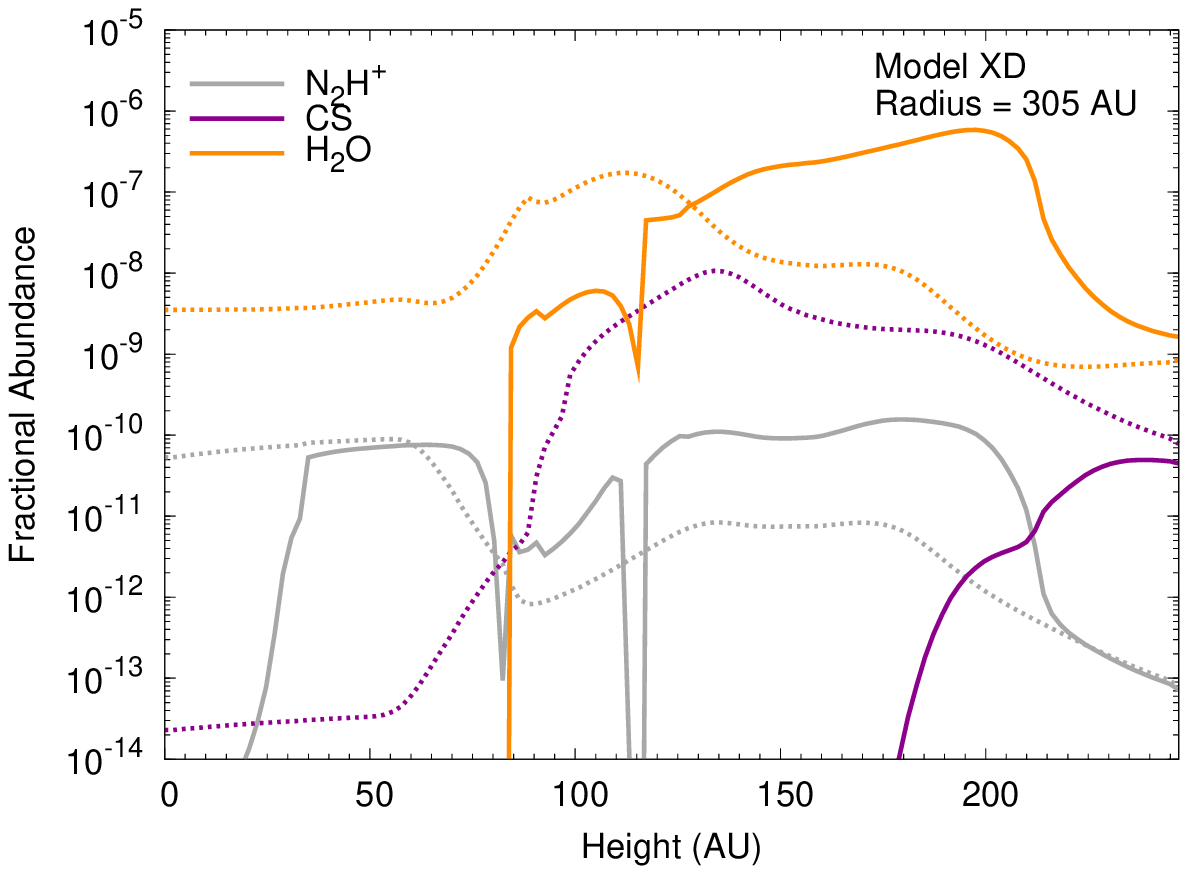}}
\caption{Fractional abundances of several gas-phase molecules and molecular ions 
as a function of disk height at a radius, $r$~=~305~AU comparing 
results from model 0 (solid lines) with each non-thermal desorption model (dotted lines), 
CRH (left), PH+CRH (middle) and XD (right).}
\label{figure6}
\end{figure*}

X-ray desorption is the least theoretically or  experimentally constrained non-thermal 
desorption mechanism we considered, hence, we have used 
conservative estimates of molecular yields and thus X-ray desorption 
rates.  
The right-hand plots in Figures~\ref{figure5} and \ref{figure6} suggest that, even using conservative 
estimates, X-ray desorption has the largest effect on gas-phase molecular abundances.  
X-rays, with their higher energy, can penetrate deeper into the disk material 
than UV photons, hence, X-rays have an effect in the disk  midplane as well as 
in the molecular region.  
At $r$~=~10~AU (top plot), the abundances of both H$_2$O and CS are enhanced in model XD
relative to the results from models 0, CRH and PH+CRH.  
X-ray desorption also acts to smooth out abundances in the upper disk, similar 
to the effects of  photodesorption.  
At 305~AU, the effects of the inclusion of X-ray desorption are most apparent.  
The fractional abundances of all molecules considered here, with the exception 
of CS, are enhanced in the  midplane of the disk, to values comparable with 
those found in the upper disk regions.  
In fact, it appears that the inclusion of X-ray desorption acts to smooth 
or homogenise the fractional abundances of gas-phase 
CO, H$_2$O and HCO$^+$ throughout the depth of the disk.  
Again, as seen in the results for  photodesorption, the gas-phase distributions 
are altered in model XD compared with model 0 due to the alteration of 
the gas-phase chemistry.  
In the upper disk, the results for models PH+CRH and XD are similar with the 
exception of the fractional abundance of CS which is enhanced in abundance between 
height of $\approx$~70~AU and $\approx$~150~AU in model XD relative to 
model PH+CRH.   

\subsection{Effects of Grain-surface Chemistry}
\label{grainsurfacechemistryeffects}

\begin{figure*}
\subfigure{\includegraphics[width=0.5\textwidth]{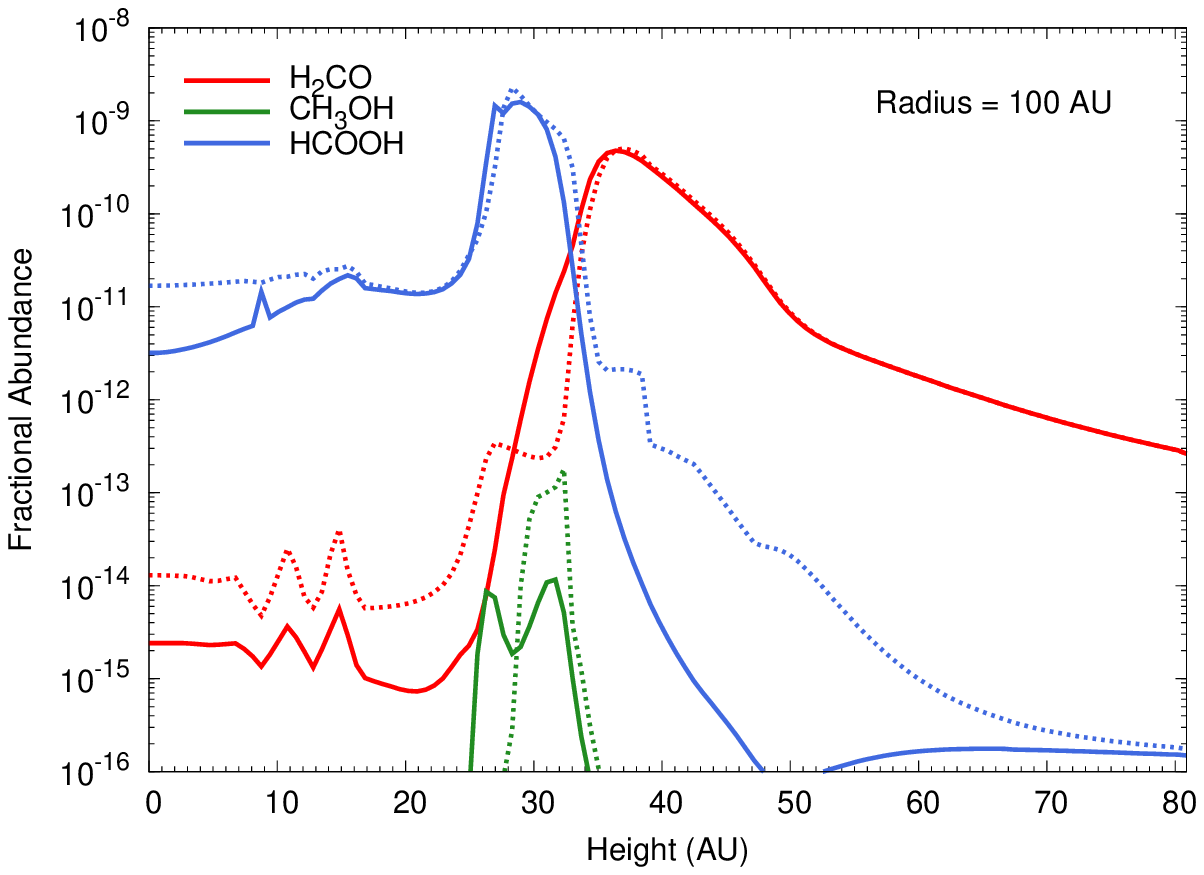}}
\subfigure{\includegraphics[width=0.5\textwidth]{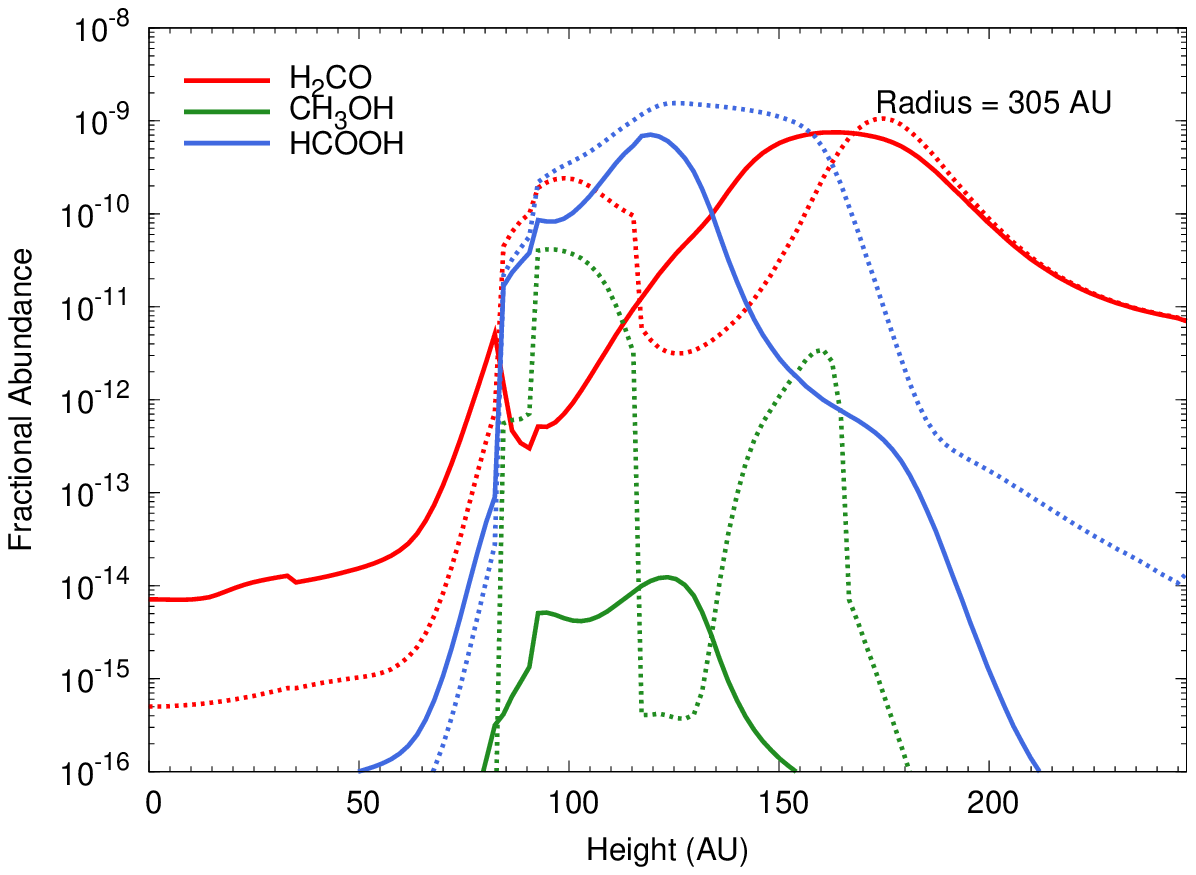}}
\subfigure{\includegraphics[width=0.5\textwidth]{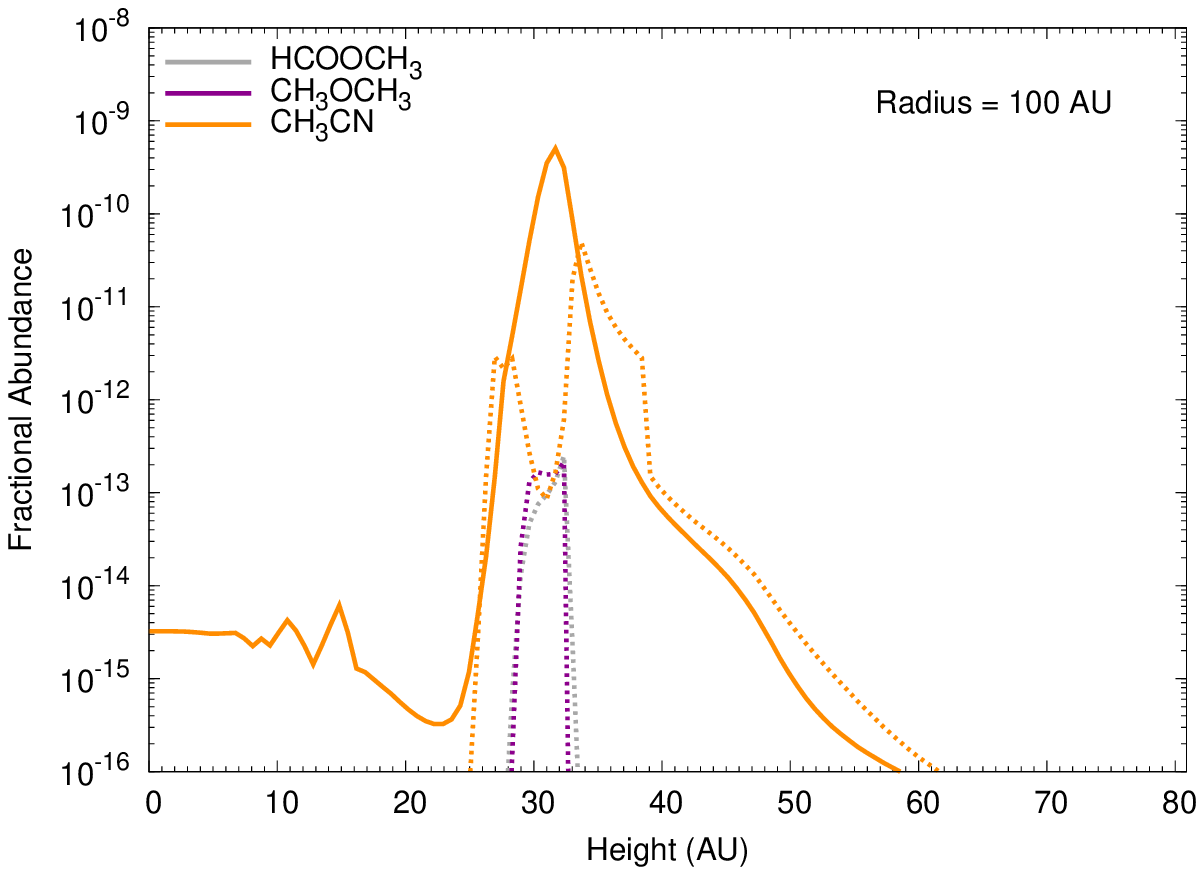}}
\subfigure{\includegraphics[width=0.5\textwidth]{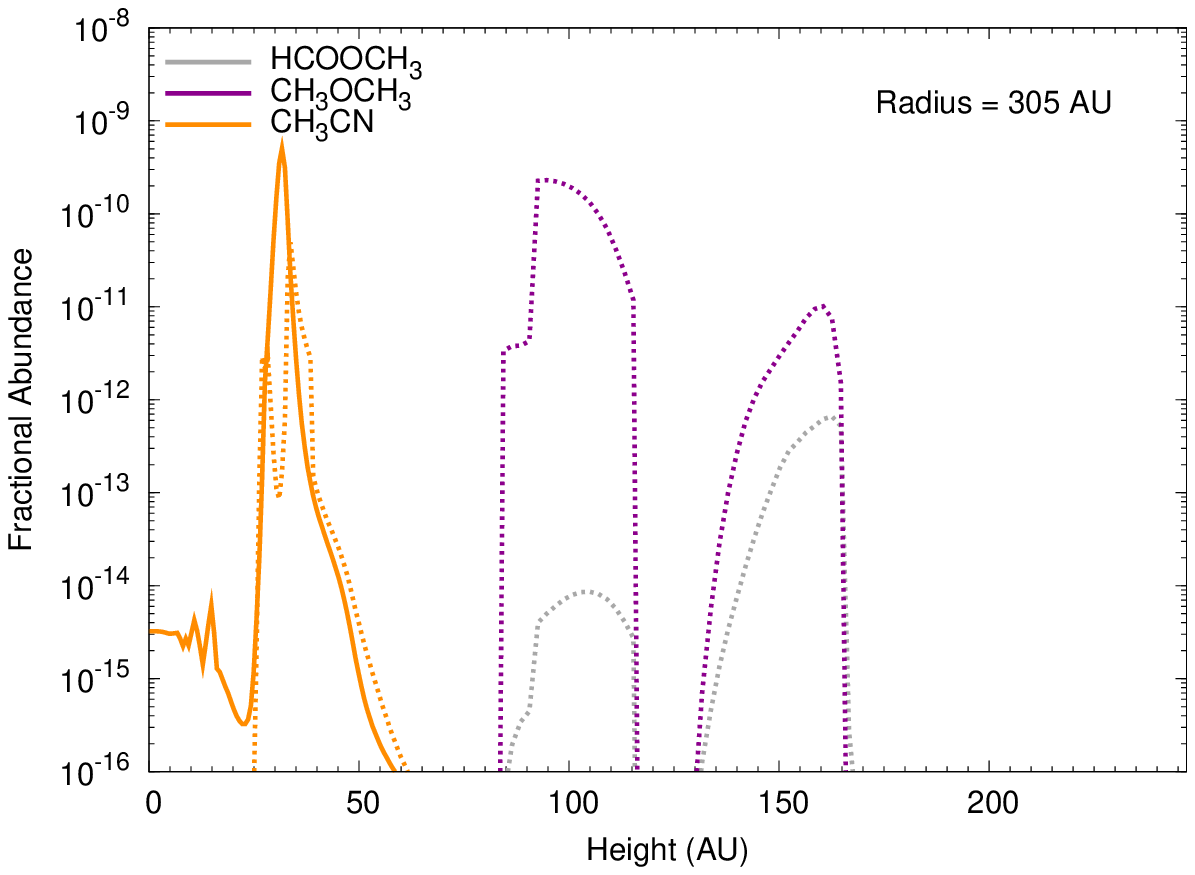}}
\caption{Fractional abundances of several small organic molecules as a function of disk 
height at radii, $r$~=~100~AU (left) and 305~AU (right) for model PH+CRH (solid lines) and 
model GR (dotted lines).}
\label{figure7}
\end{figure*}

The addition of grain surface chemistry is expected to aid the synthesis 
of complex organic molecules in regions of the disk where significant  freeze out has occurred.  
In this discussion, we look at the abundances of small organic (saturated) 
molecules in the outer disk in particular. 
In model GR (see Table~\ref{table2}), in addition to grain-surface chemistry, we also 
add cosmic-ray induced desorption and  photodesorption.  
Figure~\ref{figure7} shows the fractional abundances of several small organic molecules 
as a function of disk height at radii, $r$~=~100~AU (left) and 305~AU (right), for model 
PH+CRH (solid lines) compared with model GR (dotted lines).  
At 100~AU, the fractional abundance of HCOOH (formic acid) and H$_2$CO (formaldehyde) 
are enhanced in the disk  midplane in model GR, relative to model PH+CRH, with the abundance
of HCOOH also enhanced in the upper disk layer.  
Most of the organic molecules considered reach their peak fractional abundance between 
25 and 40~AU.  
Here, the fractional abundances of CH$_3$OH (methanol), 
HCOOCH$_3$ (methyl formate) and CH$_3$OCH$_3$ (dimethyl ether) are all enhanced 
to a value $\sim$~$10^{-13}$ in model GR, orders of magnitude larger than the respective values 
from model PH+CRH.  
At the very outer edge of our disk model, $r$~=~305~AU, the fractional abundances of 
all molecules are enhanced in model GR relative to model PH+CRH.  
Of note is the extreme enhancement seen in the abundances of 
CH$_3$OH, HCOOCH$_3$ and CH$_3$OCH$_3$, again by several orders of magnitude, to fractional 
abundances which are potentially observable ($\sim$~$10^{-11}$ to $\sim$~$10^{-10}$). 
 By fitting observed line intensities of rotational transitions in a selection of molecules 
with a simple disk model, 
\citet{thi04} estimate the column density of the relatively complex molecule, H$_2$CO, in several protoplanetary
disks as lying between $\sim$~10$^{12}$ and 10$^{13}$~cm$^{-2}$ which in the outer disk translates roughly 
to a fractional abundance of $\sim$~10$^{-10}$ as the H$_2$ column density here is 
$\sim$~10$^{22}$~cm$^{-2}$.  
Complex organic molecules in hot cores and dark clouds are routinely observed with fractional abundances 
$\gtrsim$~10$^{-11}$ (see e.g. \citet{herbst09}).  
We display the fractional abundances of the molecules discussed in this section as a function 
of disk radius and height comparing results from model PH+CRH and model GR in Figure~\ref{figure12} 
(online only).  

\subsection{Disk Ionisation Fraction}
\label{diskionisationfraction}

The ionisation fraction in protoplanetary disks is an important parameter as it is thought that 
this drives the accretion flow in the disk through the coupling of the gas with 
the strong magnetic fields generated by the system.  
The required turbulence is generated via magneto-rotational instabilities or MRI 
\citep{balbus91,hawley91}.  
For effective accretion, the ionisation fraction is required to exceed a critical 
level which is dependent on the nature of the star-disk system (see e.g. \citet{ilgner06}).  
Regions in which the ionisation fraction falls below this critical value and where 
effectively magneto-hydrodynamic accretion is switched off, are termed `dead zones'.  

\begin{figure*}
\subfigure{\includegraphics[width=0.5\textwidth]{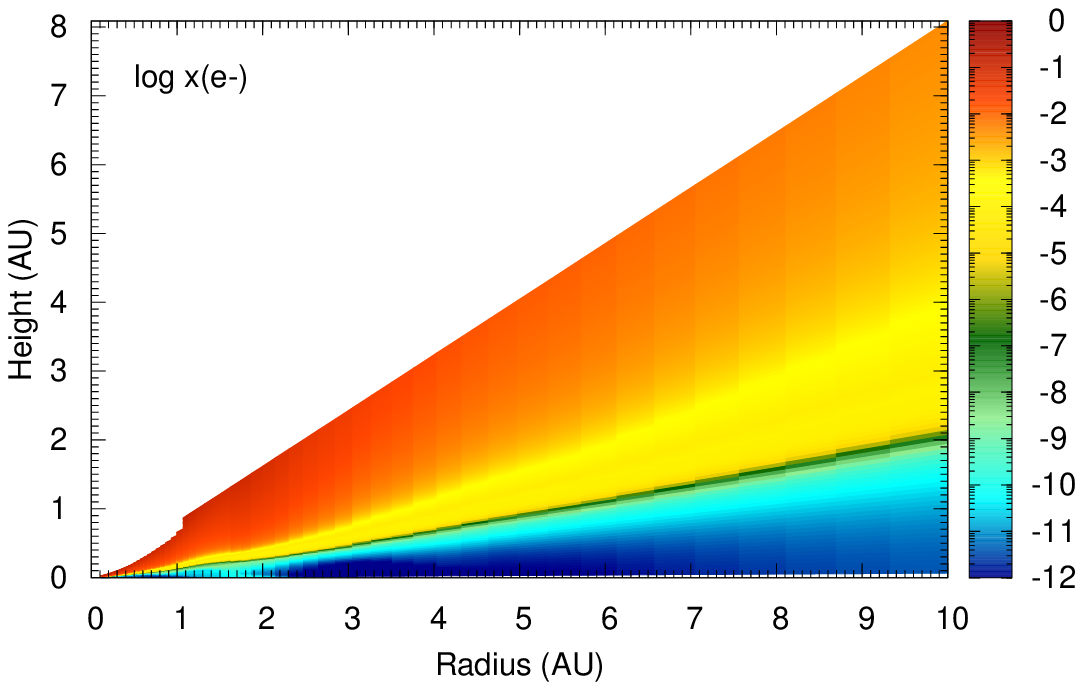}}
\subfigure{\includegraphics[width=0.5\textwidth]{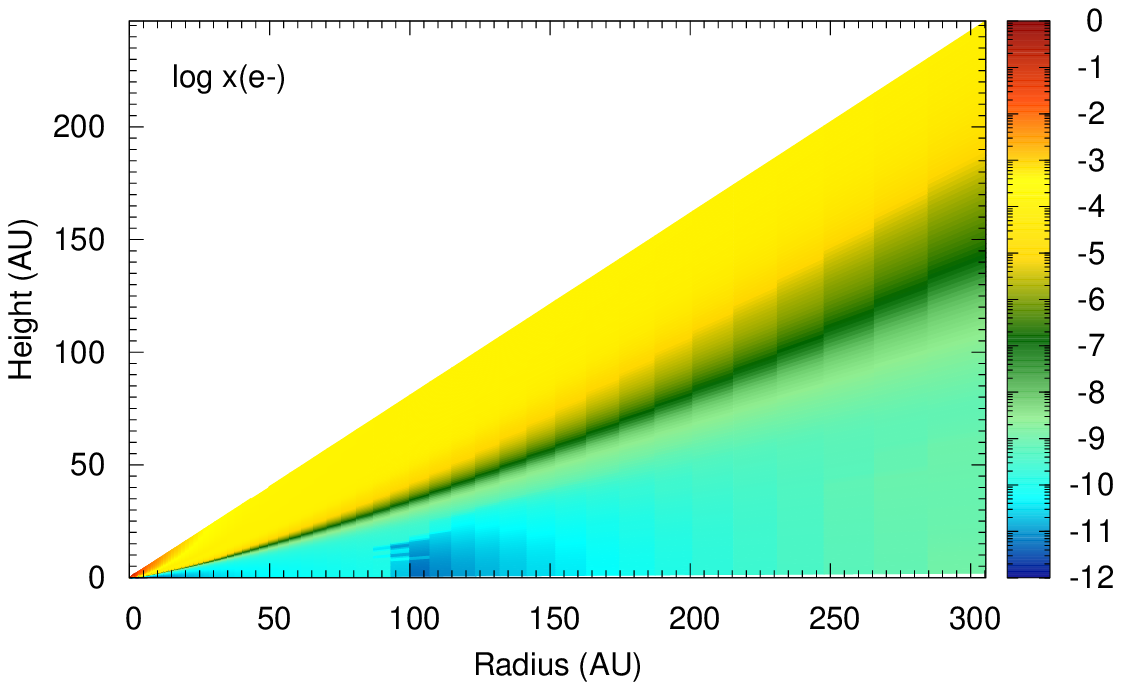}}
\caption{Fractional  abundance of electrons as a function of disk radius and
height up to maximum radii of 10~AU (left) and 305~AU (right).}
\label{figure8}
\end{figure*}

\begin{figure}
\centering
\subfigure{\includegraphics[width=0.32\textwidth]{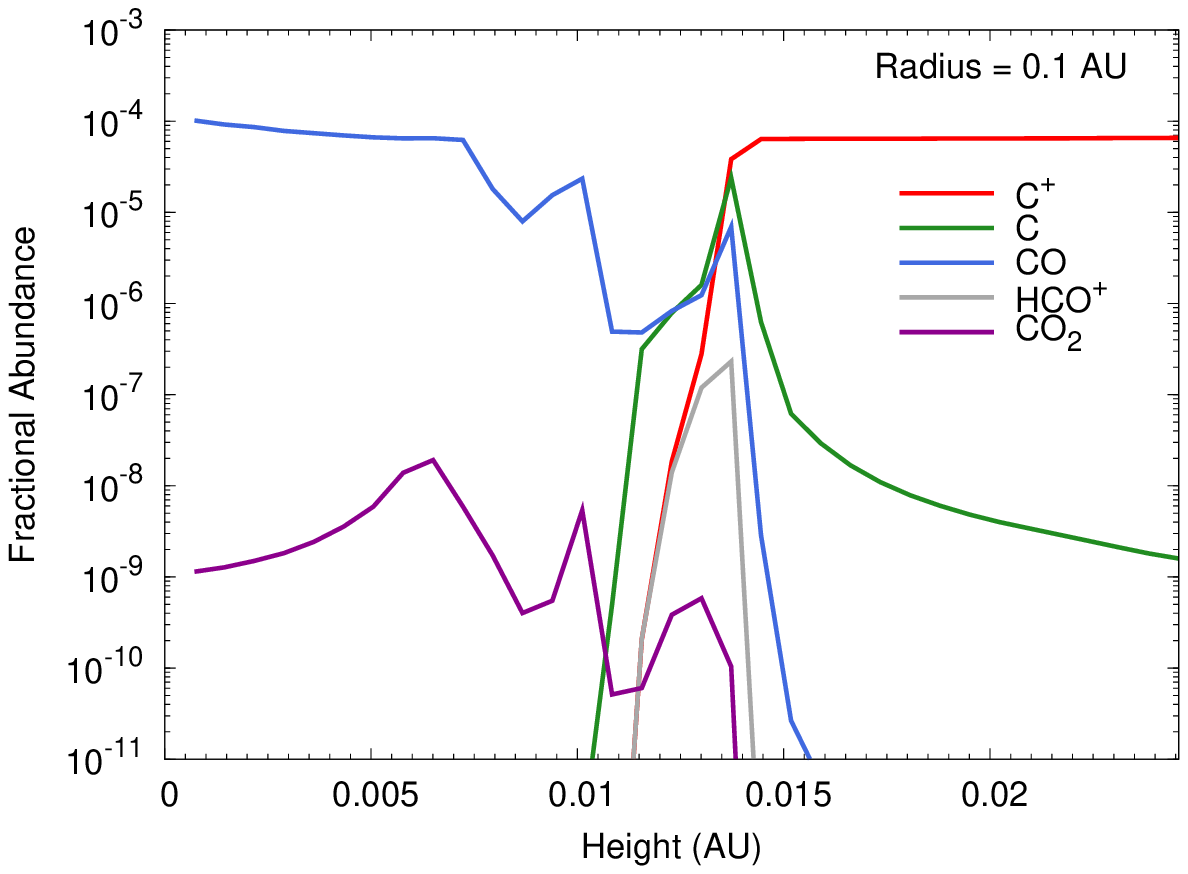}}
\subfigure{\includegraphics[width=0.32\textwidth]{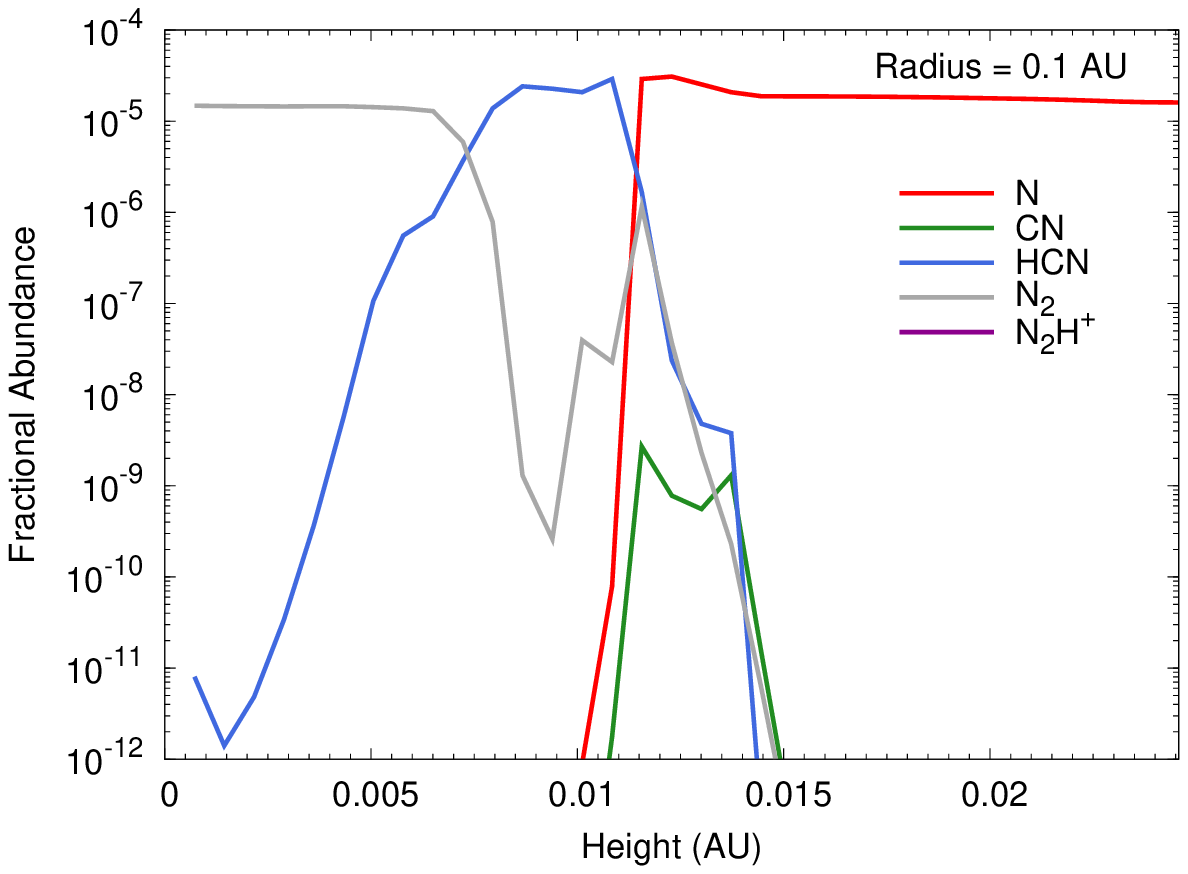}}
\subfigure{\includegraphics[width=0.32\textwidth]{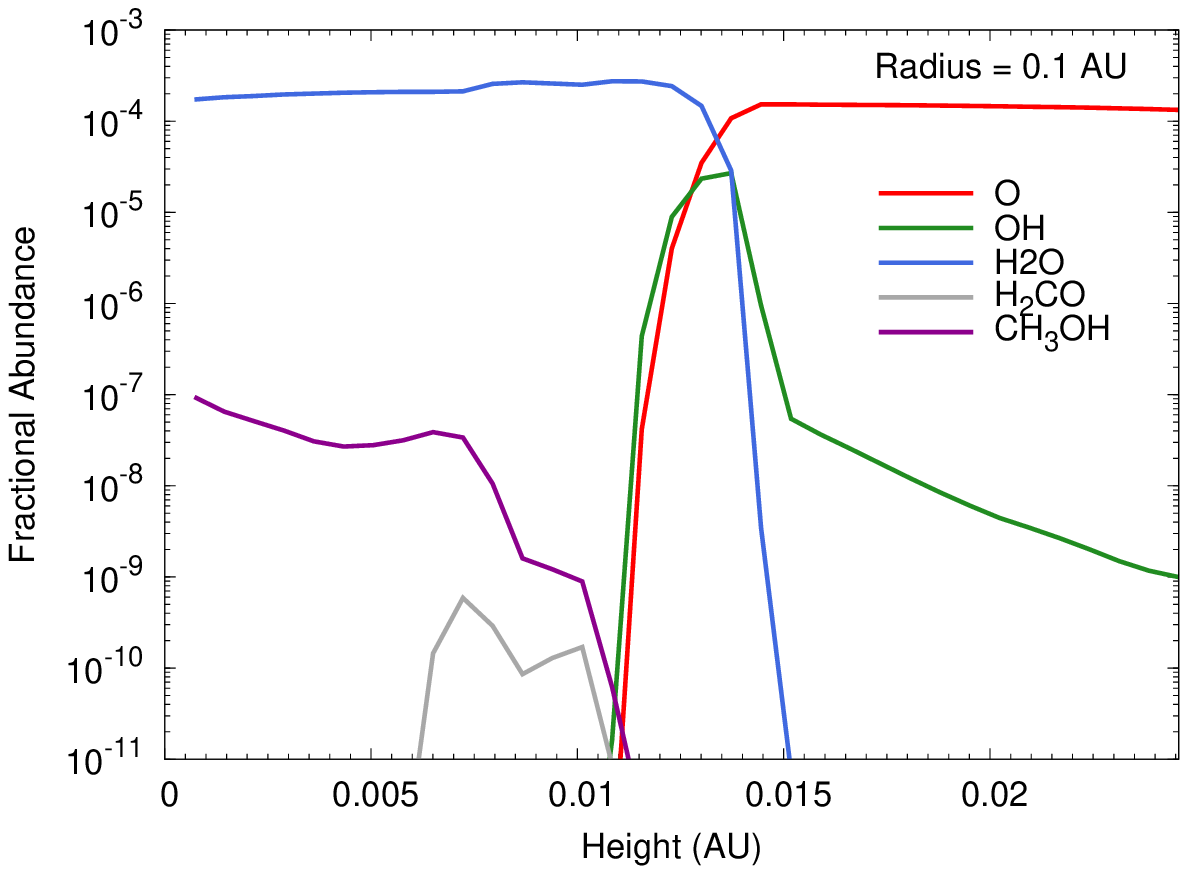}}
\subfigure{\includegraphics[width=0.32\textwidth]{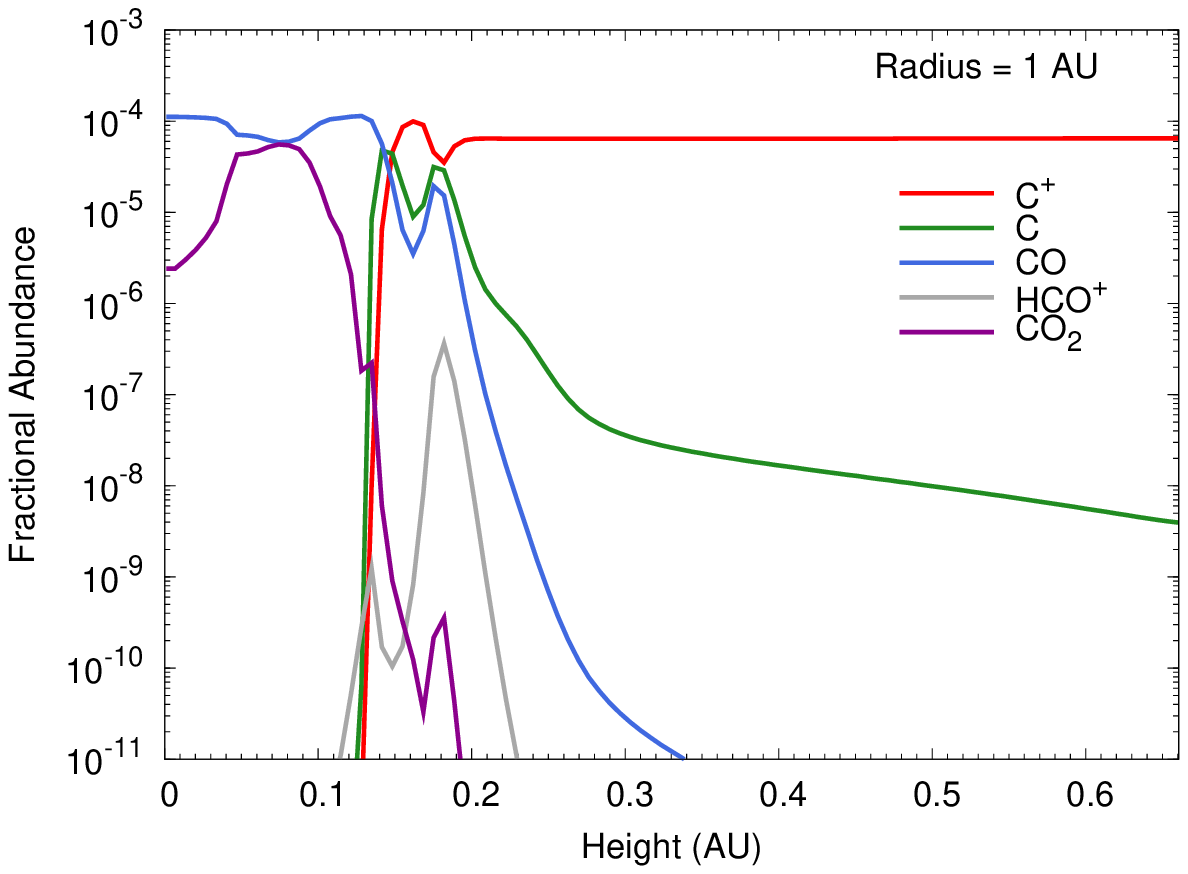}}
\subfigure{\includegraphics[width=0.32\textwidth]{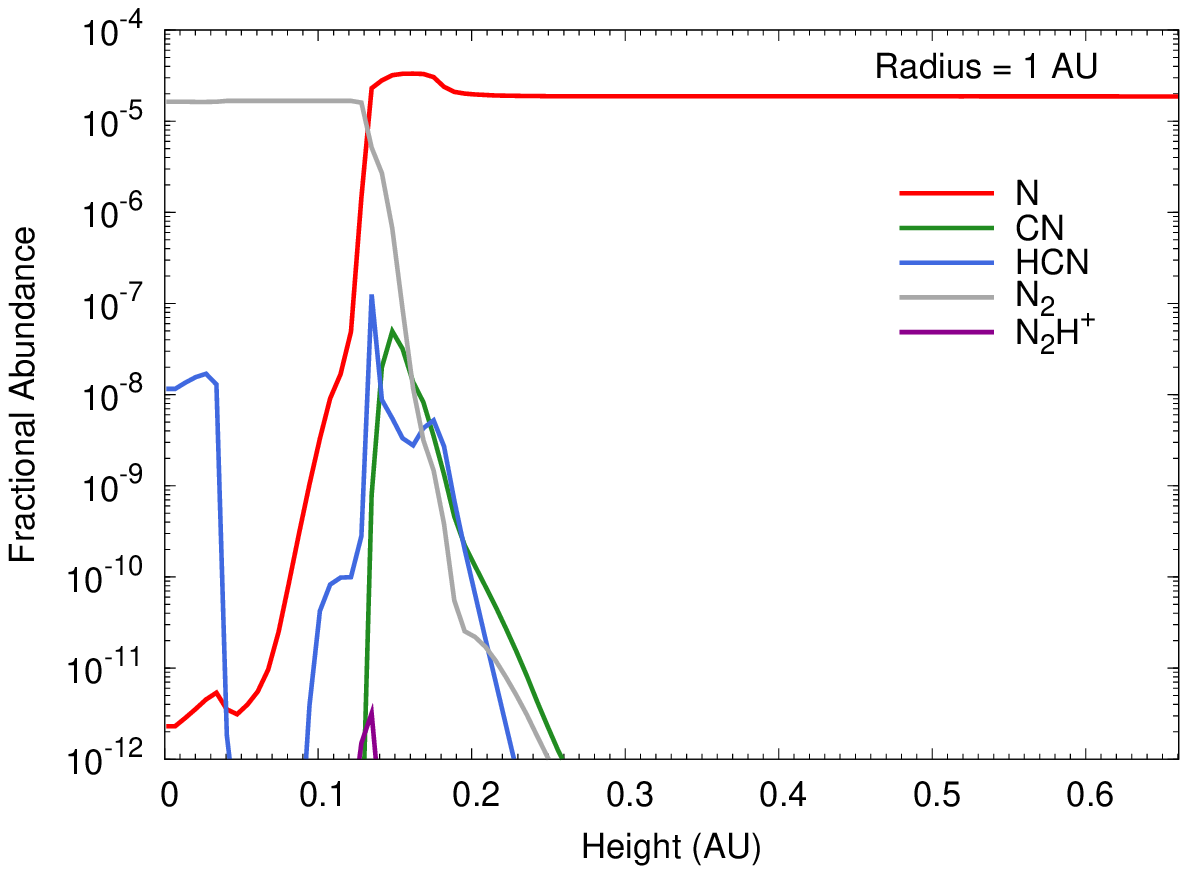}}
\subfigure{\includegraphics[width=0.32\textwidth]{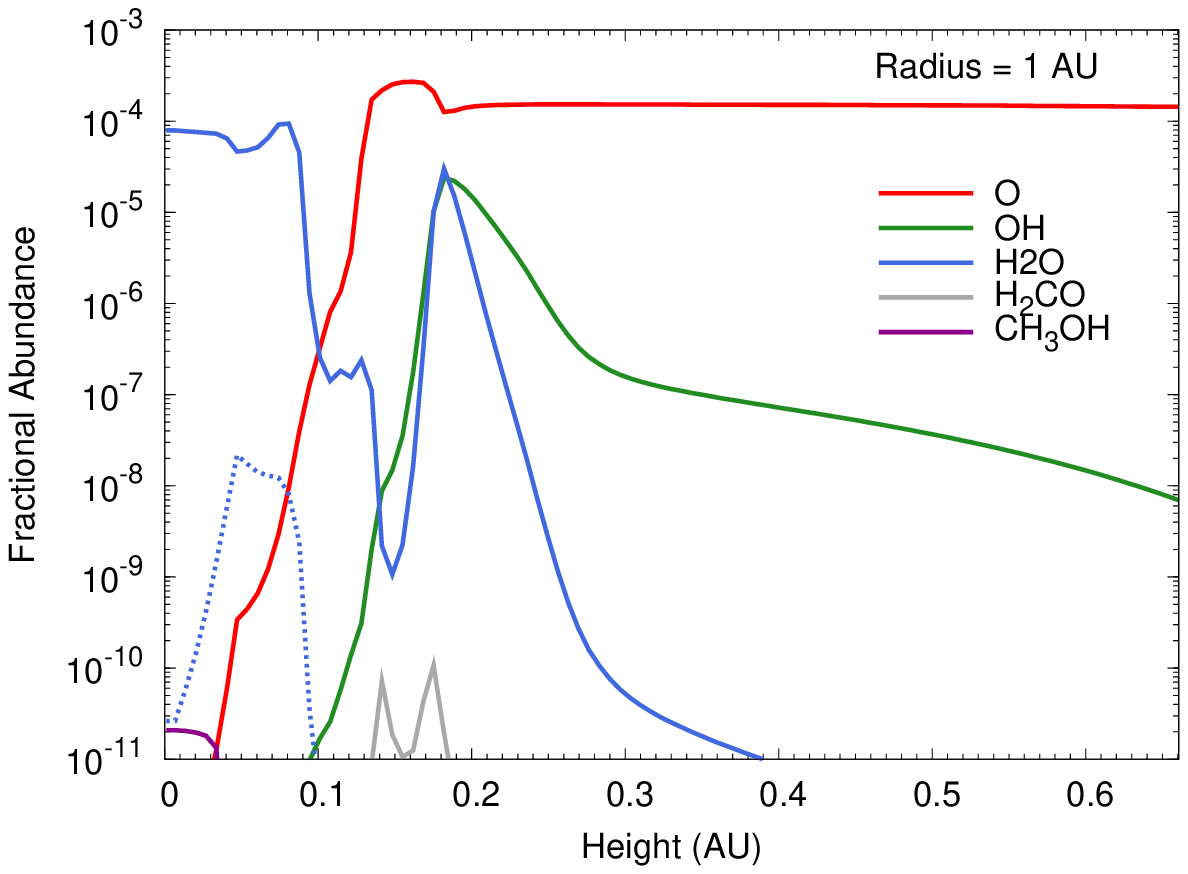}}
\subfigure{\includegraphics[width=0.32\textwidth]{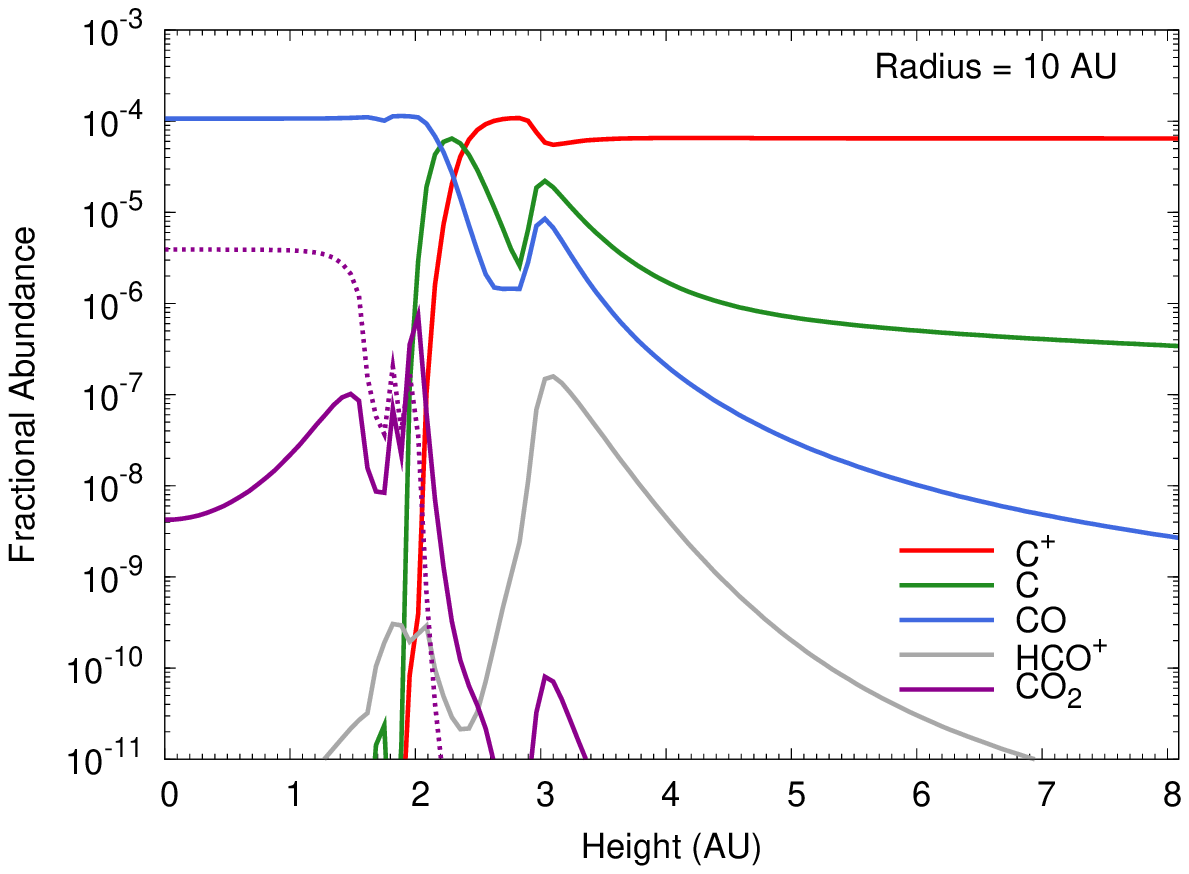}}
\subfigure{\includegraphics[width=0.32\textwidth]{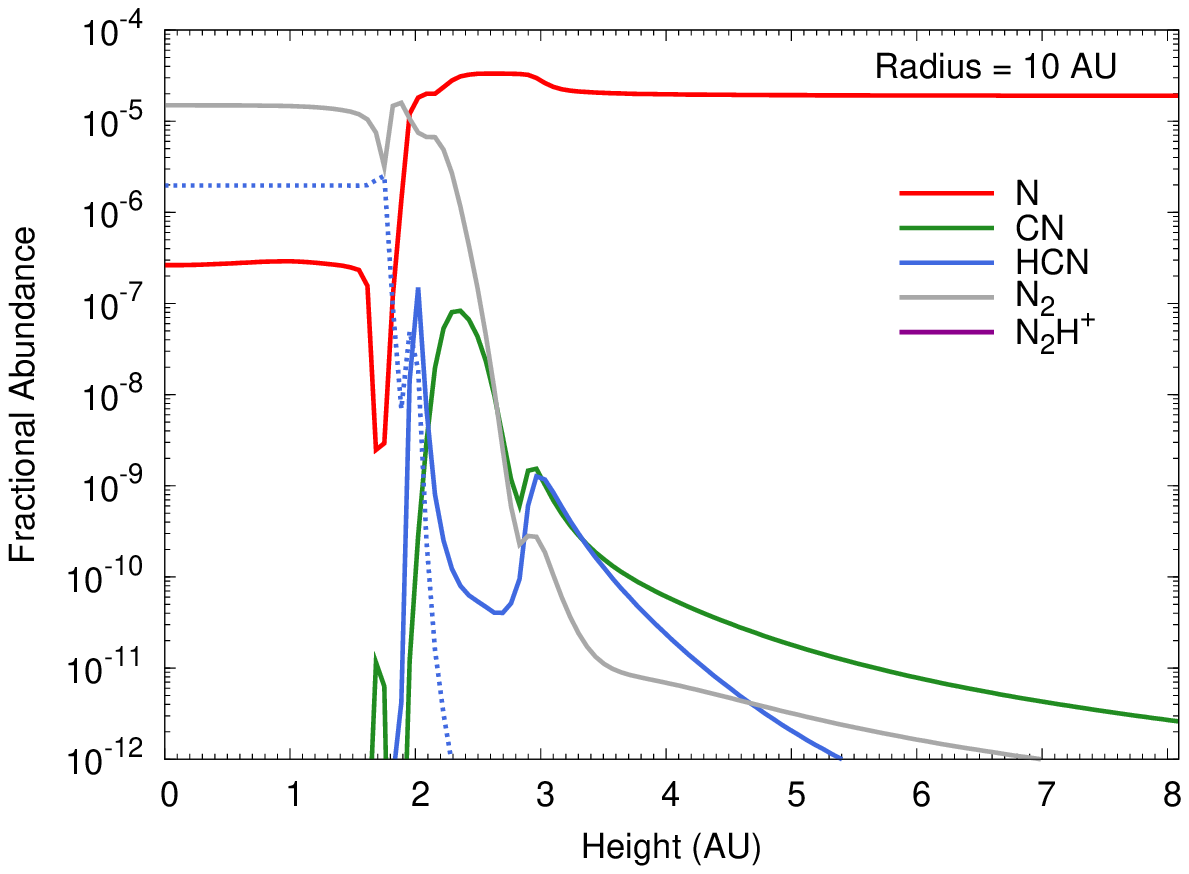}}
\subfigure{\includegraphics[width=0.32\textwidth]{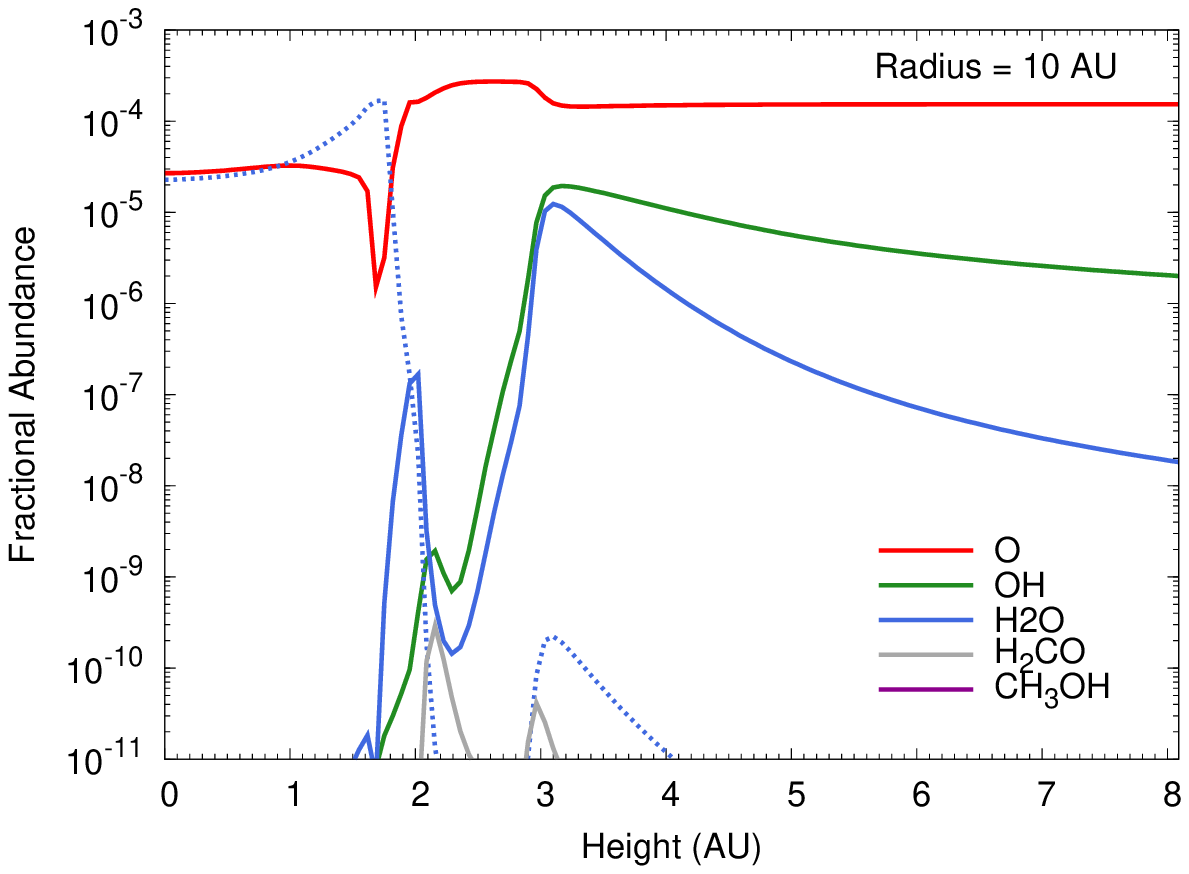}}
\subfigure{\includegraphics[width=0.32\textwidth]{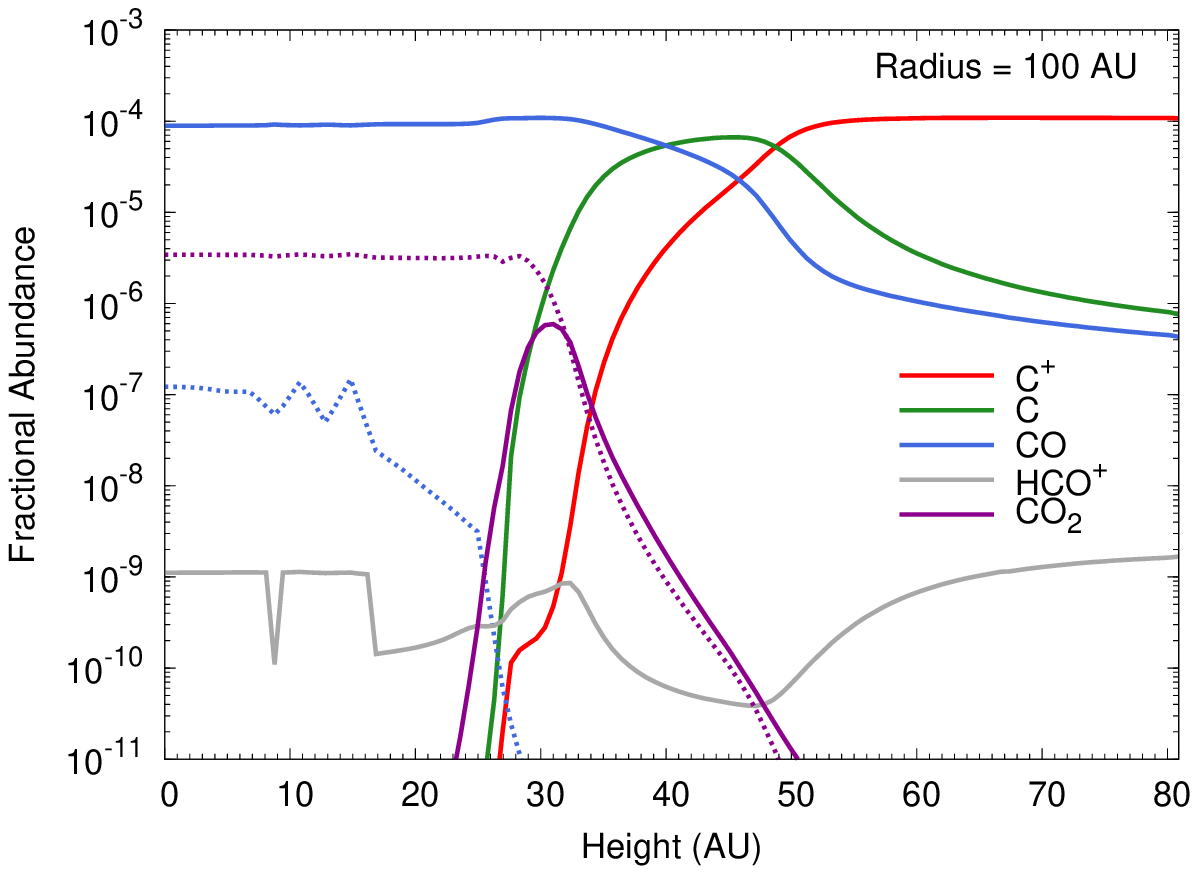}}
\subfigure{\includegraphics[width=0.32\textwidth]{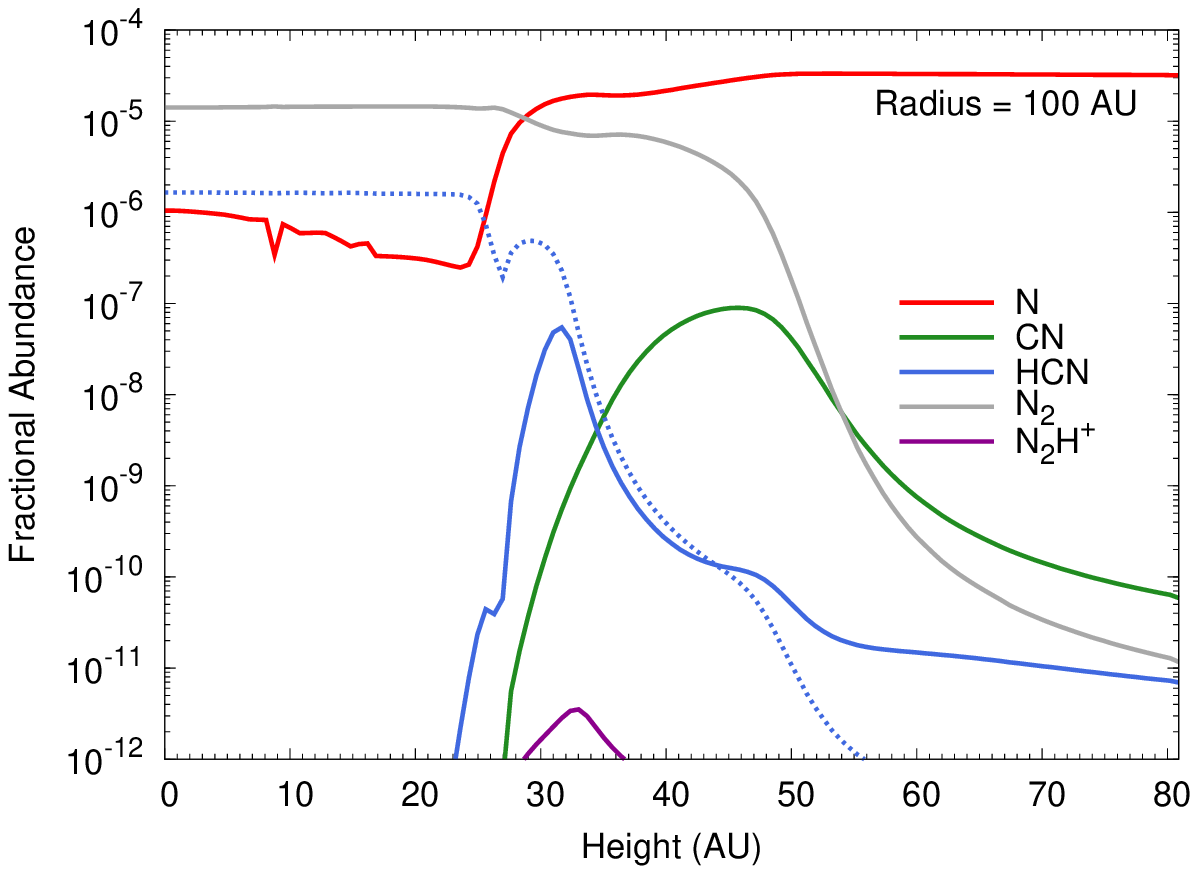}}
\subfigure{\includegraphics[width=0.32\textwidth]{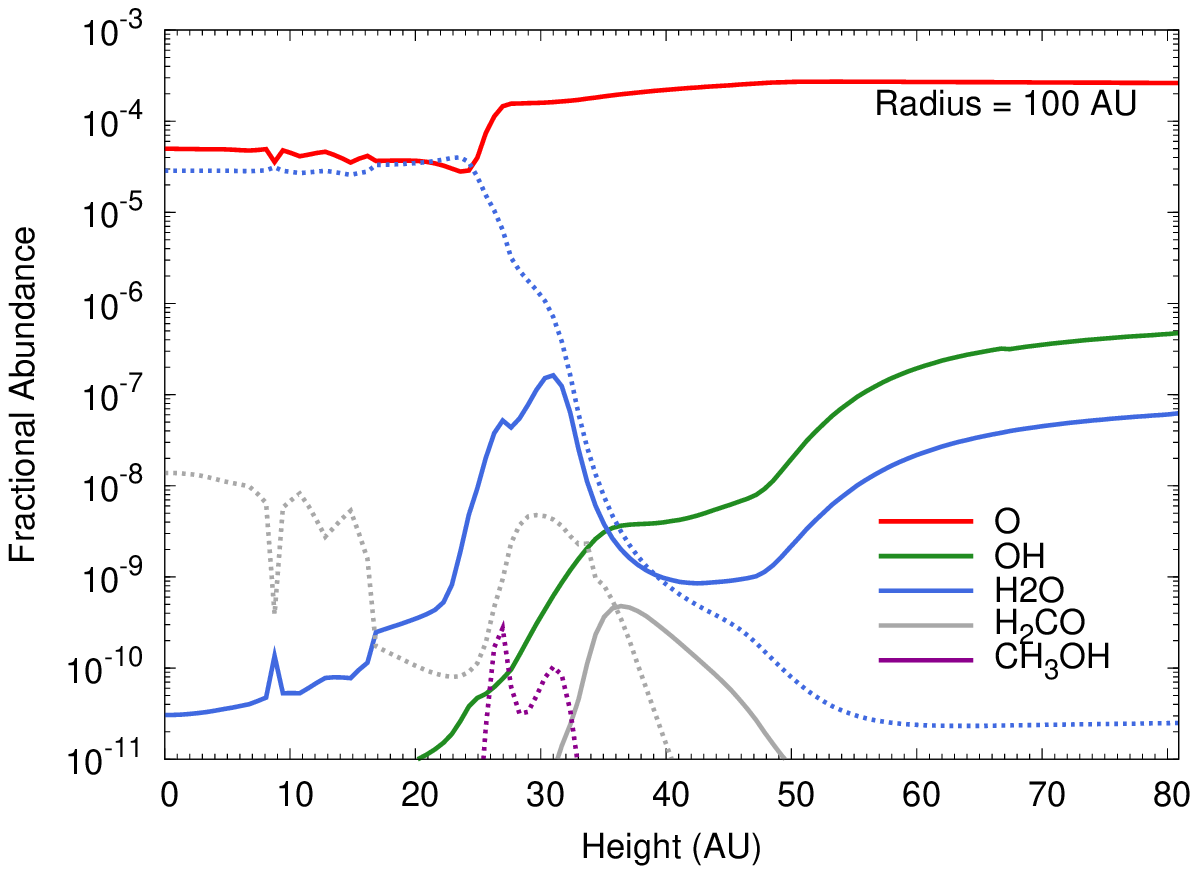}}
\caption{Fractional abundances as a function of disk height at radii, $r$~=~0.1~AU (top), 
1~AU (second), 10~AU (third) and 100~AU (bottom).  
Grain-surface (ice) abundances are represented by dotted lines. Figure available online only.}
\label{figure10}
\end{figure}

\begin{figure}
\centering
\subfigure{\includegraphics[width=0.32\textwidth]{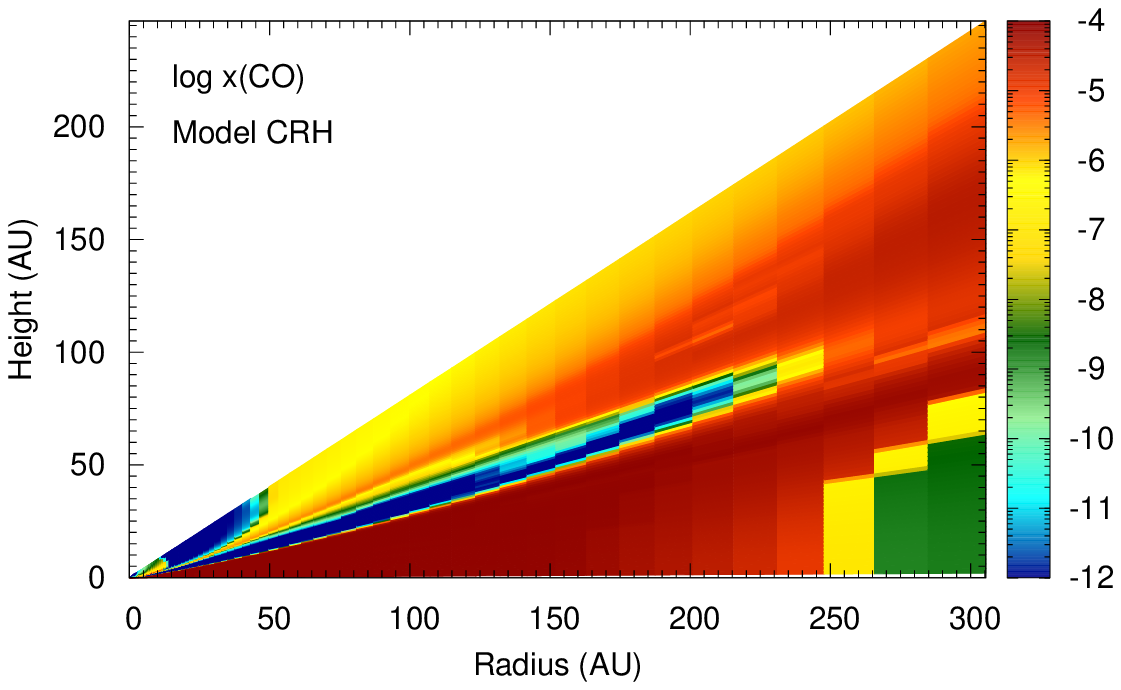}}
\subfigure{\includegraphics[width=0.32\textwidth]{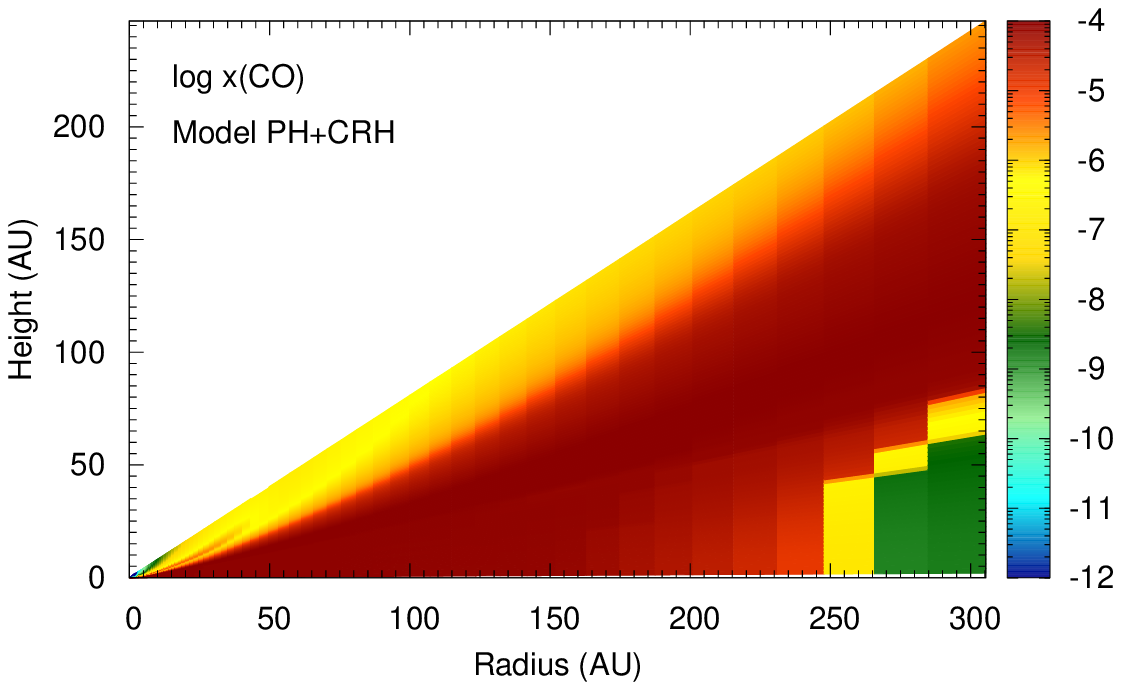}}
\subfigure{\includegraphics[width=0.32\textwidth]{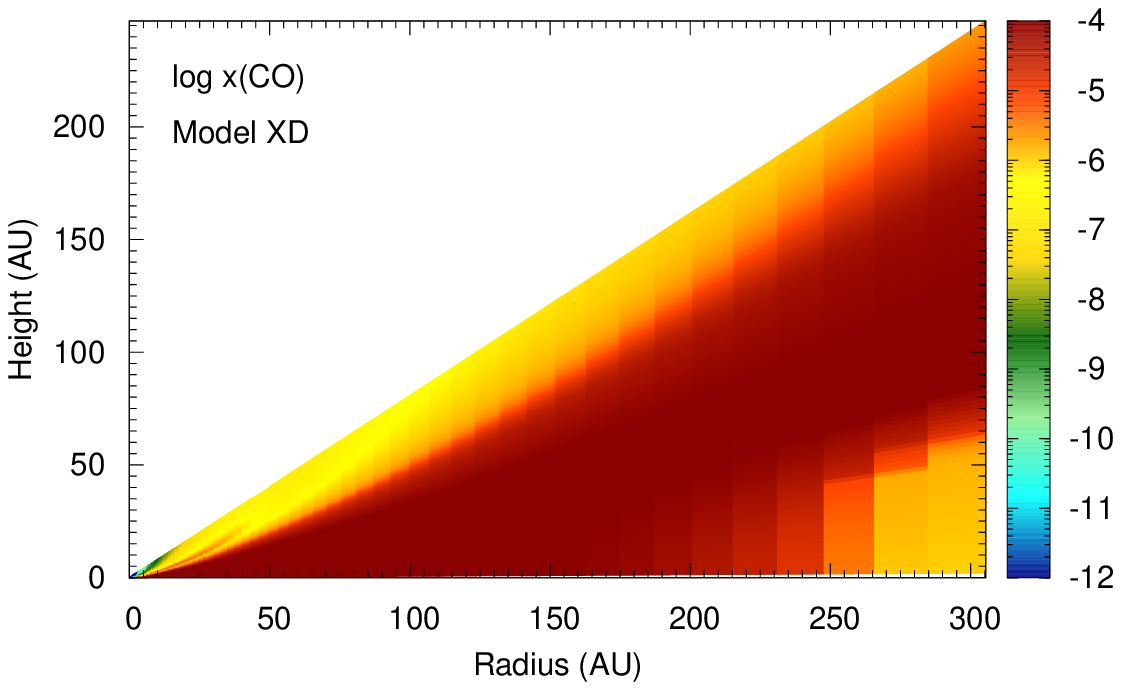}}
\subfigure{\includegraphics[width=0.32\textwidth]{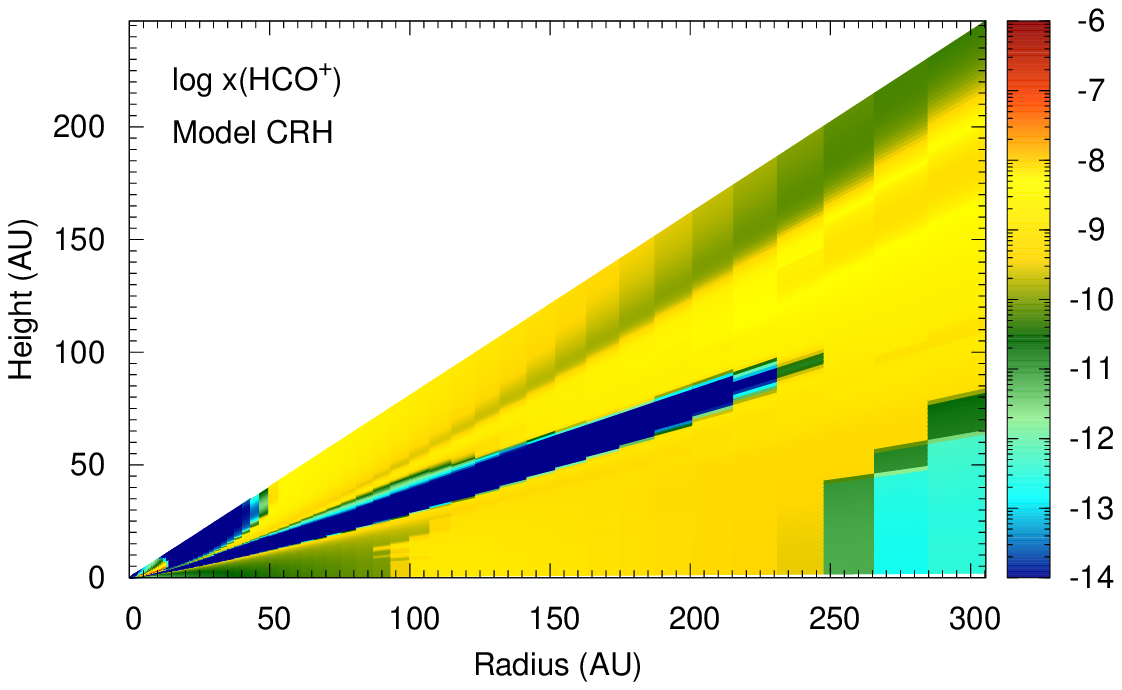}}
\subfigure{\includegraphics[width=0.32\textwidth]{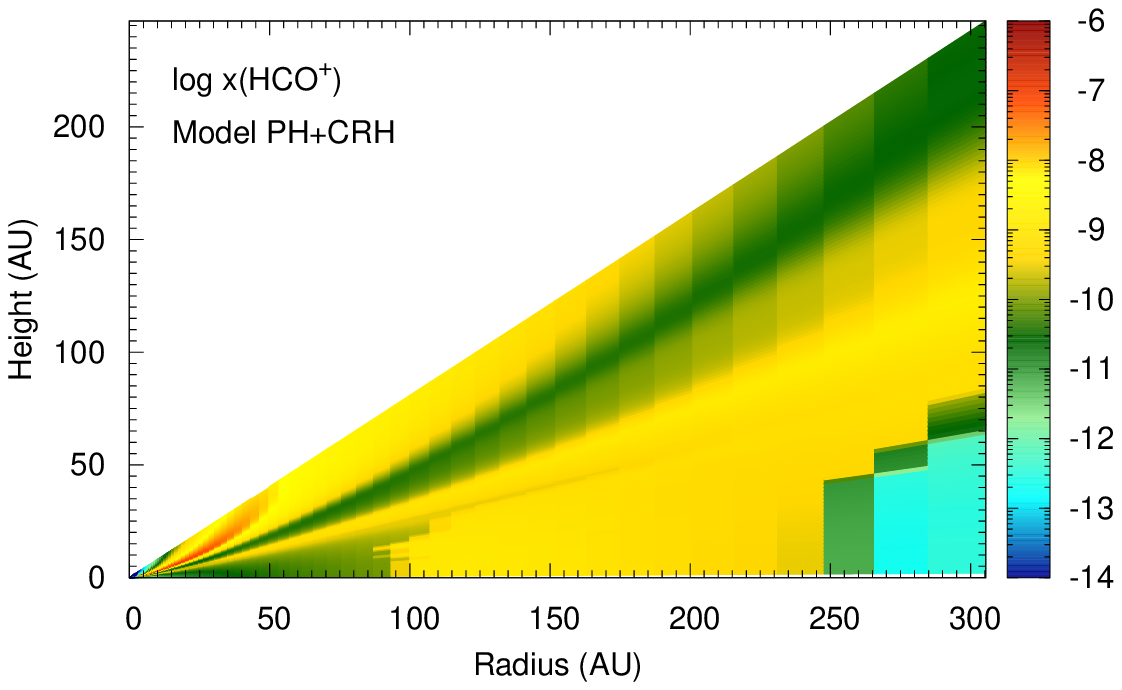}}
\subfigure{\includegraphics[width=0.32\textwidth]{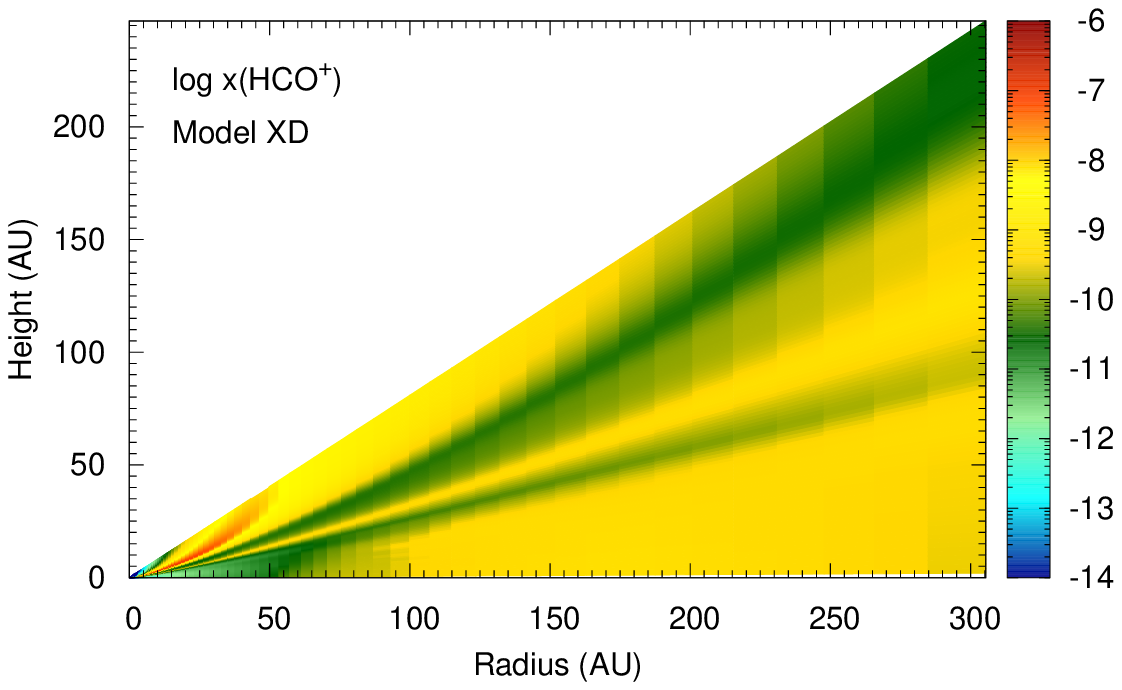}}
\subfigure{\includegraphics[width=0.32\textwidth]{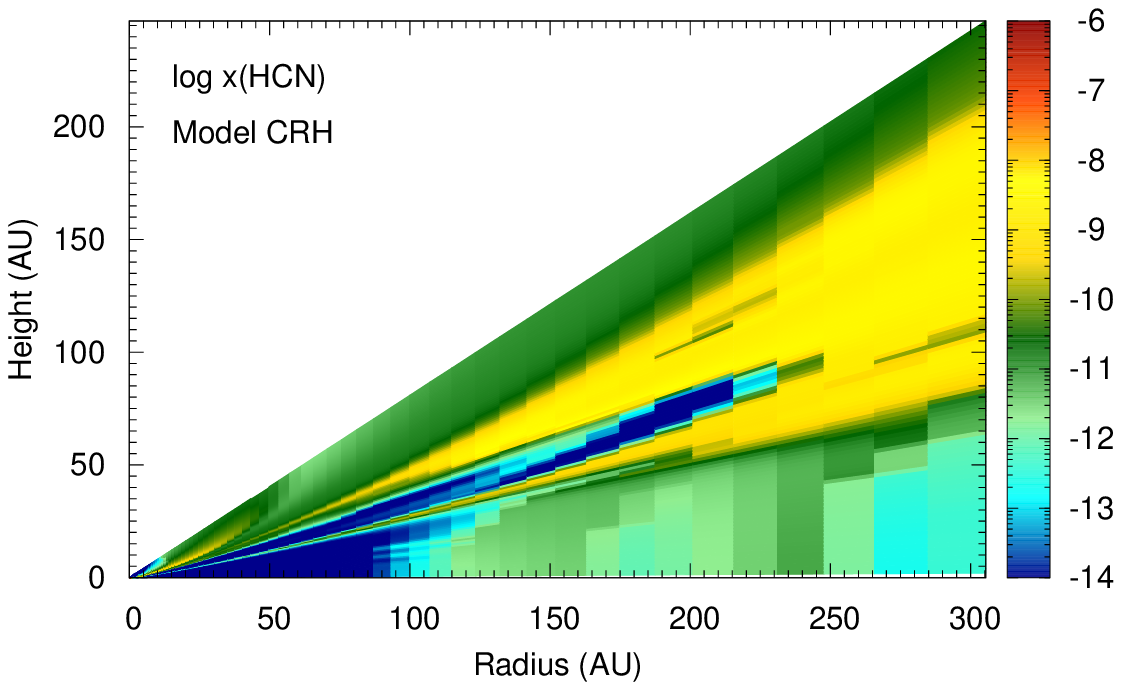}}
\subfigure{\includegraphics[width=0.32\textwidth]{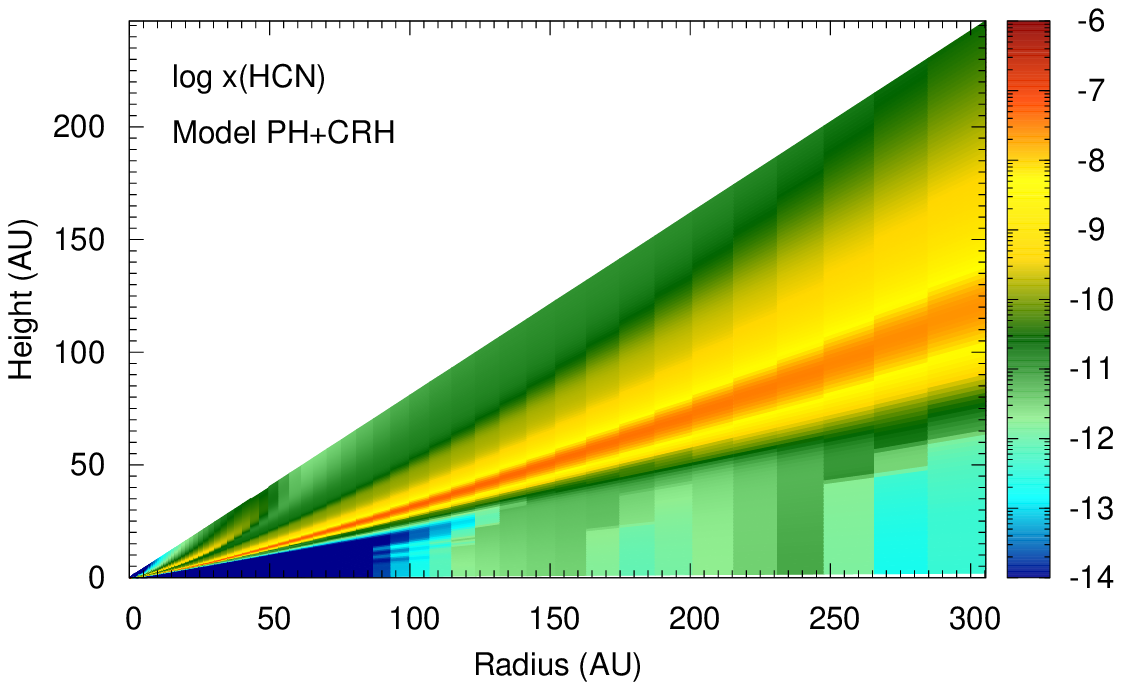}}
\subfigure{\includegraphics[width=0.32\textwidth]{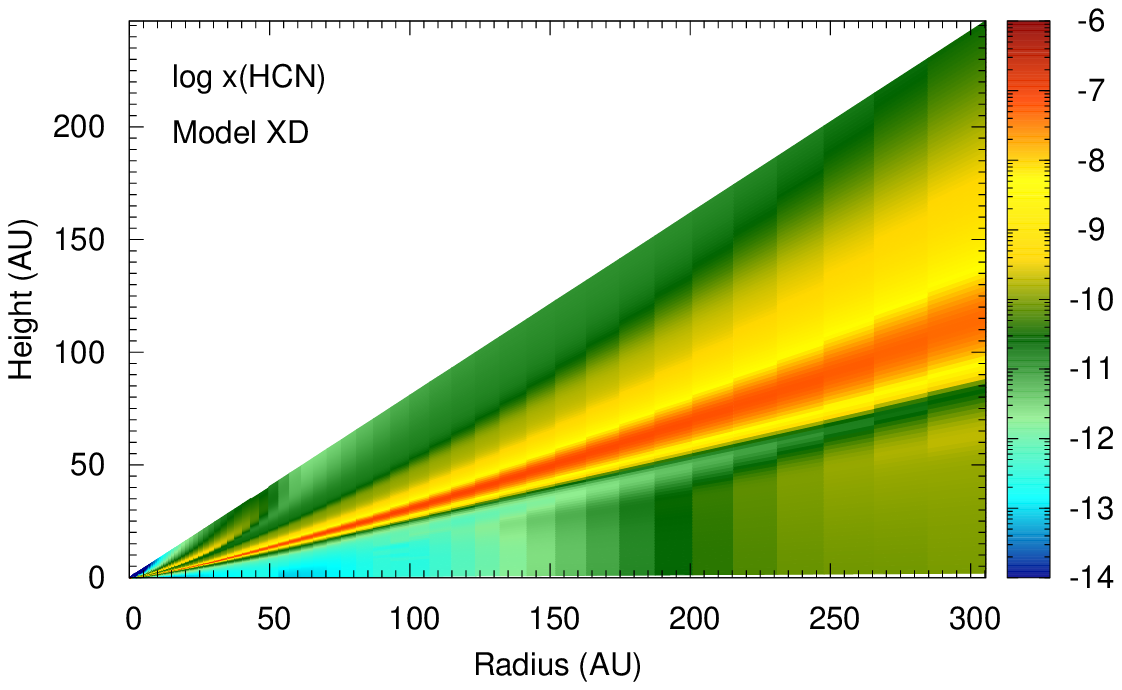}}
\subfigure{\includegraphics[width=0.32\textwidth]{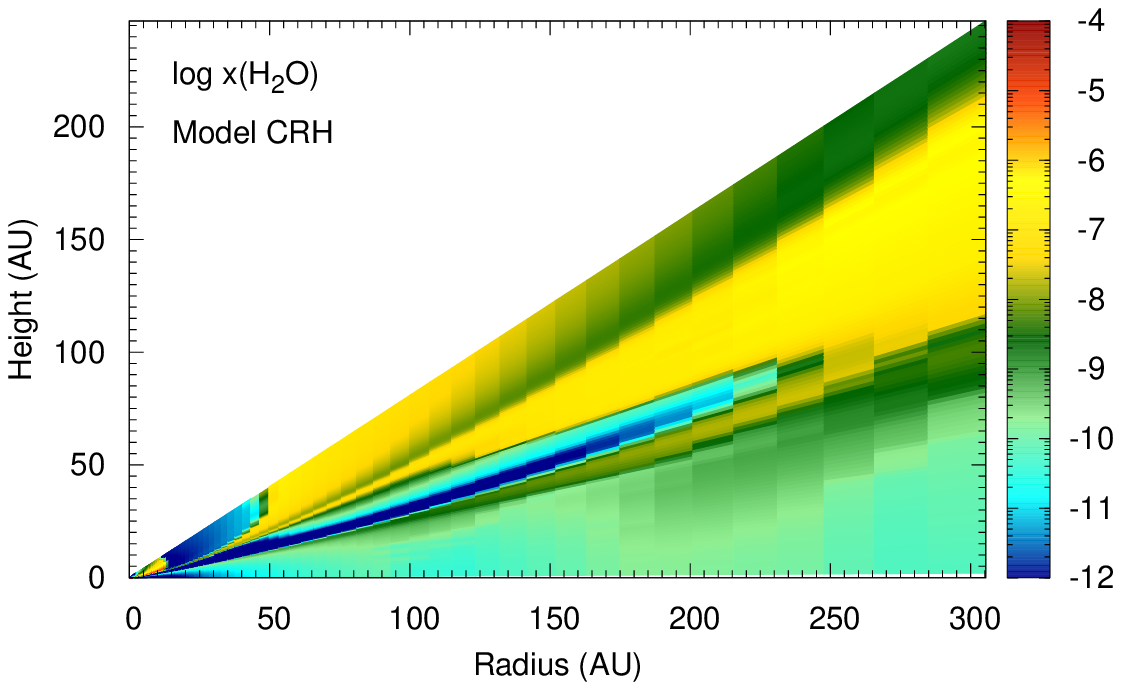}}
\subfigure{\includegraphics[width=0.32\textwidth]{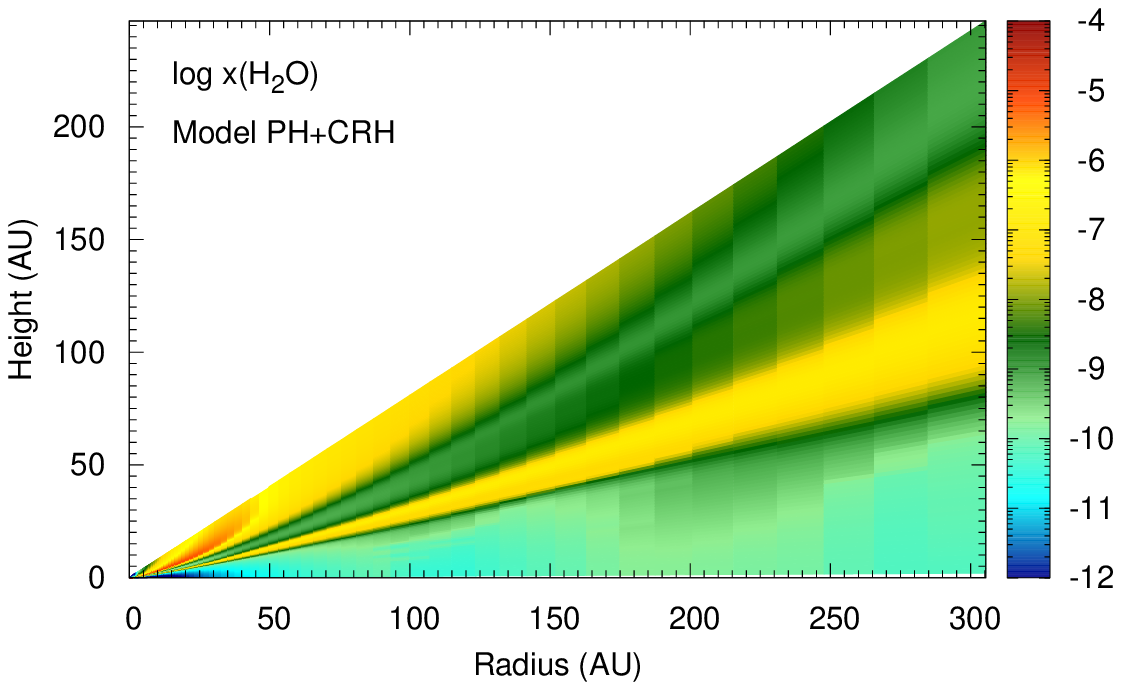}}
\subfigure{\includegraphics[width=0.32\textwidth]{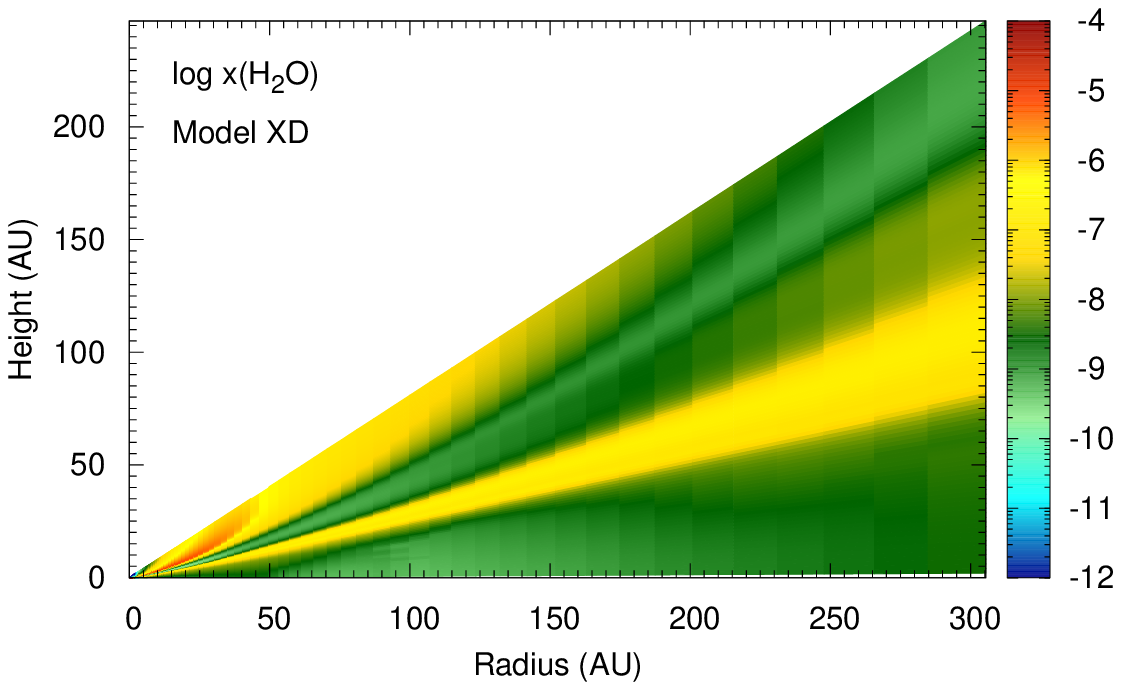}}
\subfigure{\includegraphics[width=0.32\textwidth]{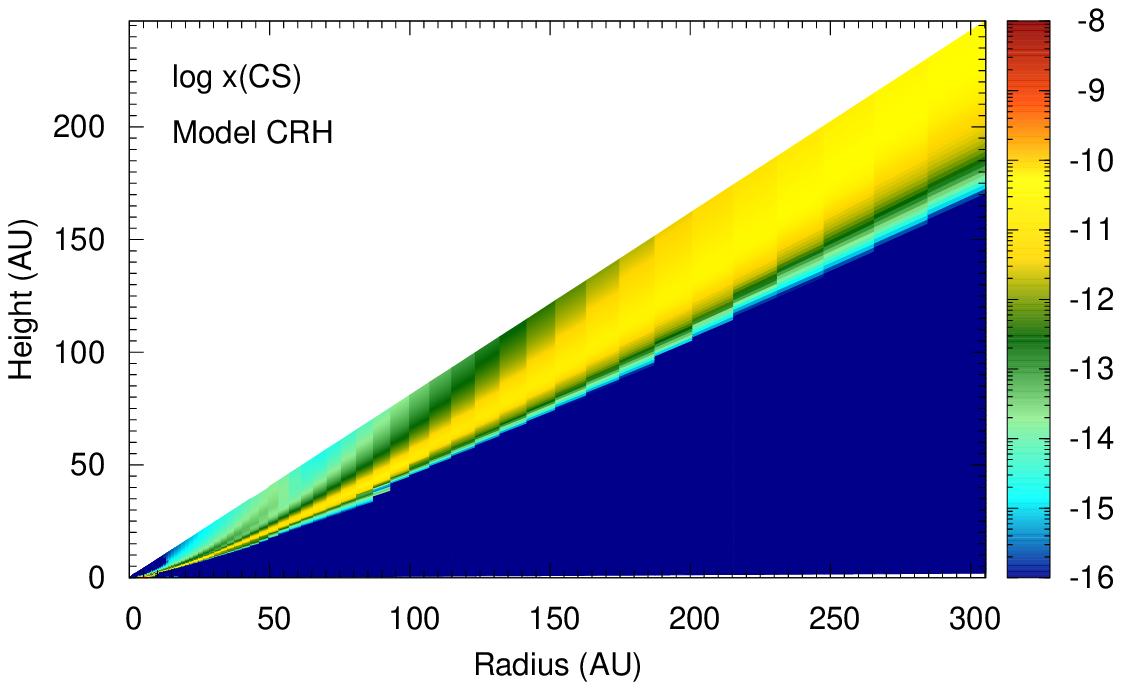}}
\subfigure{\includegraphics[width=0.32\textwidth]{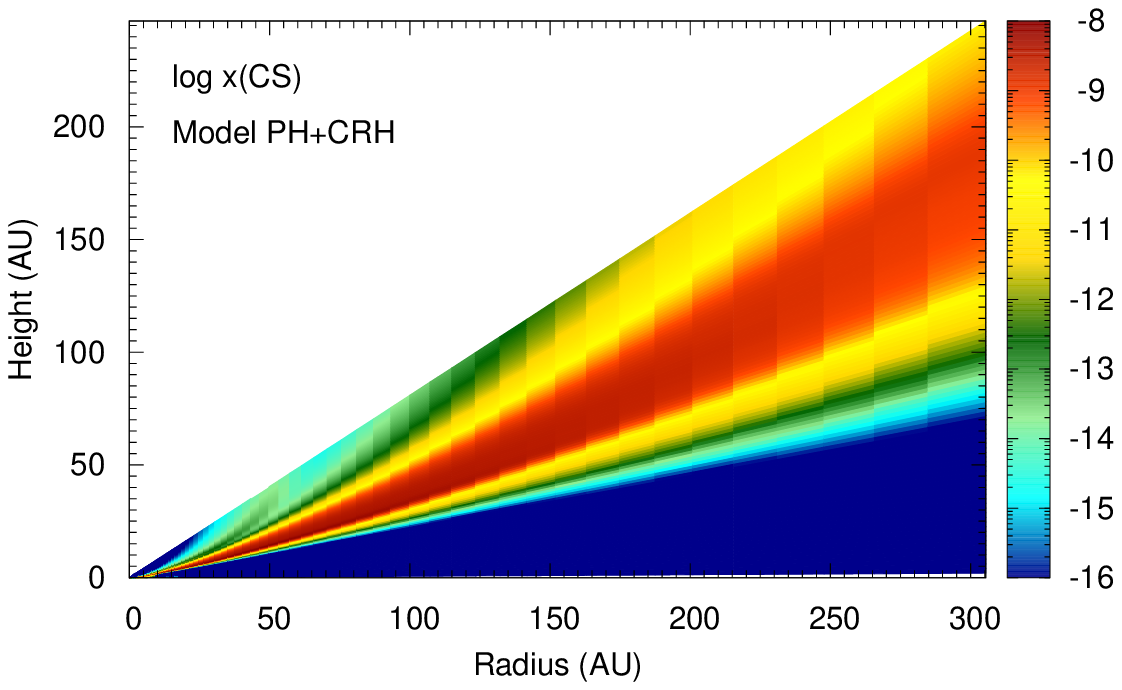}}
\subfigure{\includegraphics[width=0.32\textwidth]{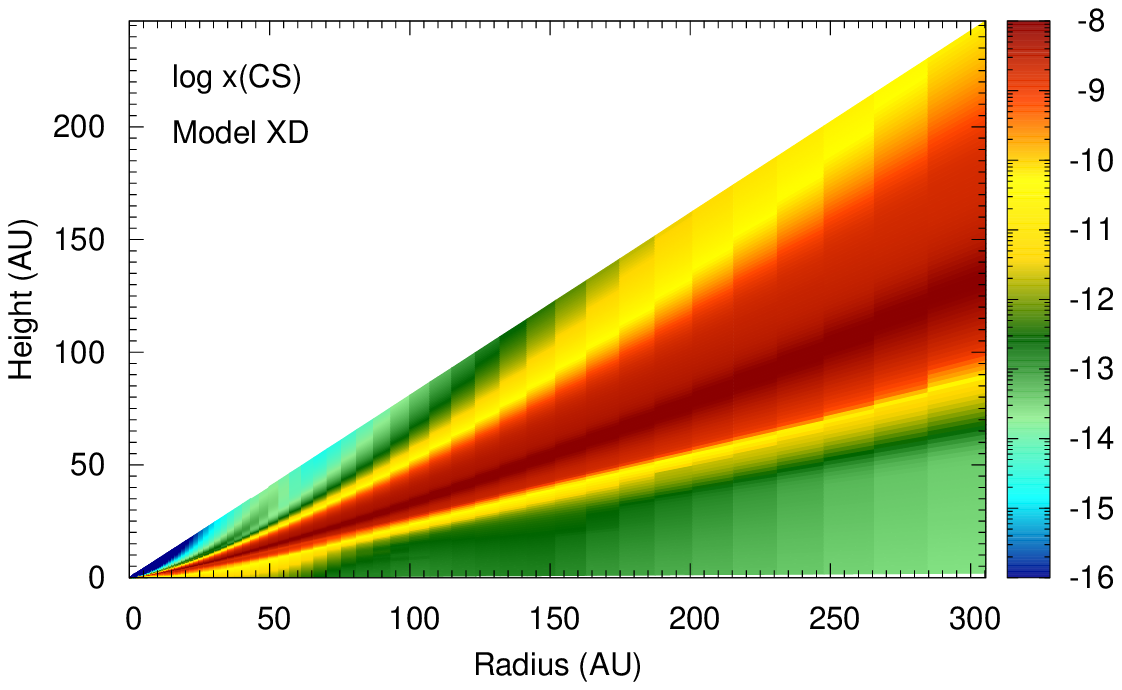}}
\subfigure{\includegraphics[width=0.32\textwidth]{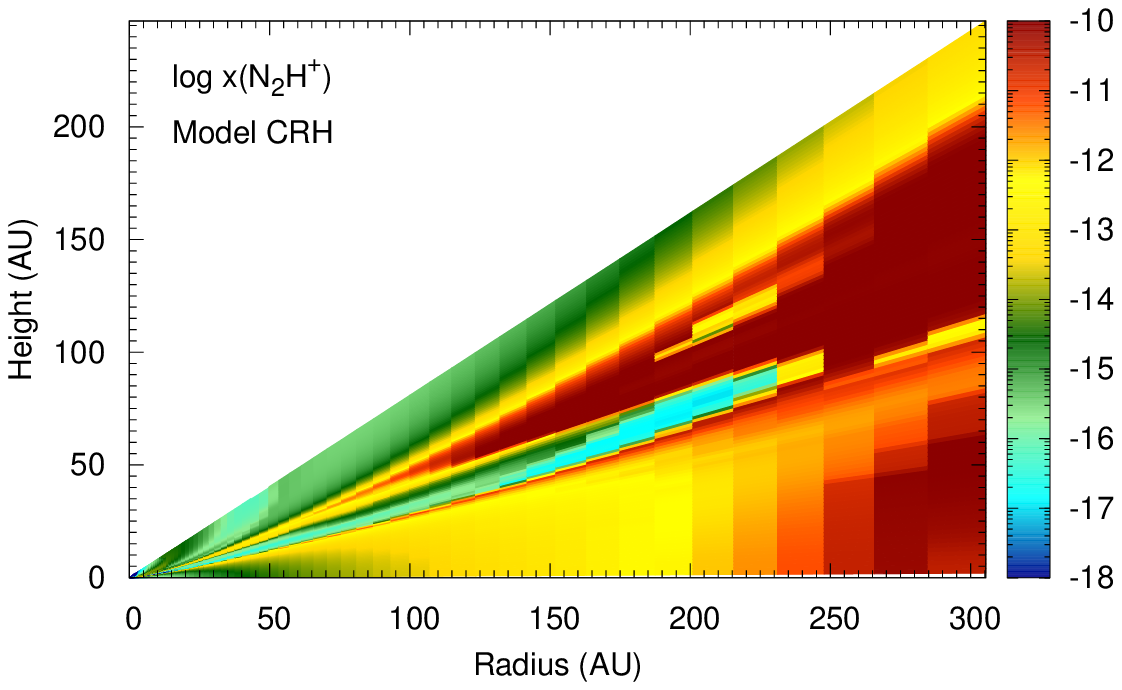}}
\subfigure{\includegraphics[width=0.32\textwidth]{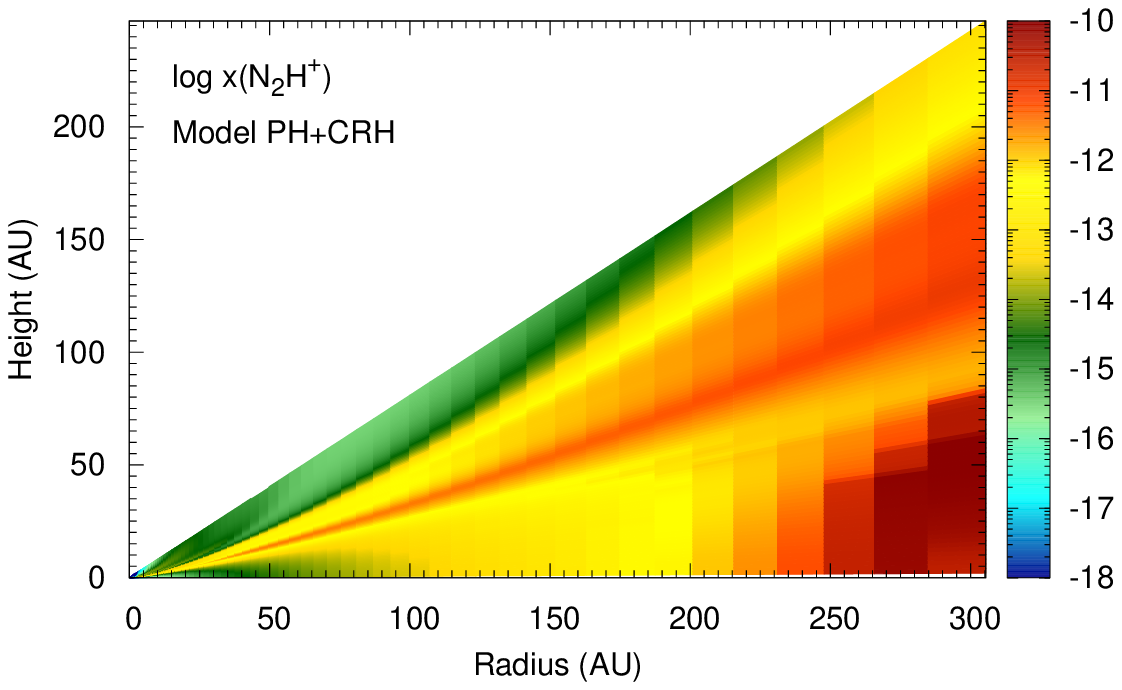}}
\subfigure{\includegraphics[width=0.32\textwidth]{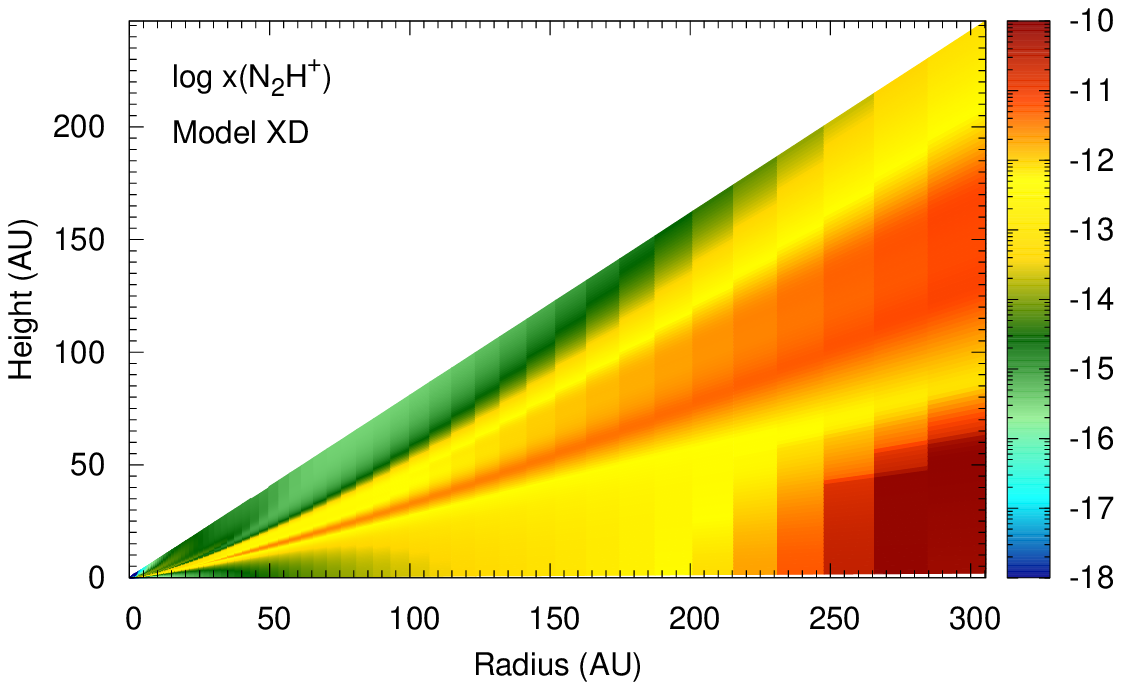}}
\caption{Fractional abundances as a function of disk radius and height for several 
gas-phase molecules for model CRH (left), PH+CRH (middle) and XD (right).  Figure available online only.}
\label{figure11}
\end{figure}

\begin{figure}
\subfigure{\includegraphics[width=0.5\textwidth]{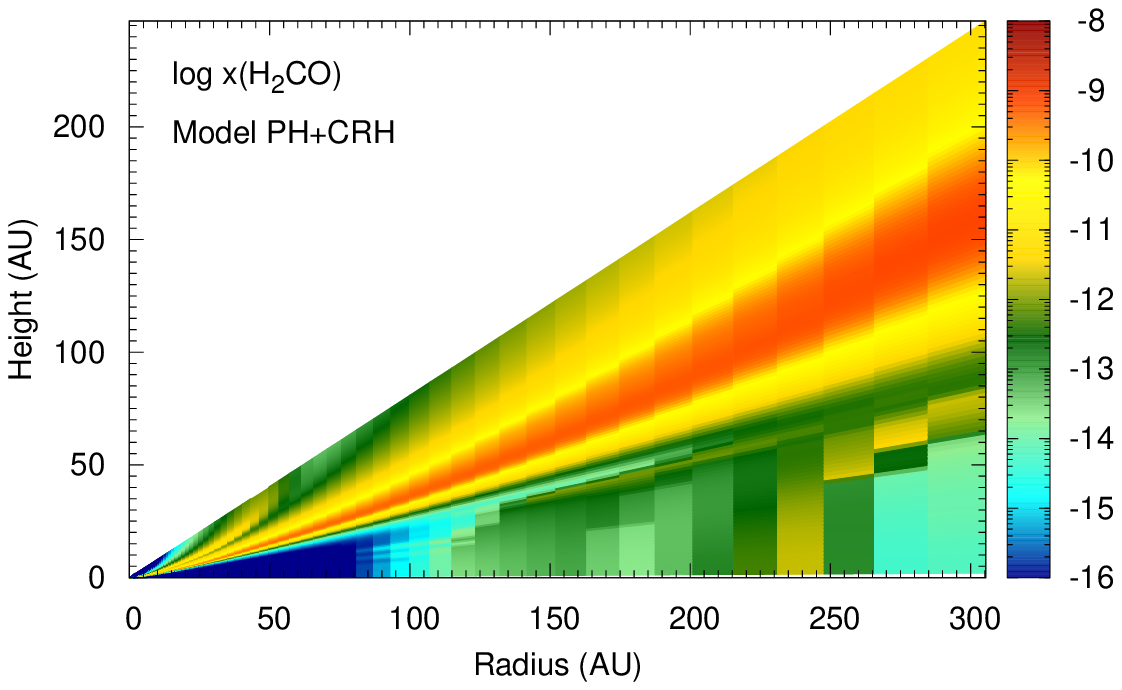}}
\subfigure{\includegraphics[width=0.5\textwidth]{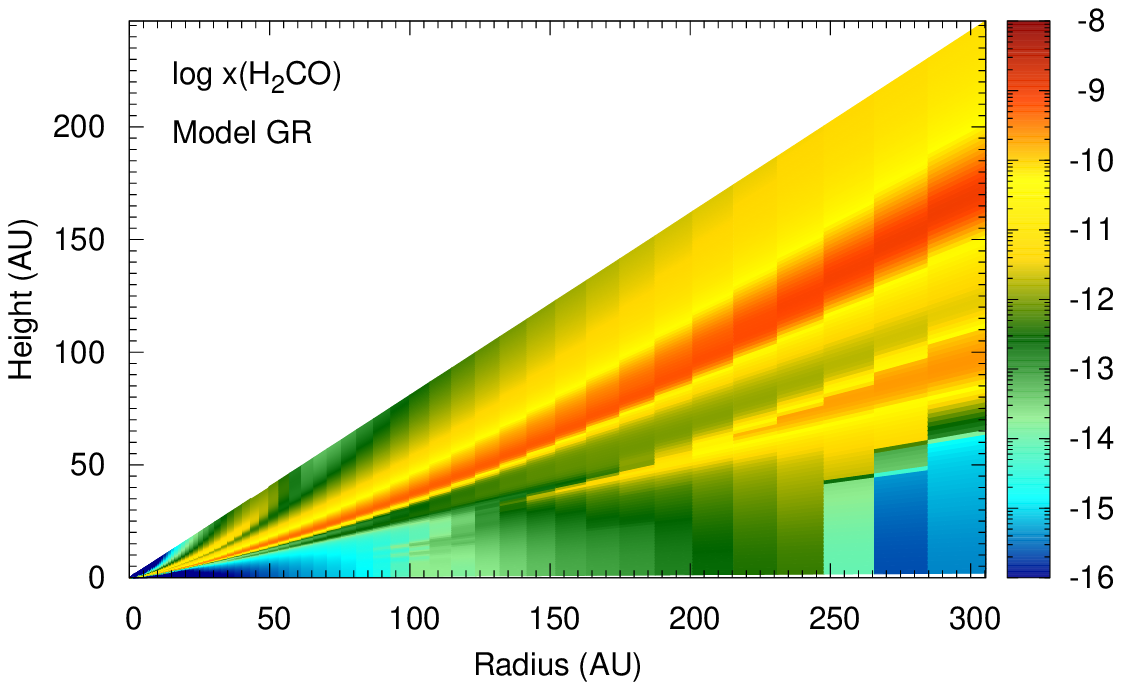}}
\subfigure{\includegraphics[width=0.5\textwidth]{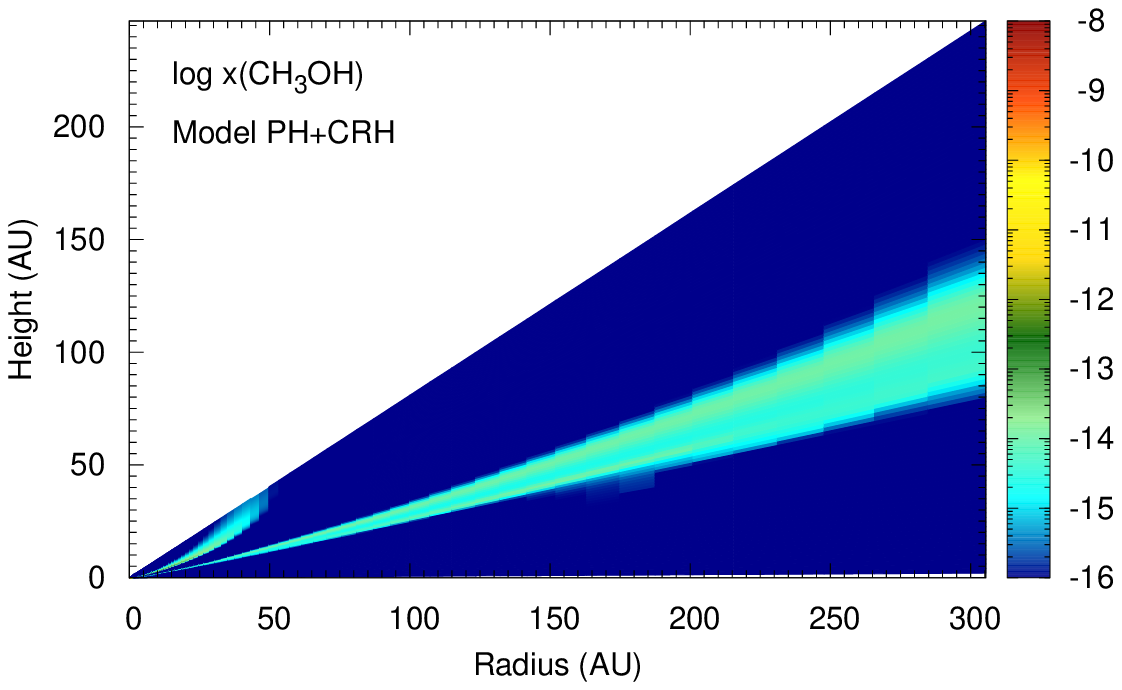}}
\subfigure{\includegraphics[width=0.5\textwidth]{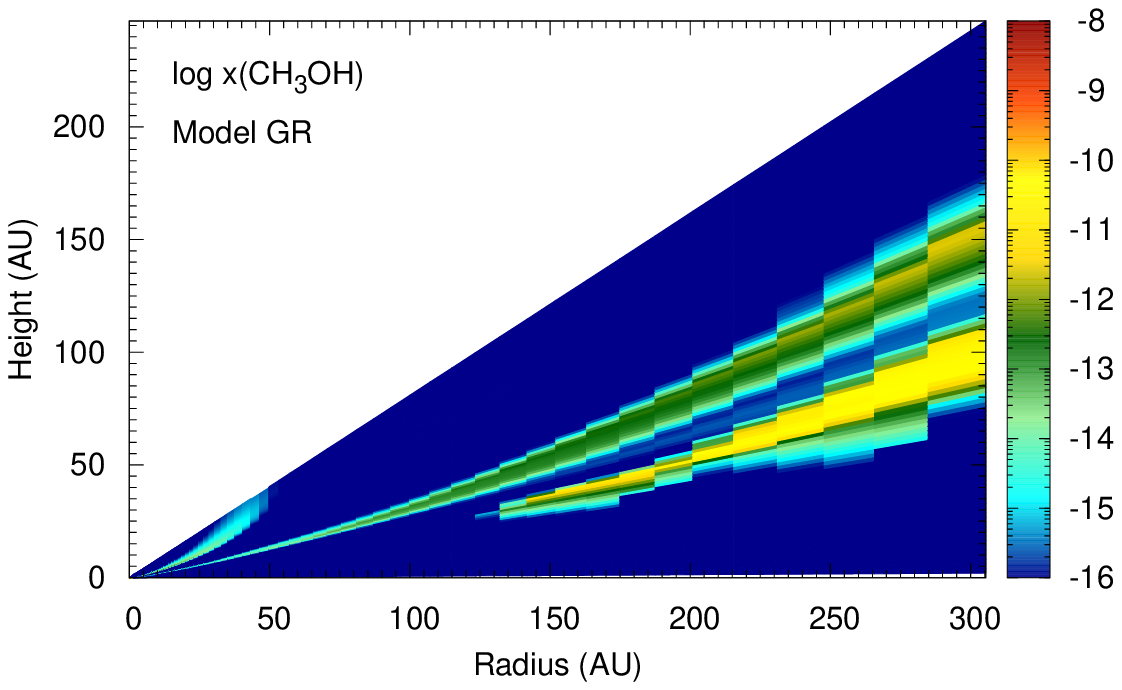}}
\subfigure{\includegraphics[width=0.5\textwidth]{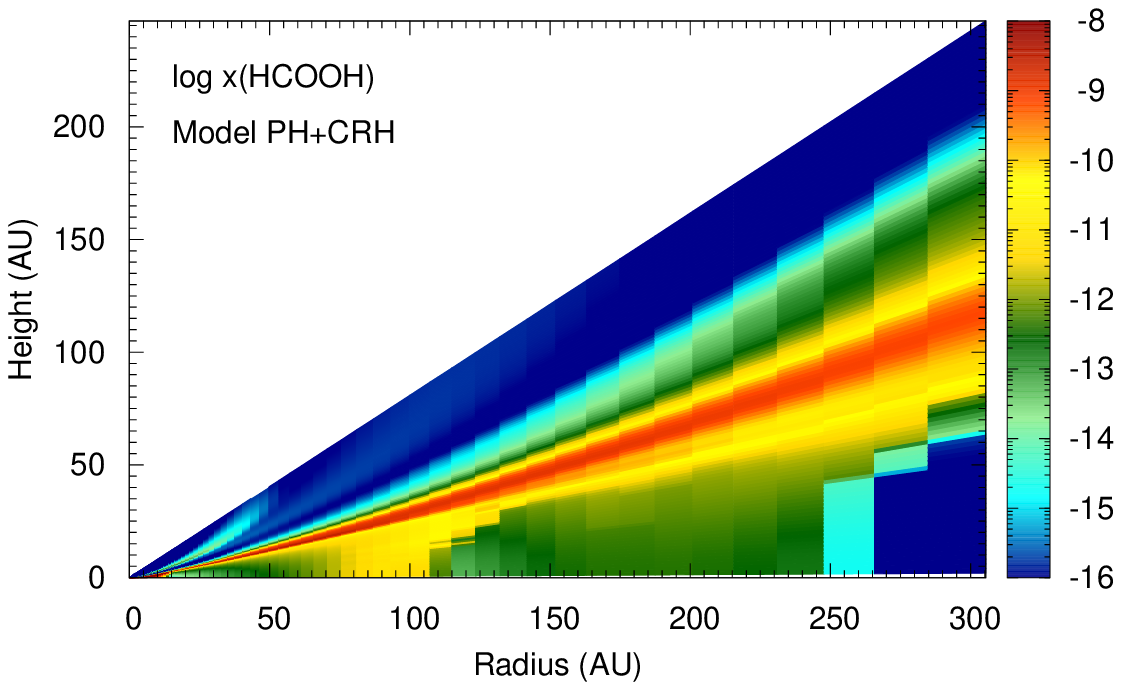}}
\subfigure{\includegraphics[width=0.5\textwidth]{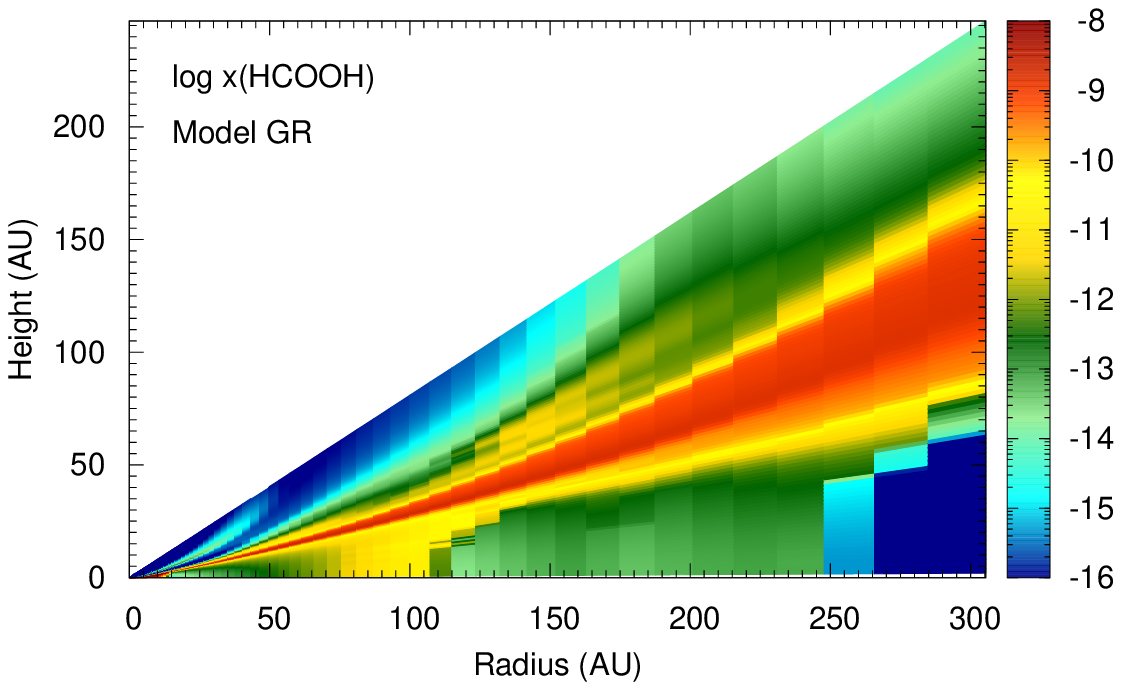}}
\subfigure{\includegraphics[width=0.5\textwidth]{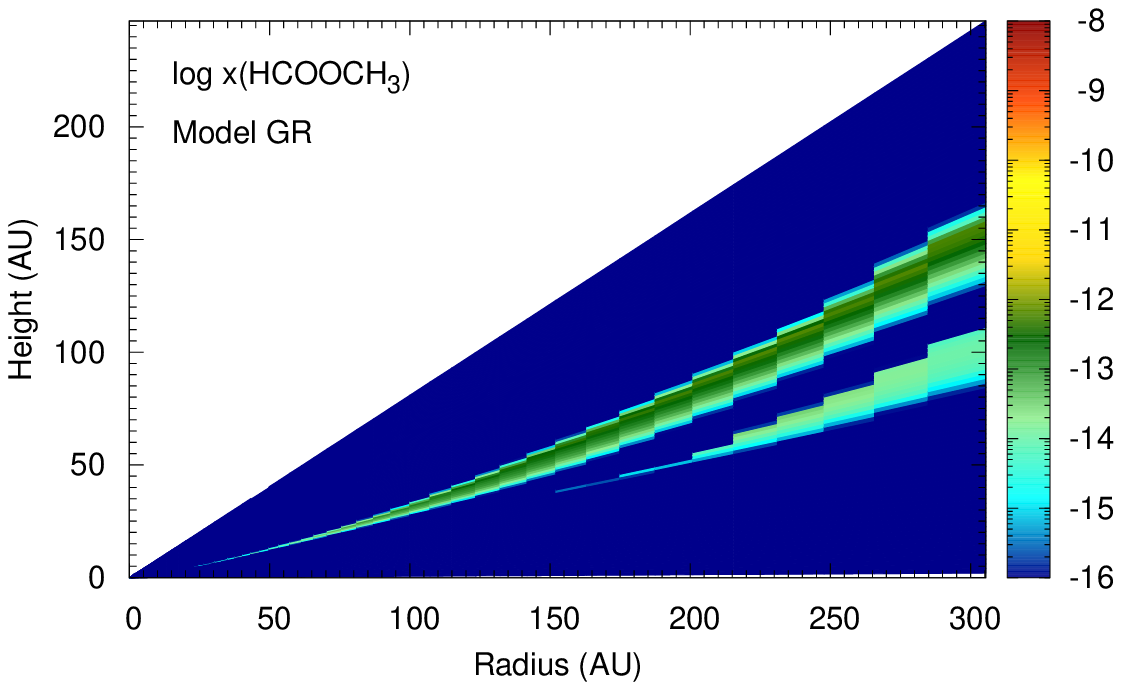}}
\subfigure{\includegraphics[width=0.5\textwidth]{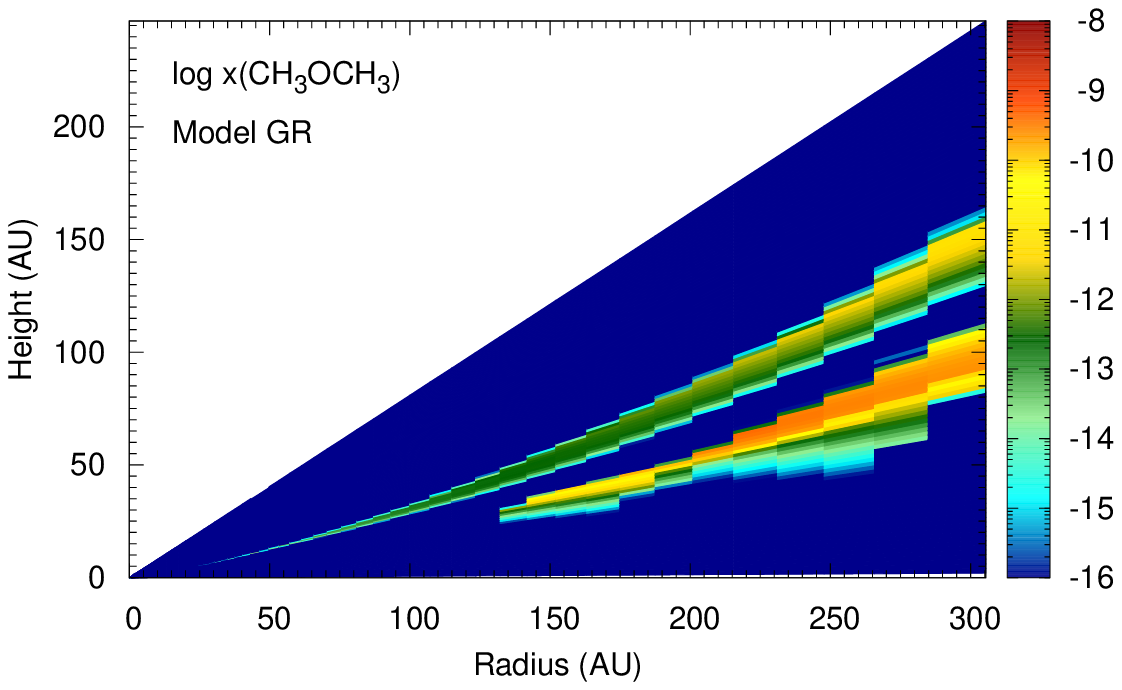}}
\caption{Fractional abundances as a function of disk radius and height for several 
organic molecules for model PH+CRH (left) and GR (right). 
We display results from model GR only for HCOOCH$_{3}$ and CH$_3$OCH$_3$ (bottom row) 
as model PH+CRH produces negligible fractional abundances ($<$~10$^{-16}$) 
for both of these molecules throughout the disk.   Figure available online only.}
\label{figure12}
\end{figure}

Figure~\ref{figure8} shows the electron fractional abundance as a function of disk radius and 
height within a radius of 10~AU (left) and 305~AU (right) using the results from model PH+CRH.    
The ionisation fraction varies between a minimum value of $\sim$~$10^{-12}$ in the densest region of 
the disk up to a value of $\sim$~0.1 in the hottest, most irradiated surface region closest to 
the star.  
 We find that we attain similar electron abundances throughout the disk regardless of 
chemical model.  The ionisation threshold required for effective accretion is related to the magnetic 
Reynold's number \citep{gammie96} which must be determined in advance of addressing the location of 
any `dead zones' in our particular star-disk system.
The question of whether accretion is suppressed in our disk model will be considered in detail 
in a subsequent publication in which we also investigate the effects of the recalculation 
of photo-rates and direct X-ray ionisation on the disk chemical structure and ionisation 
fraction.  

\subsection{Radial Column Densities}
\label{columndensities}

The column density, $N_i$, at each radius, $r$, for each species, $i$, 
is calculated by integrating the number density over the depth of the disk i.e.\
\begin{equation}
N_i(r) = \int^{z_{+\infty}}_{z_{-\infty}} n_i(r,z)\; \mathrm{d}z \qquad \mathrm{cm^{-2}}.  
\label{columdensity}
\end{equation}
The radial column densities provide an excellent means to trace the radial mass 
distribution in the disk and also to compare directly results from each of our chemical models to 
determine the species sensitive to each chemical process.  
In Table~\ref{table3} we list the column densities of various important molecules 
at radii of 1~AU, 10~AU, 100~AU and 305~AU for each chemical model and we display 
the column densities of many of the molecules discussed thus far, 
as a function of radius, up to maximum 
radii of 10~AU (left) and 305~AU (right) in Figure~\ref{figure13} (online only). 

At a radius of 1~AU, the column densities are relatively insensitive to the choice 
of chemical model, unsurprising given that most molecules are in the  gas phase
at this radius and the chemistry is dominated in the  midplane by neutral-neutral 
reactions and in the surface by photo-chemistry.  

At 10~AU, the effects of the inclusion of non-thermal desorption become apparent.  
The column densities for those models which include  photodesorption 
are consistently higher for all molecules with the column densities of 
HCN, CN and CS particularly 
sensitive.  
X-ray desorption has a similar effect although we also see a dramatic increase in the 
column densities of H$_2$O and CO$_2$ due to the penetrative power of X-rays in this 
region.  
This is due to the relatively strong X-ray field at this radius coupled with 
the low column density of intervening absorbing material (from the disk surface 
to the  midplane).
Grain-surface chemistry has only a mild effect at 10~AU on the column densities of 
the listed molecules.  
For the molecular ions, HCO$^+$ and N$_2$H$^+$, the addition of  photodesorption  
causes a rise in  the column densities of both molecules while the addition 
of X-ray desorption produces a fall in the column density of the former 
and a rise in that of the latter.  

At 100~AU, we see some of the same behaviour as at 10~AU with  photodesorption 
increasing the column densities of CO, HCN, CN, CS, C$_2$H, H$_2$CO, H$_2$O, CO$_2$ and 
C$_2$H$_2$ with HCN, CS and CO$_2$ particularly affected.  
X-ray desorption further enhances the column densities of these species.  
Here, we begin to see the effects of grain-surface chemistry with 
HCN, H$_2$O and CO$_2$ casualties of the increased synthesis of more complex species.  

Finally, in the outer disk, the effects of cosmic-ray induced desorption 
on the column density of N$_2$H$^{+}$ become apparent, so that this molecule is 
a potential observable tracer of this desorption mechanism.  
We can also see how both  photodesorption and X-ray desorption act to counteract 
the depletion of CO onto dust grains in the outer cold disk  midplane thus 
enhancing its overall column density.    
In the outer disk,  observable in the column densities is the detrimental effect 
that the addition of grain-surface chemistry 
has on the column densities of HCN and CS in particular, as N and S atoms are incorporated 
into larger, more complex species, via grain-surface reactions. 
We can also see the dramatic effect on methanol due to the inclusion of 
grain-surface chemistry with its column density enhanced by around three orders of 
magnitude.  

\begin{deluxetable}{lccccc}
\tablecaption{Column Densities\label{table3}}
\tablewidth{0pt}
\tablehead{\colhead{Species} & \colhead{0} & \colhead{CRH} & \colhead{PH+CRH} & \colhead{XD} &
\colhead{GR}}
\startdata
\cutinhead{1 AU}
H$_2$       & 1.9(25) & 1.9(25) & 1.9(25) & 1.9(25) & 1.9(25)\\
CO          & 2.3(21) & 2.3(21) & 2.3(21) & 2.3(21) & 2.3(21) \\
HCO$^+$     & 1.7(14) & 1.7(14) & 1.7(14) & 1.7(14) & 1.7(14) \\
HCN         & 2.0(17) & 2.0(17) & 2.0(17) & 2.0(17) & 2.0(17) \\
CN          & 2.7(14) & 2.7(14) & 2.7(14) & 2.7(14) & 2.7(14) \\
CS          & 4.5(12) & 4.5(12) & 4.5(12) & 4.5(12) & 4.5(12) \\
C$_2$H      & 4.1(14) & 4.1(14) & 4.1(14) & 4.1(14) & 4.1(14) \\
H$_2$CO     & 1.3(12) & 1.3(12) & 1.3(12) & 1.3(12) & 1.3(12) \\
N$_2$H$^+$  & 5.5(10) & 5.5(10) & 5.3(10) & 5.2(10) & 5.3(10) \\
OH          & 1.7(16) & 1.7(16) & 1.7(16) & 1.7(16) & 1.7(16) \\
H$_2$O      & 1.7(21) & 1.7(21) & 1.7(21) & 1.6(21) & 1.7(21) \\
CO$_2$      & 4.6(20) & 4.6(20) & 4.6(20) & 4.6(20) & 4.6(20) \\
C$_2$H$_2$  & 1.8(15) & 1.8(15) & 1.8(15) & 1.8(15) & 1.7(15) \\
CH$_3$OH    & 2.6(14) & 2.6(14) & 2.6(14) & 2.6(14) & 2.6(14) \\
\cutinhead{10 AU}
H$_2$       & 2.6(24) & 2.6(24) & 2.6(24) & 2.6(24) & 2.6(24) \\
CO          & 3.6(20) & 3.6(20) & 3.6(20) & 3.8(20) & 3.2(20) \\
HCO$^+$     & 4.9(13) & 4.9(13) & 5.4(13) & 3.5(13) & 4.4(13) \\
HCN         & 2.7(11) & 2.7(11) & 7.1(14) & 8.0(14) & 7.0(14) \\
CN          & 1.6(13) & 1.6(13) & 3.9(14) & 3.9(14) & 2.1(14) \\
CS          & 2.0(12) & 1.5(12) & 1.0(14) & 2.5(14) & 3.4(13) \\
C$_2$H      & 3.6(13) & 3.6(13) & 2.1(14) & 2.1(14) & 1.4(14) \\
H$_2$CO     & 3.1(11) & 3.0(11) & 1.4(12) & 1.4(12) & 1.4(12) \\
N$_2$H$^+$  & 1.4(09) & 1.4(09) & 1.4(10) & 1.4(10) & 1.6(10) \\
OH          & 8.7(15) & 8.7(15) & 8.8(15) & 8.8(15) & 8.8(15) \\
H$_2$O      & 2.6(15) & 2.6(15) & 4.3(15) & 2.0(16) & 4.3(15) \\
CO$_2$      & 4.4(16) & 4.4(16) & 4.9(16) & 1.3(18) & 4.0(16) \\
C$_2$H$_2$  & 1.1(14) & 1.1(14) & 8.7(13) & 2.8(13) & 1.5(14) \\
CH$_3$OH    & 2.1(07) & 2.1(07) & 7.7(07) & 1.6(08) & 8.5(07) \\
\cutinhead{100 AU}
H$_2$       & 2.0(23) & 2.0(23) & 2.0(23) & 2.0(23) & 2.0(23) \\
CO          & 2.2(19) & 2.2(19) & 2.3(19) & 2.8(19) & 2.2(19) \\
HCO$^+$     & 2.2(14) & 2.2(14) & 2.2(14) & 8.3(13) & 4.9(13) \\
HCN         & 1.6(12) & 1.6(12) & 2.1(14) & 3.7(14) & 2.3(13) \\
CN          & 1.2(13) & 1.2(13) & 2.6(14) & 2.6(14) & 2.4(14) \\
CS          & 7.0(09) & 7.0(09) & 3.2(13) & 9.7(13) & 1.7(13) \\
C$_2$H      & 1.1(13) & 1.1(13) & 1.2(14) & 1.2(14) & 1.1(14) \\
H$_2$CO     & 7.3(11) & 7.2(11) & 1.9(12) & 1.9(12) & 1.7(12) \\
N$_2$H$^+$  & 5.0(10) & 5.2(10) & 3.7(10) & 3.1(10) & 7.0(10) \\
OH          & 4.4(14) & 4.4(14) & 2.2(14) & 2.2(14) & 2.8(14) \\
H$_2$O      & 1.2(14) & 1.2(14) & 1.1(15) & 2.6(15) & 7.3(14) \\
CO$_2$      & 3.8(12) & 3.8(12) & 3.4(15) & 1.2(16) & 7.7(14) \\
C$_2$H$_2$  & 2.8(12) & 2.8(12) & 1.9(13) & 2.0(13) & 1.6(13) \\ 
CH$_3$OH    & 5.6(06) & 5.6(06) & 7.5(07) & 1.5(08) & 4.8(08) \\
\cutinhead{305 AU}
H$_2$       & 5.7(22) & 5.7(22) & 5.7(22) & 5.7(22) & 5.7(22) \\
CO          & 5.2(17) & 5.2(17) & 1.2(18) & 1.8(18) & 1.1(18) \\
HCO$^+$     & 3.2(13) & 3.2(13) & 1.2(13) & 2.6(13) & 3.6(13) \\
HCN         & 2.8(13) & 2.8(13) & 9.1(13) & 1.8(14) & 7.4(12) \\
CN          & 9.9(13) & 9.9(13) & 1.6(14) & 1.7(14) & 1.6(14) \\
CS          & 2.0(10) & 2.0(10) & 3.1(12) & 4.3(13) & 8.3(11) \\
C$_2$H      & 3.1(13) & 3.1(13) & 8.0(13) & 8.0(13) & 7.2(13) \\
H$_2$CO     & 6.3(12) & 6.3(12) & 2.1(12) & 2.5(12) & 2.4(12) \\
N$_2$H$^+$  & 2.5(12) & 5.1(12) & 5.0(12) & 3.7(12) & 1.2(13) \\
OH          & 7.9(14) & 8.0(14) & 8.0(13) & 1.7(14) & 2.0(14) \\
H$_2$O      & 1.4(15) & 1.4(15) & 7.8(14) & 1.4(15) & 1.2(15) \\
CO$_2$      & 9.4(13) & 9.5(13) & 1.9(15) & 3.4(15) & 7.9(14) \\
C$_2$H$_2$  & 1.3(13) & 1.3(13) & 1.3(13) & 3.1(13) & 2.3(13) \\ 
CH$_3$OH    & 1.6(08) & 1.6(08) & 5.6(07) & 7.7(08) & 1.6(11)
\enddata
\tablecomments{$a(b)$ means $a \times 10^{b}$}
\end{deluxetable}

\subsection{Comparison with Other Models}
\label{comparison}

A direct comparison with other chemical models of protoplanetary disks is difficult as no 
two models are identical in either their physical basis or their chemical networks.  
Given the plethora of preceding work in this field, we limit this short discussion 
to more recent models which are comparable to ours in chemical complexity.  

The work presented here builds upon previous investigations into the importance of 
cosmic-ray induced desorption and  photodesorption in disks by \citet{willacy00} and \citet{willacy07}, 
the latter of which uses the model of \citet{dalessio01} for their physical framework.  
We obtain encouragingly similar results to \citet{willacy07} in particular, although, 
her primary objective was the investigation into deuterated species in disks.  
As such, the reaction network of the non-deuterated species was truncated to accommodate 
the additional reactions involving deuterium and deuterium-containing molecules.  
Our work also differs in that our physical model includes X-ray heating and also explicitly 
determines the gas temperature which can decouple from the dust-grain temperature 
in regions where cooling via gas-grain collisions is inefficient.  
\citet{willacy09} also use their deuterated reaction network and adapt their model 
to investigate the chemical structure of a disk within 30~AU of the central star.  
Again, we achieve similar results although differences in the set of 
molecular desorption energies used manifests as differences in the positions of `snow lines' for 
molecules such as HCN.   
Also differing prescriptions for 
the UV radiation field in the disk leads to different distributions of radicals such as C$_2$H and CN.  

 Photodesorption and cosmic-ray induced desorption are now routinely included in modern chemical 
models of protoplanetary disks (e.g.\ \citet{woitke09, henning10}).  
X-ray desorption and grain-surface chemistry have both been included in work by 
by other groups (e.g.\ \citet{semenov08,henning10}) although not explicitly investigated given 
the theoretical uncertainty behind the exact mechanism of X-ray desorption for the 
former process and the usual truncation of chemical networks for the latter.  
Our work presented here and subsequent follow-up publications  on 
X-ray desorption and grain-surface chemistry, we hope, will go some way to 
addressing this.    

\section{SUMMARY}
\label{summary}

In this work, we have presented a selection of results from our high-resolution combined 
chemical and physical model of a protoplanetary disk surrounding a typical T Tauri star, constructed 
in order to trace the physical and thus chemical structure on small scales.  
We use a protoplanetary disk model in which the gas density and temperature distributions 
are obtained self-consistently and UV and X-ray irradiation by the central star is calculated 
by solving the radiative transfer equation.  
To this we applied a large comprehensive 
chemical network including gas-phase chemistry, gas-grain interactions and grain-surface 
chemistry.  
We investigated the effects of each non-thermal desorption mechanism thought to 
be important in disks: cosmic-ray induced desorption,  photodesorption and X-ray desorption.  
We also added a large grain-surface network to investigate the effectiveness of grain-surface 
reactions on the synthesis of relatively complex organic molecules. 

Using the results from model PH+CRH (see Table~\ref{table2}) extracted at a time 
of $10^{6}$~years, we find that the disk chemical structure closely mirrors the disk physical structure 
with the  freeze out of molecules onto dust grains creating an icy mantle 
in the cold, dense  midplane and an abundance of molecules 
in a layer above the  midplane created through sublimation and the resulting 
rich gas-phase chemistry.  
In the disk surface, the molecular abundances drop as the UV and X-ray radiation fields peak in strength 
dissociating molecules and ionising both molecules and atoms.  
The resulting disk ionisation fraction increases with increasing disk height. 
There is similar stratification in the radial direction as the increasing temperature drives the 
evaporation of molecules.  
The temperature dependence of the desorption energies results in a unique `snow-line' for each molecules 
with more volatile molecules returned to the gas in the  midplane at larger radii than 
less volatile ones.  
In particular, both HCN and H$_2$O remain frozen onto dust grains to within $\approx$~1 to 2 AU 
of the central star.   

The addition of cosmic-ray induced desorption has only a small effect on 
the gas-phase abundances in the outer disk  midplane although since our model is truncated at 
305~AU we would expect that this effect will continue at radii beyond this value.  
 Photodesorption, the most experimentally constrained of the mechanisms considered here, 
has a larger effect although only in the molecular and surface regions of the disk where there is 
an appreciable UV flux.  
It is especially effective at enhancing the gas-phase abundances of non-volatile molecules such as 
H$_2$O.  
X-ray desorption, has the largest effect, smoothing the abundances of gas-phase species throughout the 
height of the disk and acting to homogenise the fractional abundances.  
However, X-ray desorption is the least theoretically or experimentally constrained and thus our 
results must be treated with caution, pending further investigation.  

The addition of grain-surface chemistry also yields some encouraging results worthy 
of revisiting and further study.  
In the outer disk, where the  freeze out of molecules is most prevalent, the abundances 
of relatively complex organic molecules e.g.\ CH$_3$OH, HCOOCH$_3$ and CH$_3$OCH$_3$ are enhanced 
to potentially observable values when grain-surface chemistry and  photodesorption are included 
in our model.  
Thus, the observation of rotational transitions of these, and related, species in protoplanetary 
disks should provide an excellent means of testing grain-surface chemistry theory.  
Due to limitations in existing facilities, the most complex molecule observed, as yet, in disks is H$_2$CO. 
ALMA, however, will have the sensitivity and spectral resolution necessary to 
observe rotational transitions in these minor species. 
 Indeed, preliminary synthetic spectra we have calculated suggest that the rotational 
transition lines of methanol are enhanced above the detection threshold of ALMA when grain-surface 
chemistry is included in our model (Walsh et al., in preparation).

We have shown that running models 
of this nature, in which we test experimental data and theory, 
varying the different chemical ingredients, are necessary for disentangling the different physical influences 
on the chemical content of protoplanetary disks.  
The influence of the physical conditions and processes on the molecular content is also a strong function 
of radius and so high-resolution models, which trace the chemical structure on small scales, are also preferred.  
In this brief overview of our model, we have shown that X-ray desorption and grain-surface chemistry can have 
a powerful effect on the molecular content of disks and we intend to expand upon the work presented here with 
follow-up papers on both chemical processes.  
Although in this paper we have shown that the distribution and radial column densities of particular 
molecules are sensitive to the inclusion or omission of certain chemical processes, in order to test the 
viability of using these molecules as tracers we must compute the radiative transfer in the disk and directly 
compare our results with observations.  This work will be reported in a subsequent paper (Walsh et al., in
preparation).  

\acknowledgments

 We wish to thank an anonymous referee for his or her constructive comments 
which helped improve our paper. 
C.\ Walsh acknowledges DEL for a studentship and JSPS for the award of a short-term 
fellowship to conduct research in Japan.  H.\ Nomura acknowledges the JGC-S Scholarship 
Foundation, the Grant-in-Aid for Scientific Research 21740137 and the 
Global COE Program ``The Next Generation of Physics, Spun from Universality and 
Emergence'' from MEXT, Japan.  
Astrophysics at QUB is supported by a grant from the STFC.

\newpage

\begin{figure}
\centering
\subfigure{\includegraphics[width=0.45\textwidth]{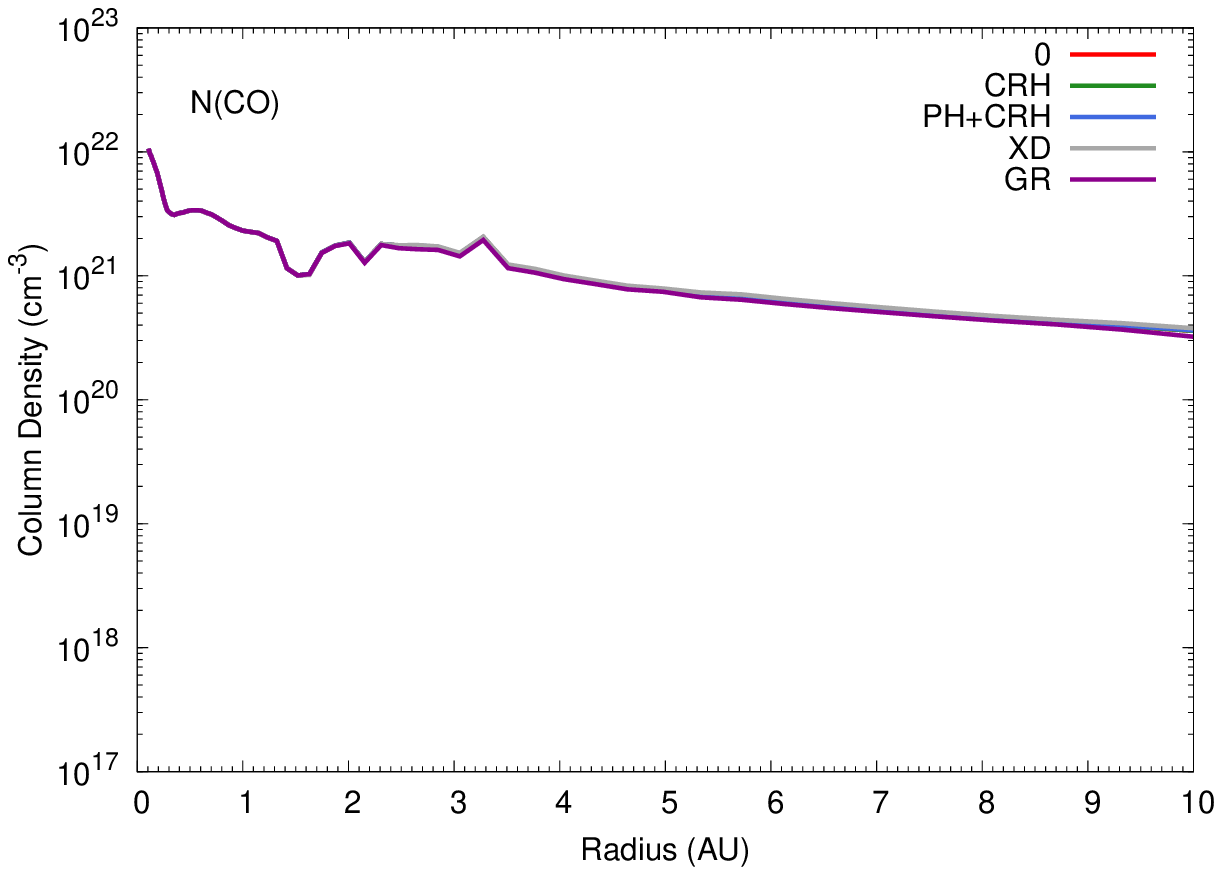}}
\subfigure{\includegraphics[width=0.45\textwidth]{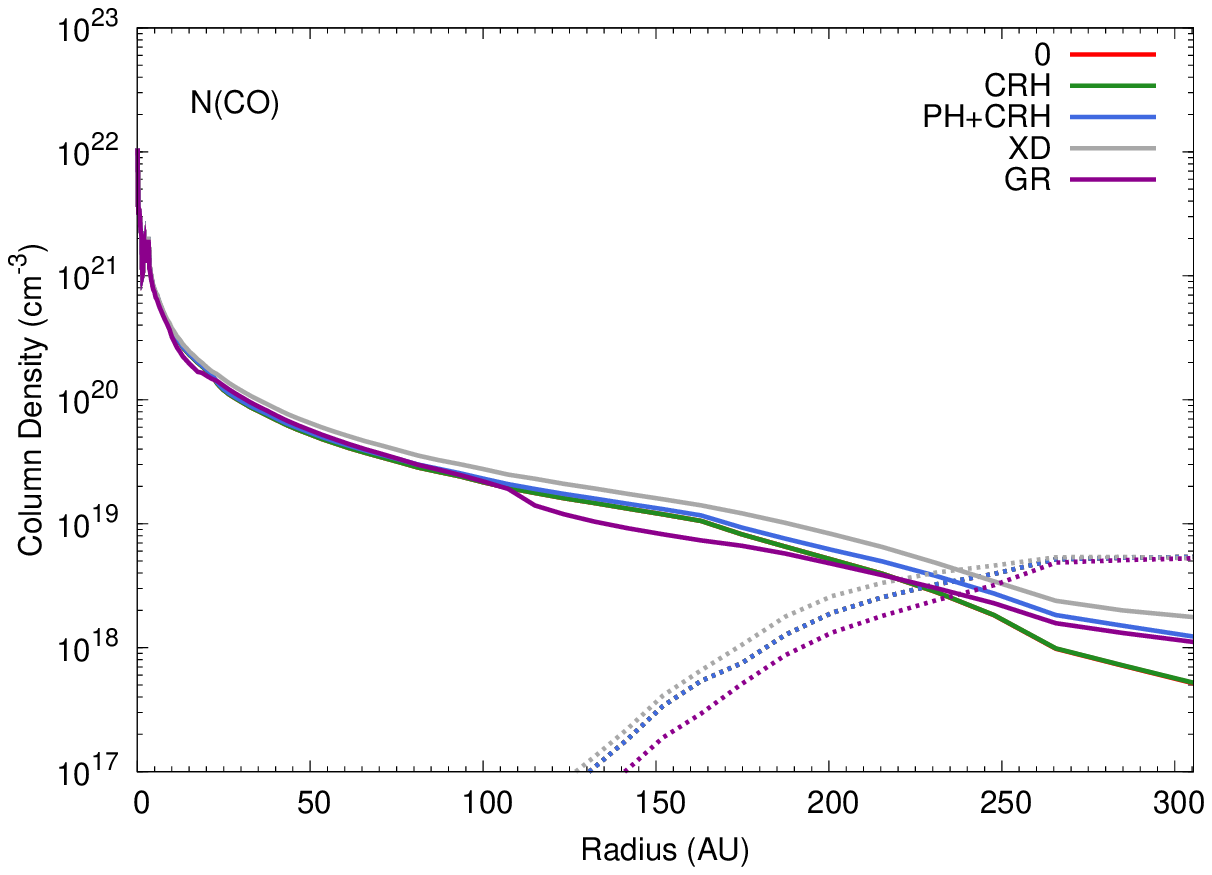}}
\subfigure{\includegraphics[width=0.45\textwidth]{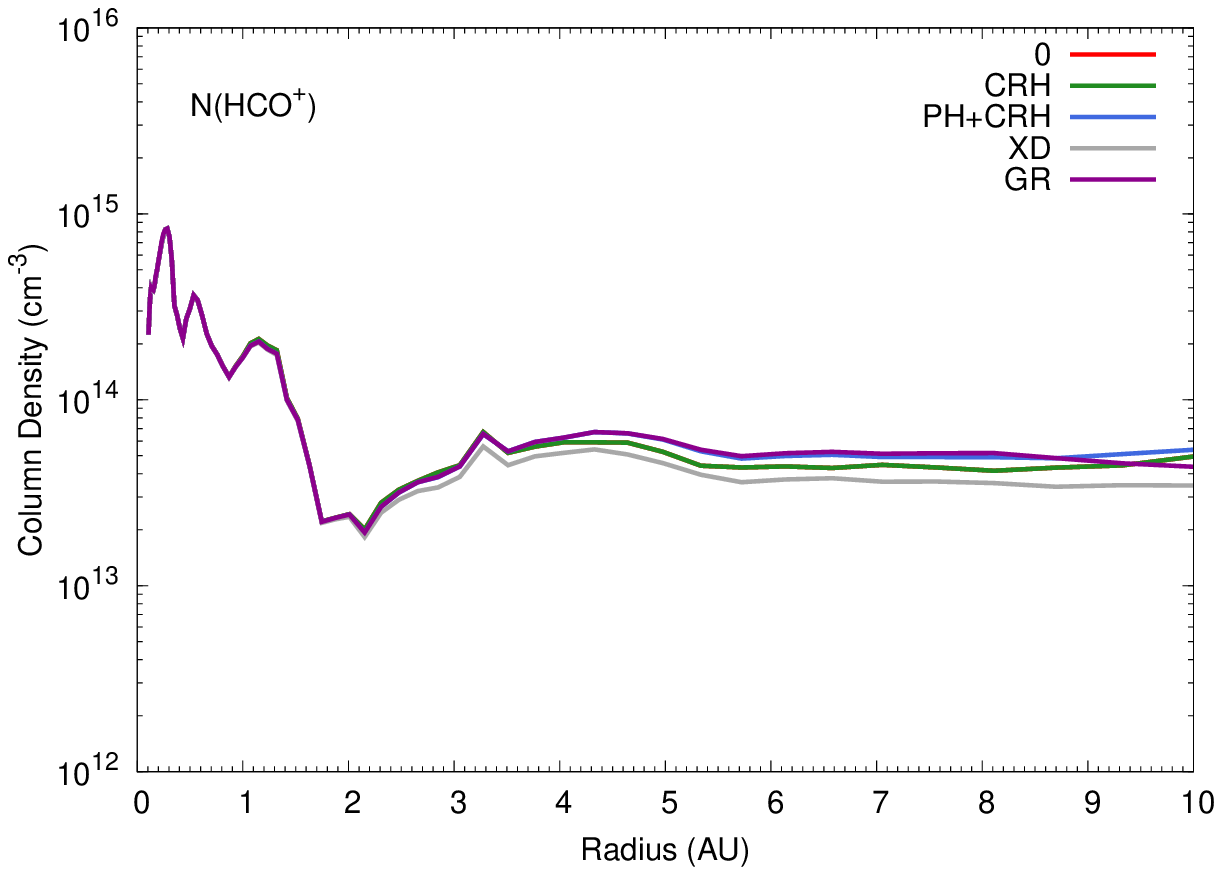}}
\subfigure{\includegraphics[width=0.45\textwidth]{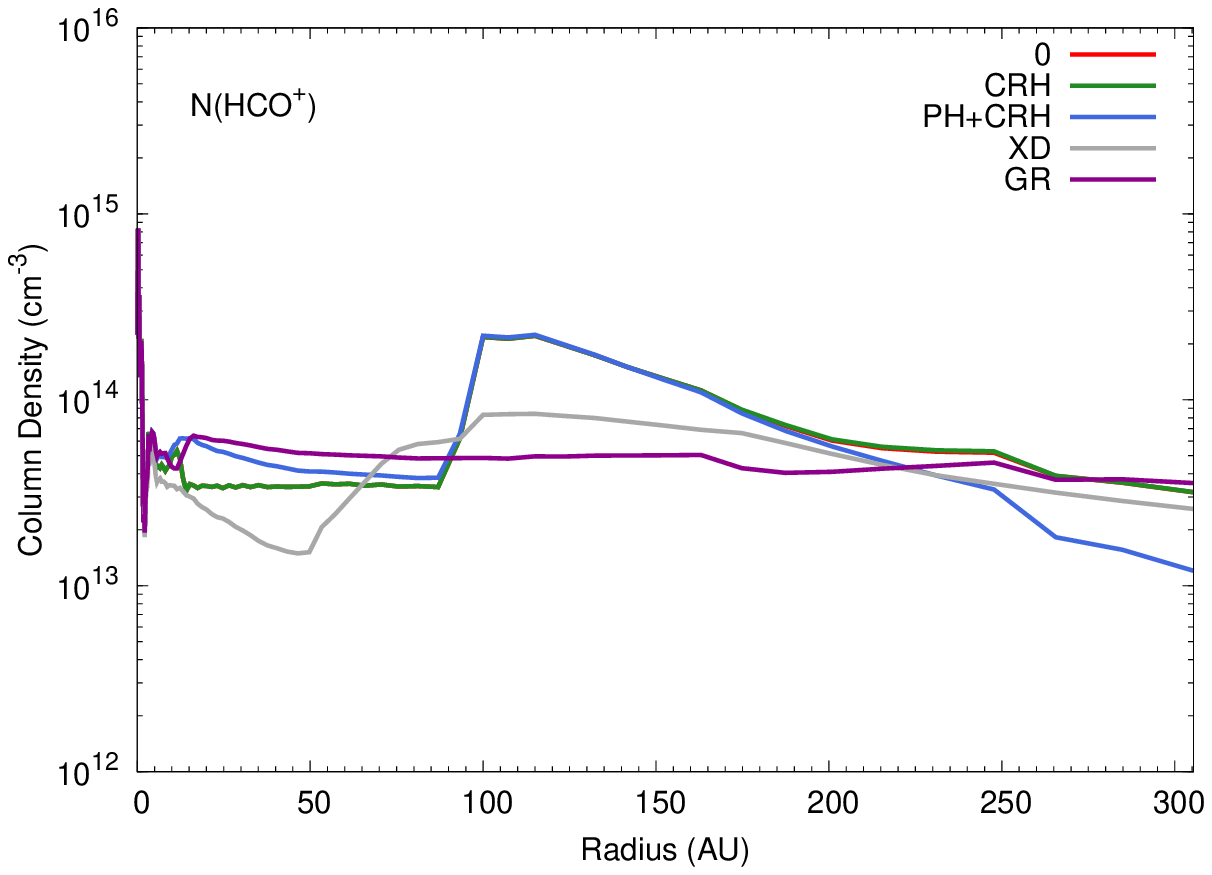}}
\subfigure{\includegraphics[width=0.45\textwidth]{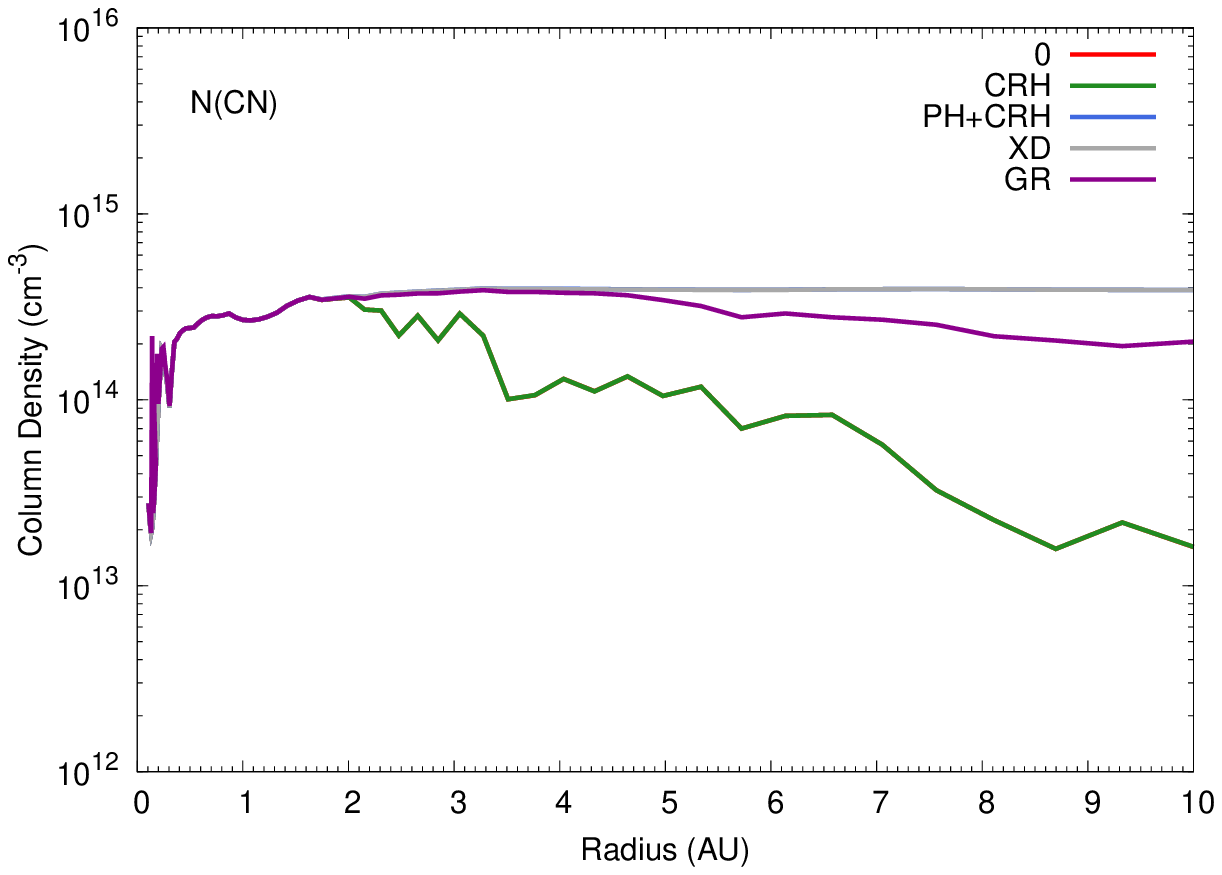}}
\subfigure{\includegraphics[width=0.45\textwidth]{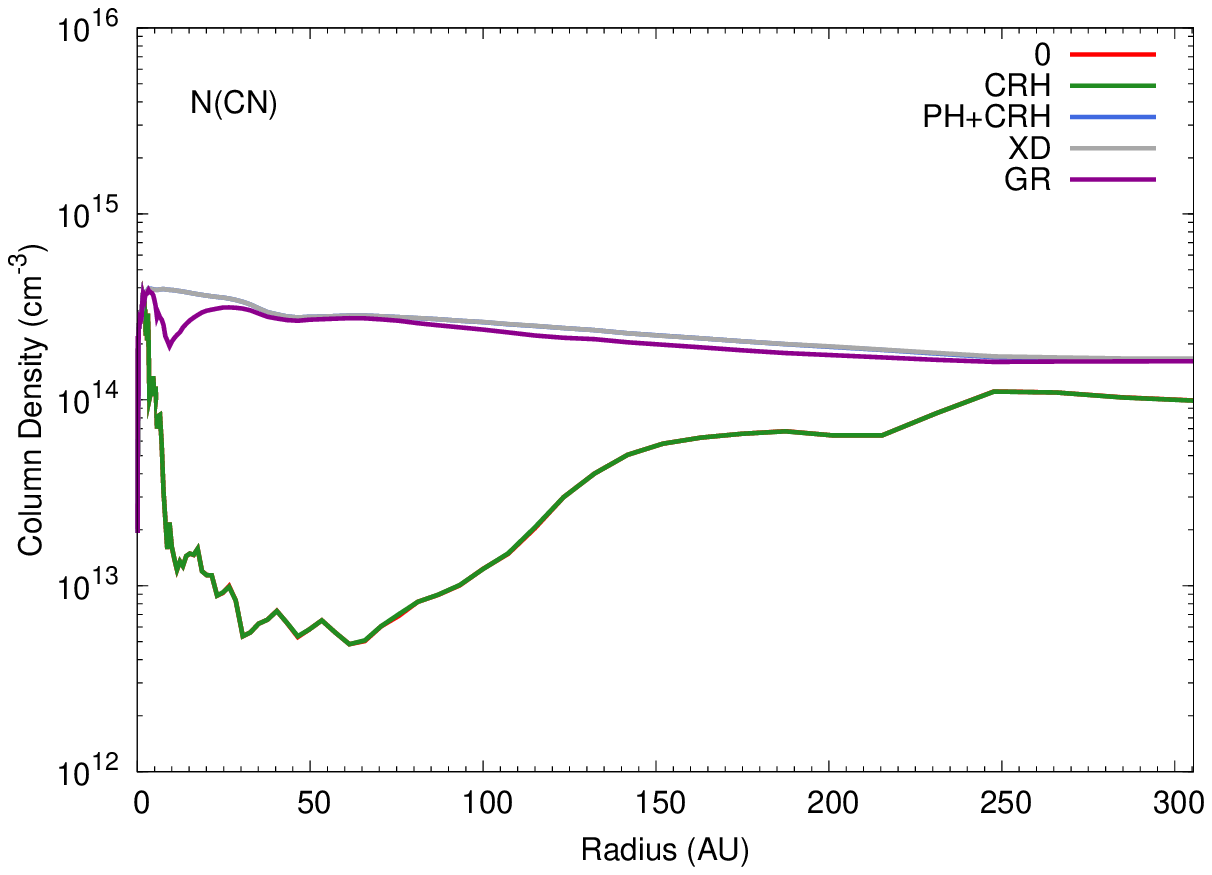}}
\subfigure{\includegraphics[width=0.45\textwidth]{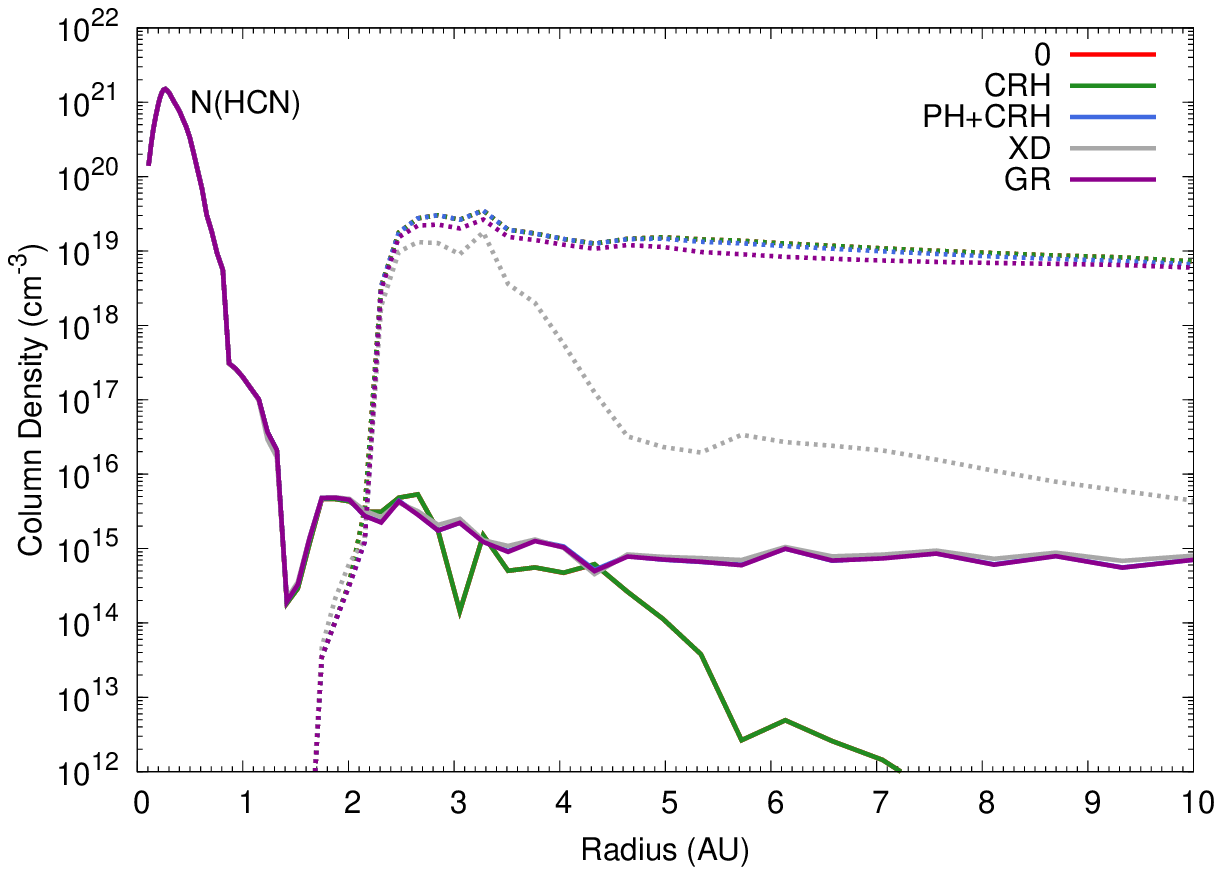}}
\subfigure{\includegraphics[width=0.45\textwidth]{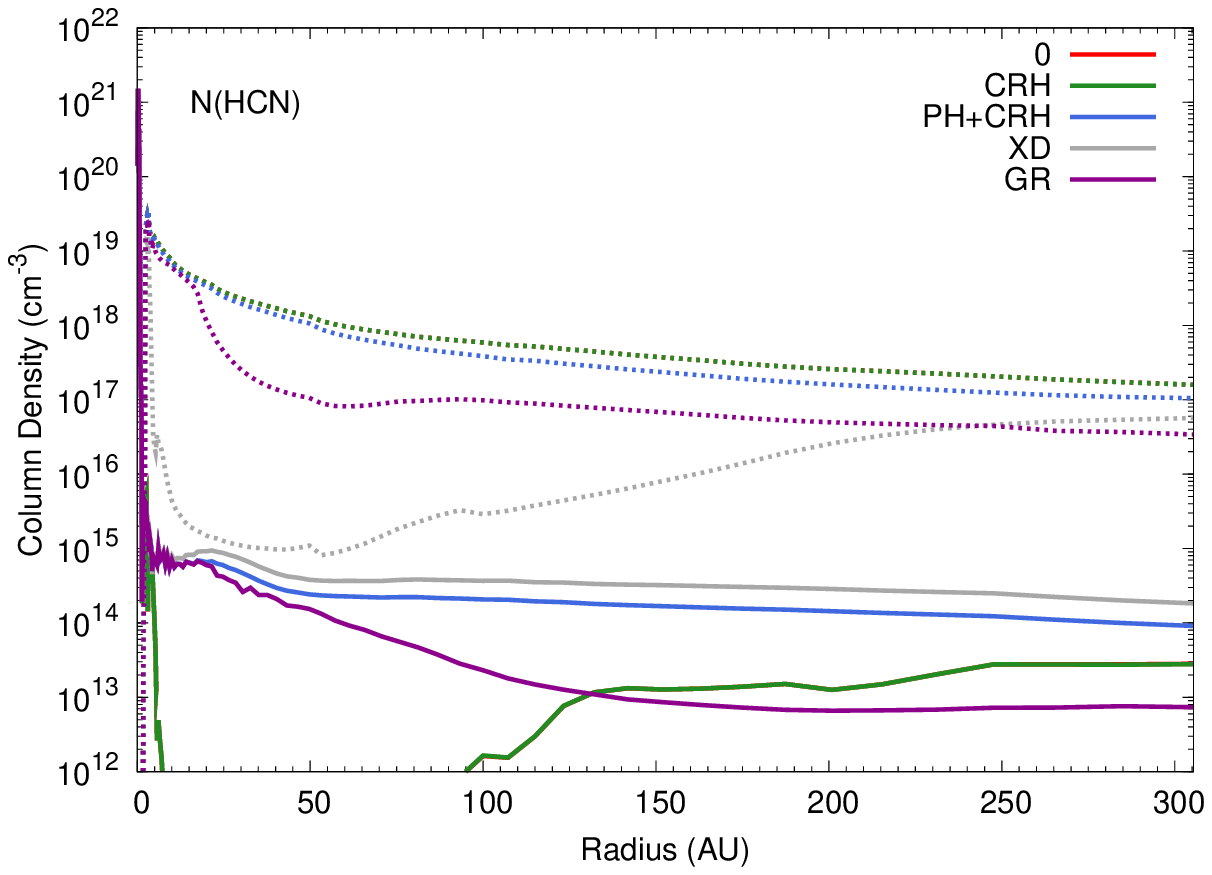}}
\captcont{Radial column densities of various gas-phase molecules up to maximum radii of 10~AU (left) and 
305~AU (right) for each chemical model.  
Grain-surface (ice) column densities are represented by dotted lines.  
 Note that results from models 0 and CRH  (red and green lines, respectively) 
 are very similar hence for many molecules 
these lines are virtually indistiguishable. 
 Note also for CO$_2$, in particular, the results from models 0, 
CRH and PH+CRH within 10~AU are almost identical.  Figure available online only.}
\end{figure}

\begin{figure}
\centering
\subfigure{\includegraphics[width=0.45\textwidth]{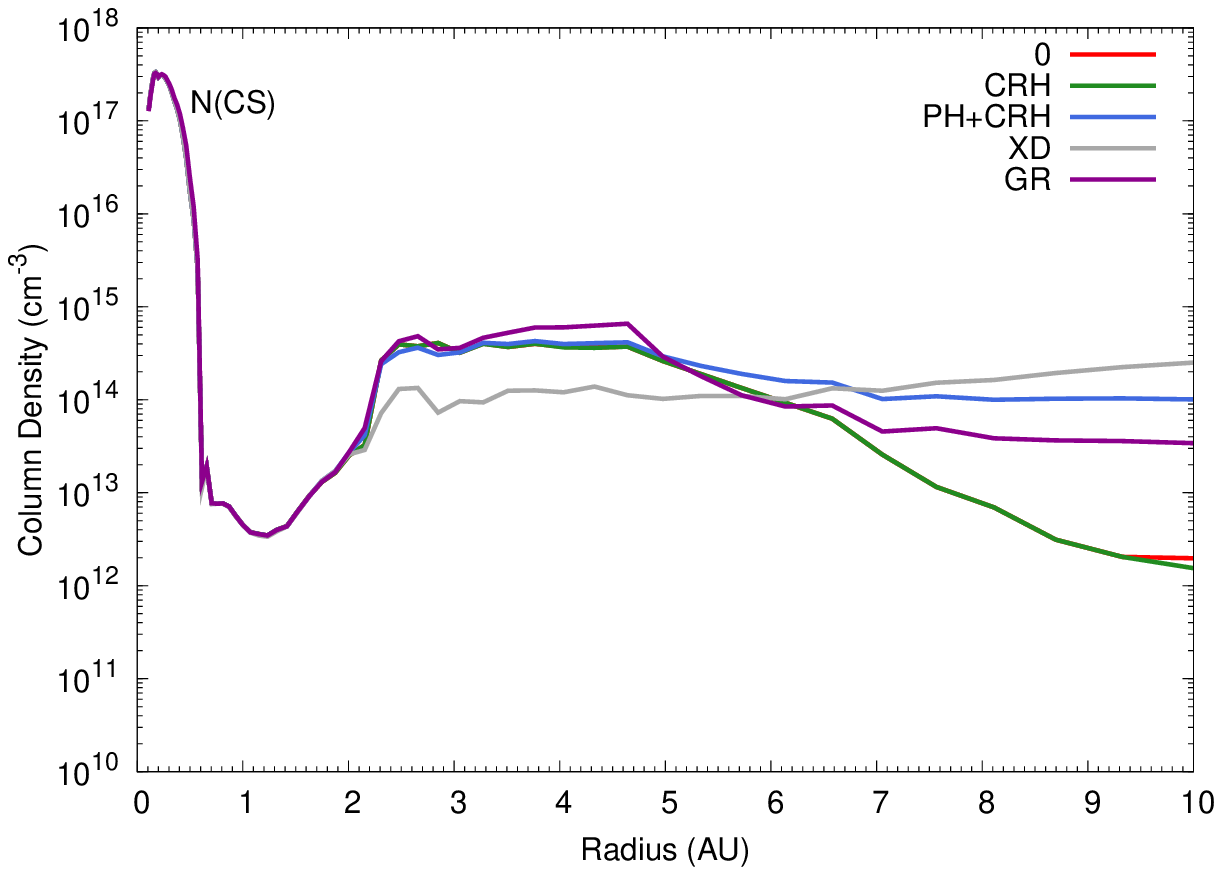}}
\subfigure{\includegraphics[width=0.45\textwidth]{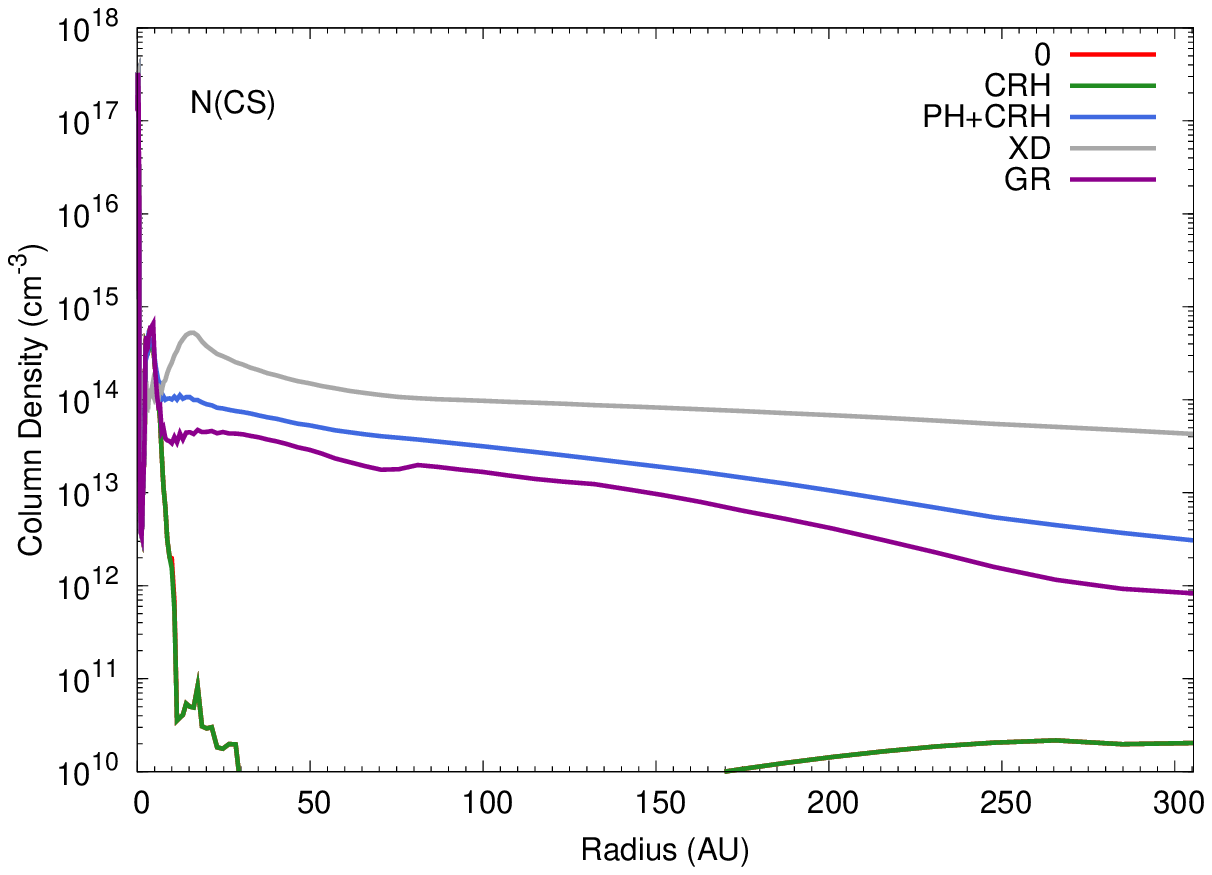}}
\subfigure{\includegraphics[width=0.45\textwidth]{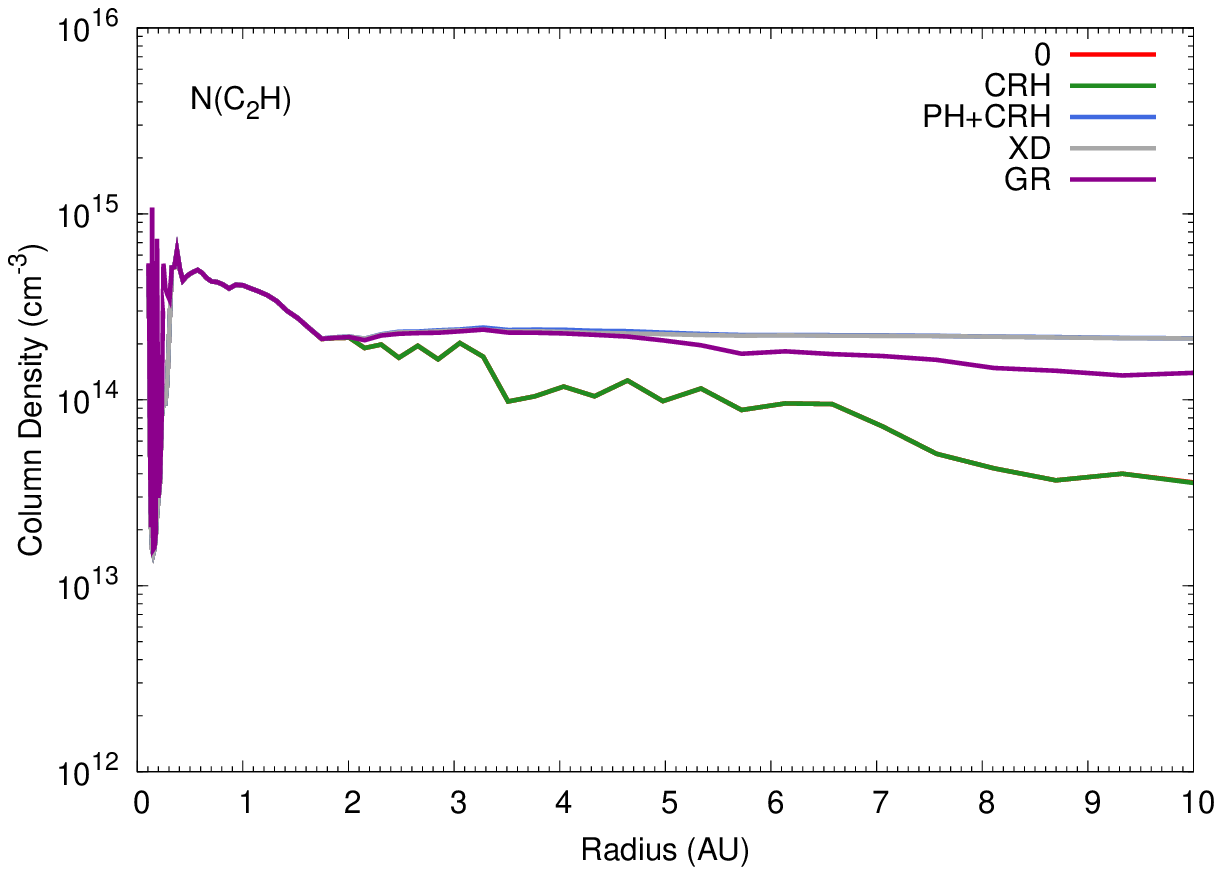}}
\subfigure{\includegraphics[width=0.45\textwidth]{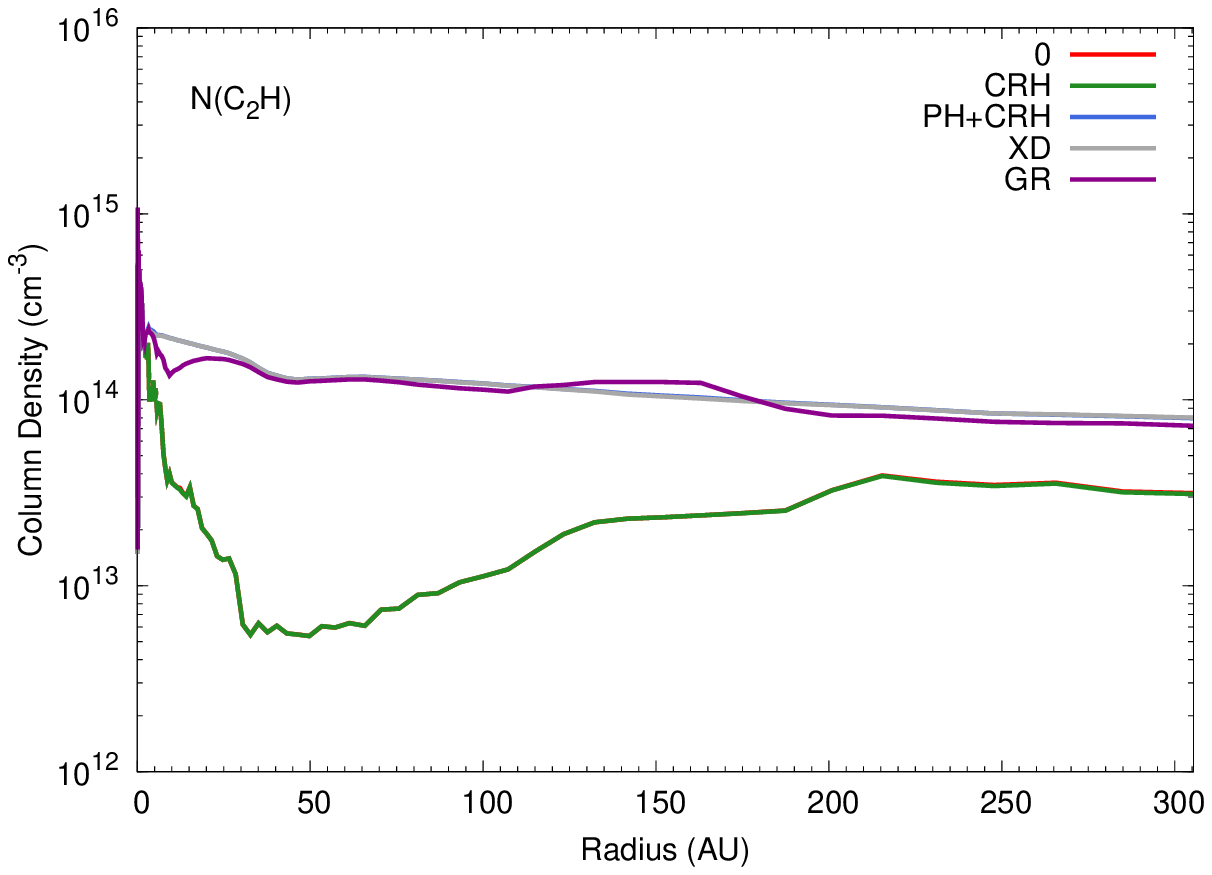}}
\subfigure{\includegraphics[width=0.45\textwidth]{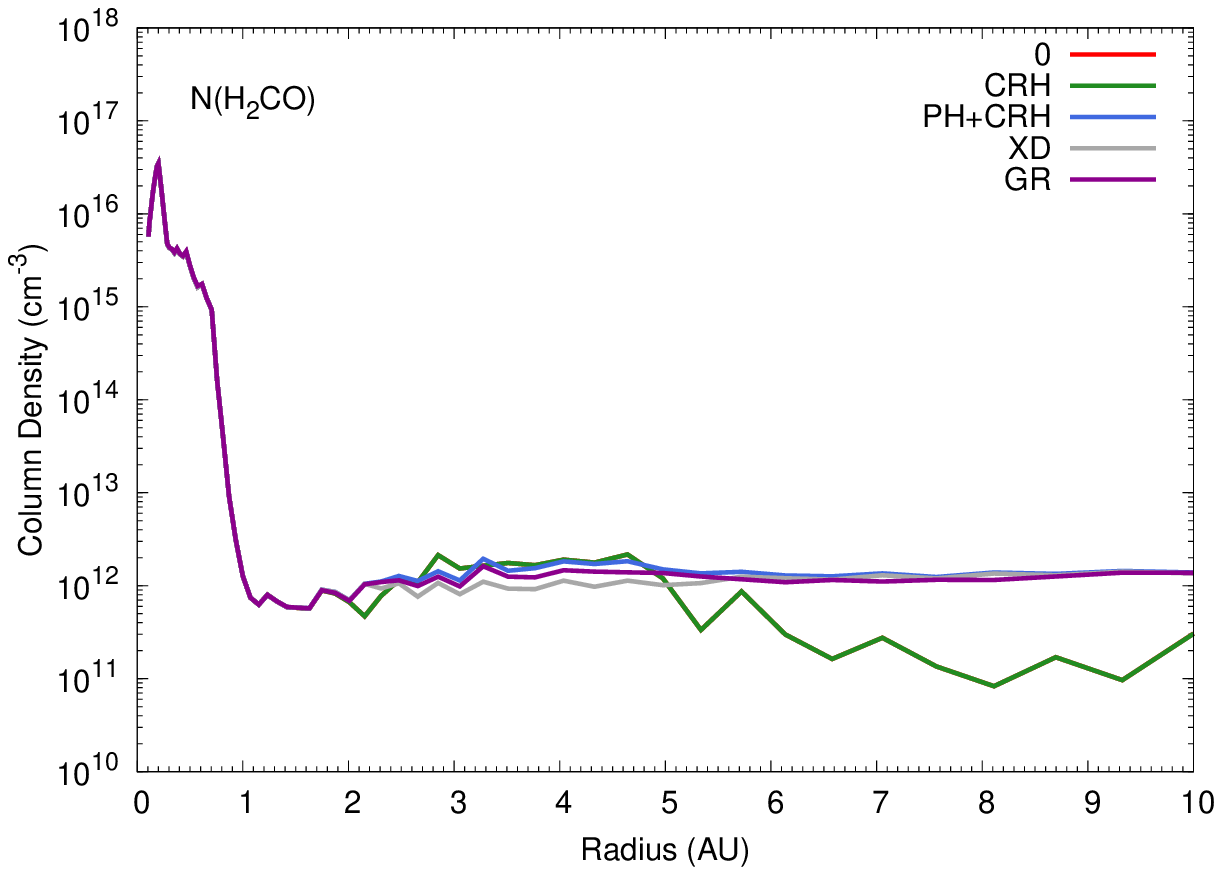}}
\subfigure{\includegraphics[width=0.45\textwidth]{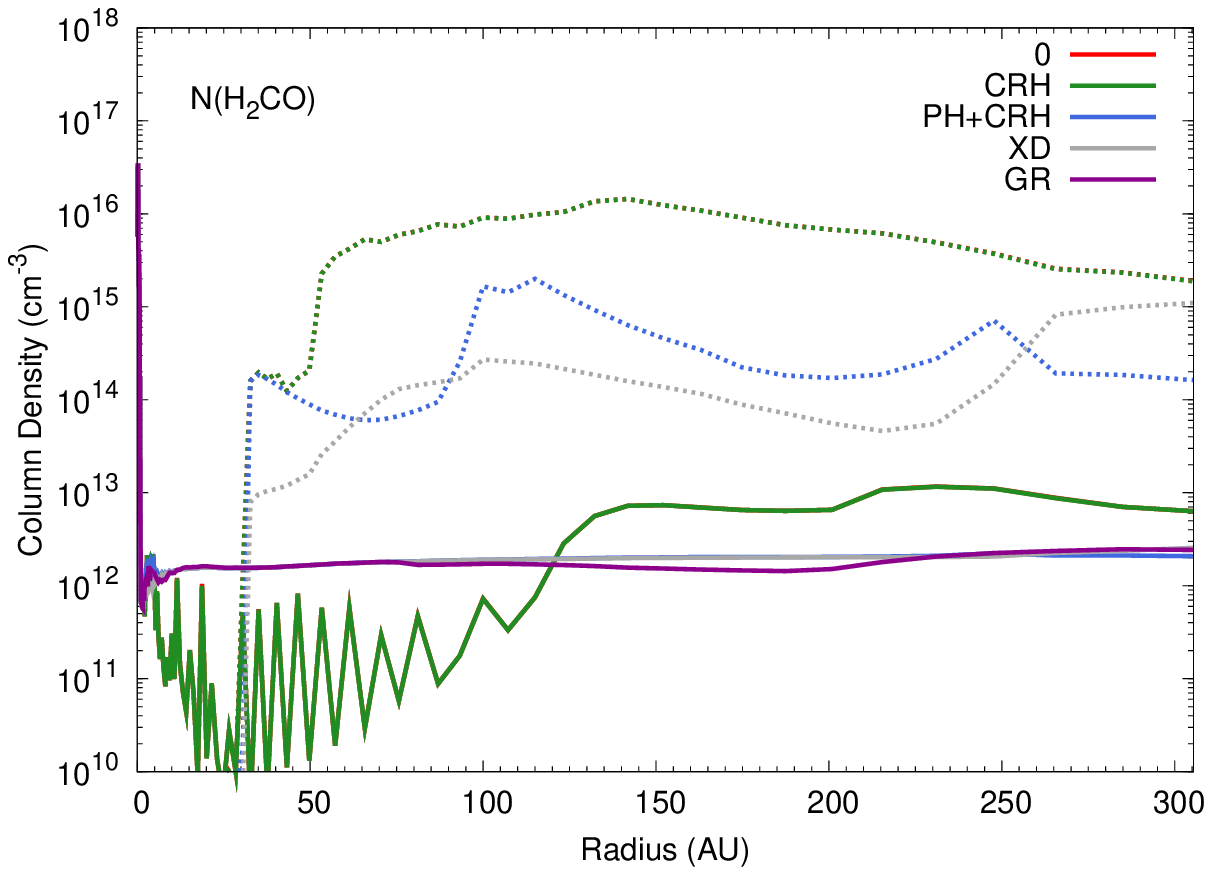}}
\subfigure{\includegraphics[width=0.45\textwidth]{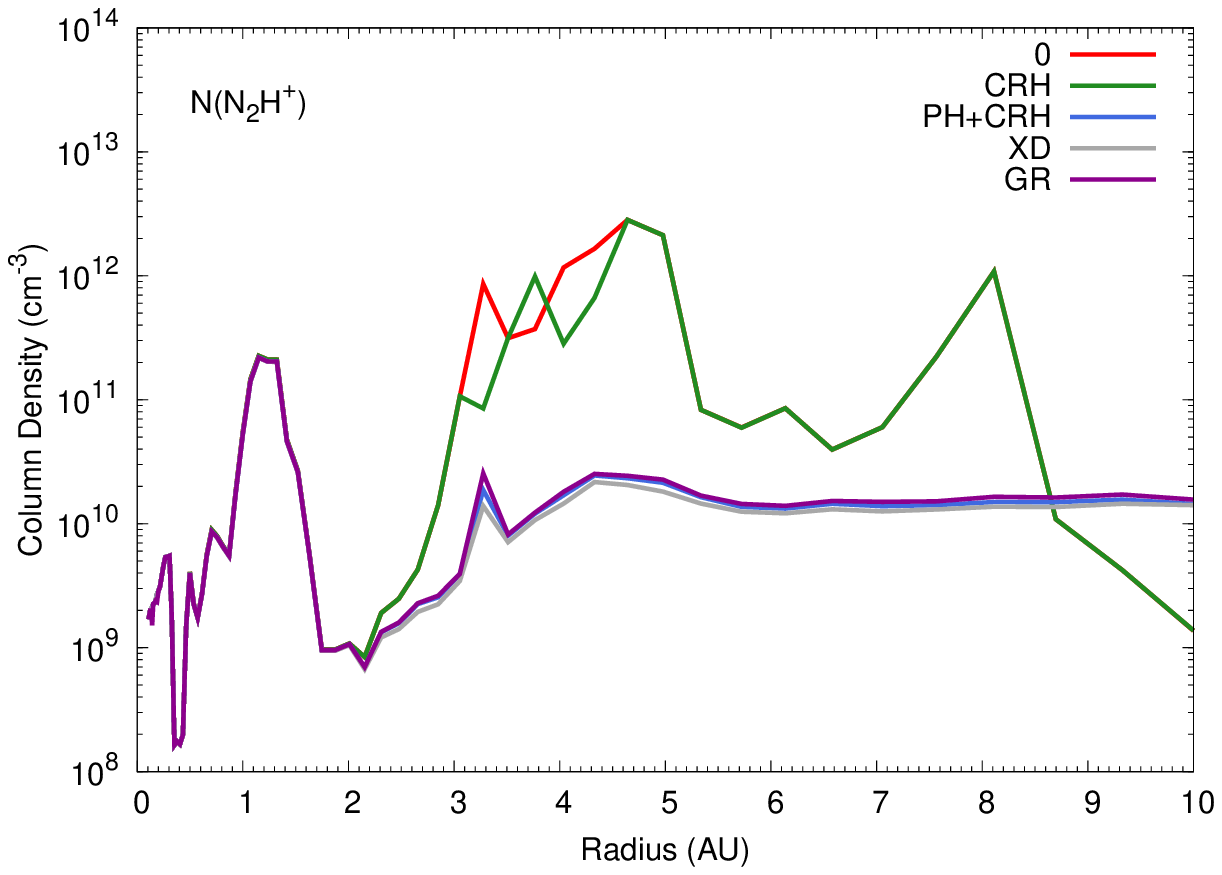}}
\subfigure{\includegraphics[width=0.45\textwidth]{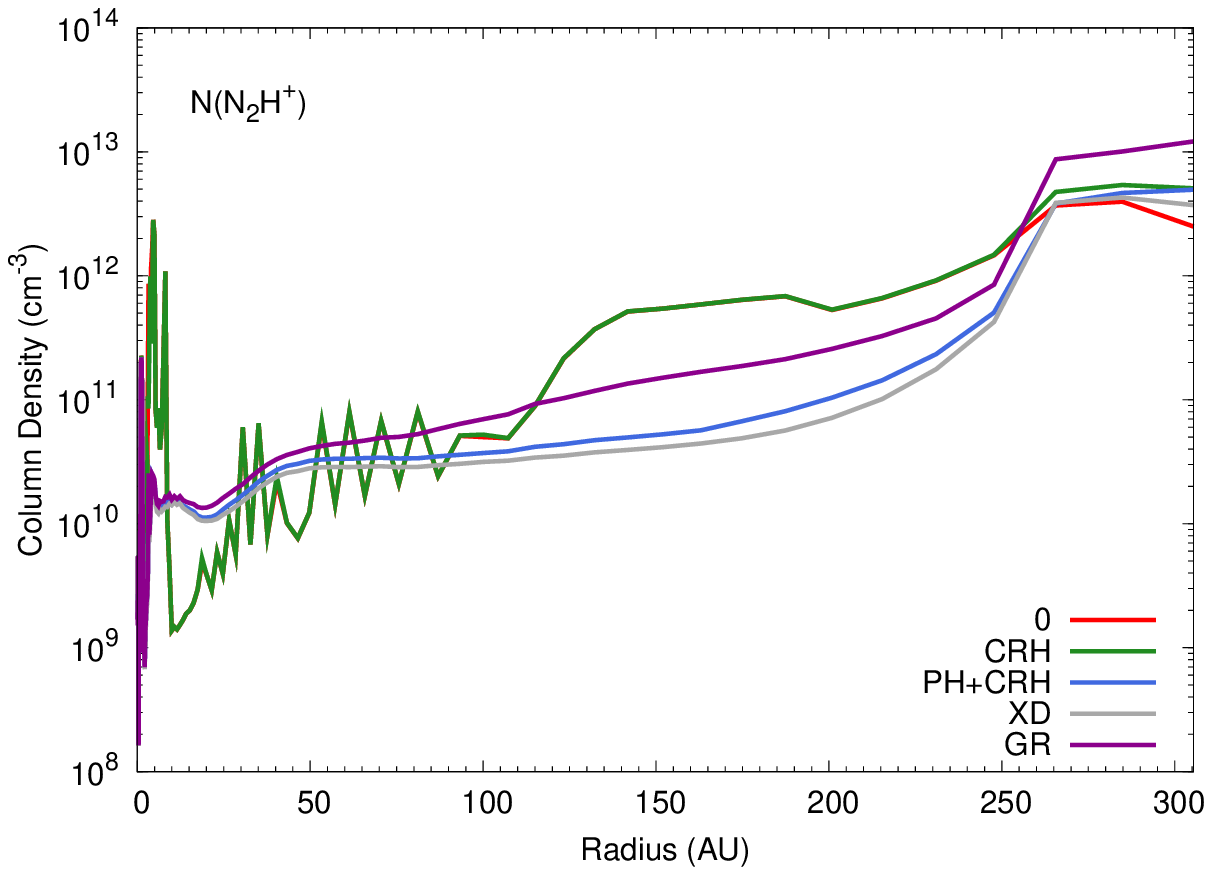}}
\captcont{(Continued.)}
\end{figure}

\begin{figure}
\centering
\subfigure{\includegraphics[width=0.45\textwidth]{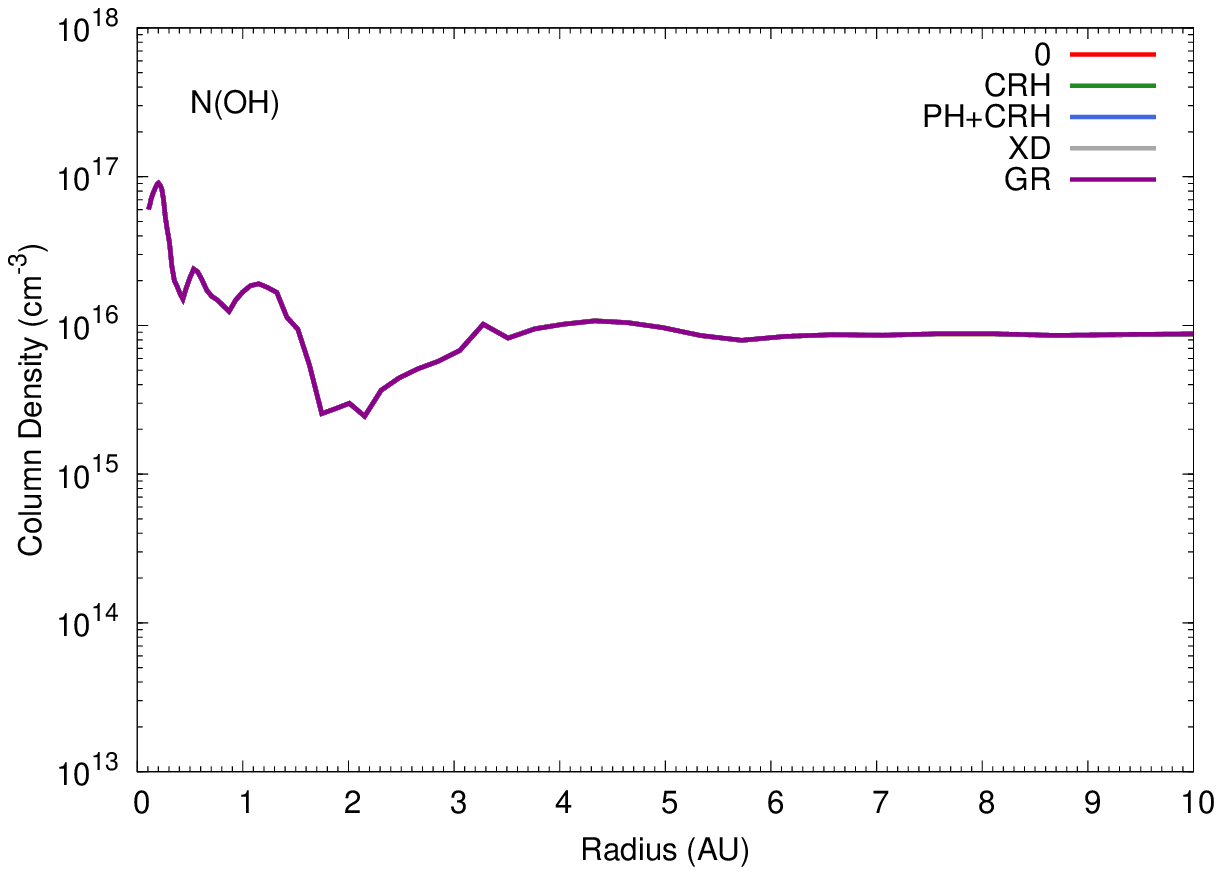}}
\subfigure{\includegraphics[width=0.45\textwidth]{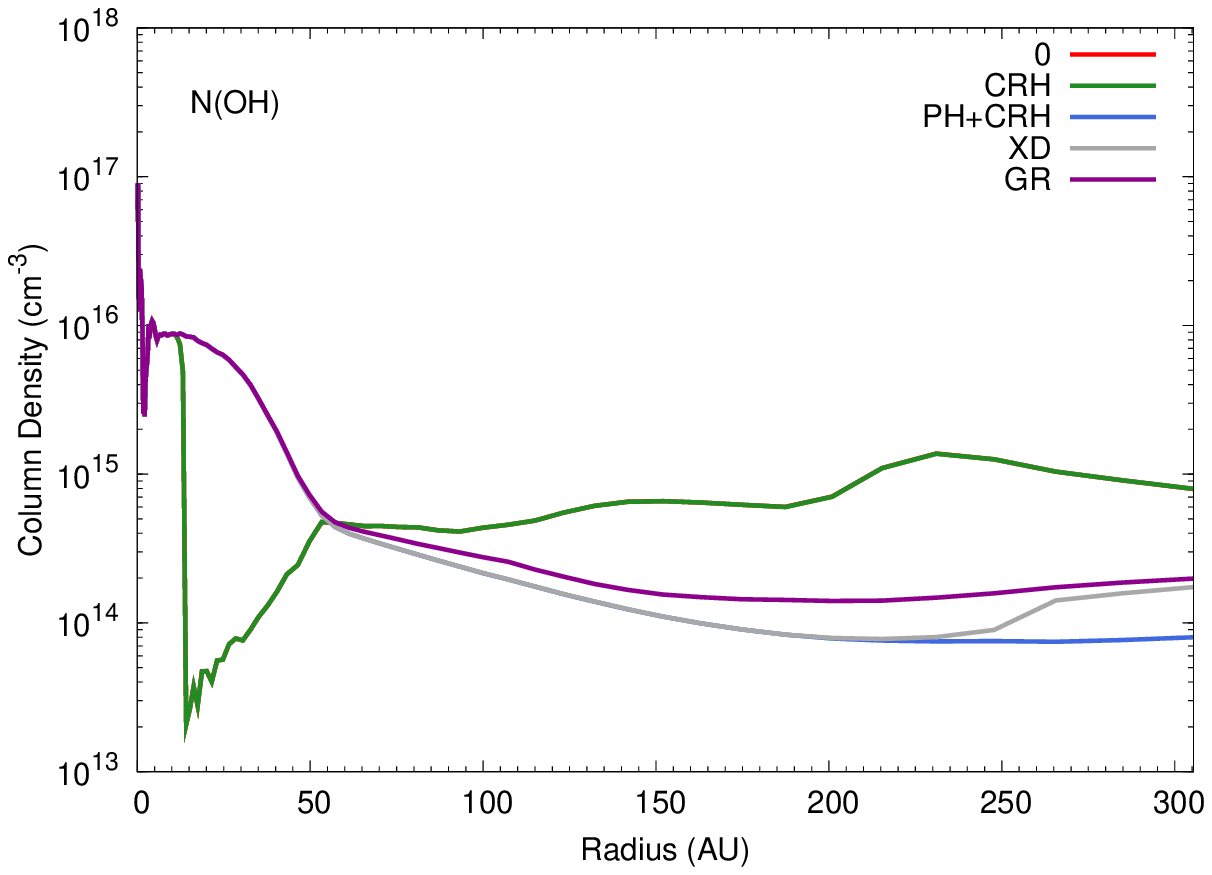}}
\subfigure{\includegraphics[width=0.45\textwidth]{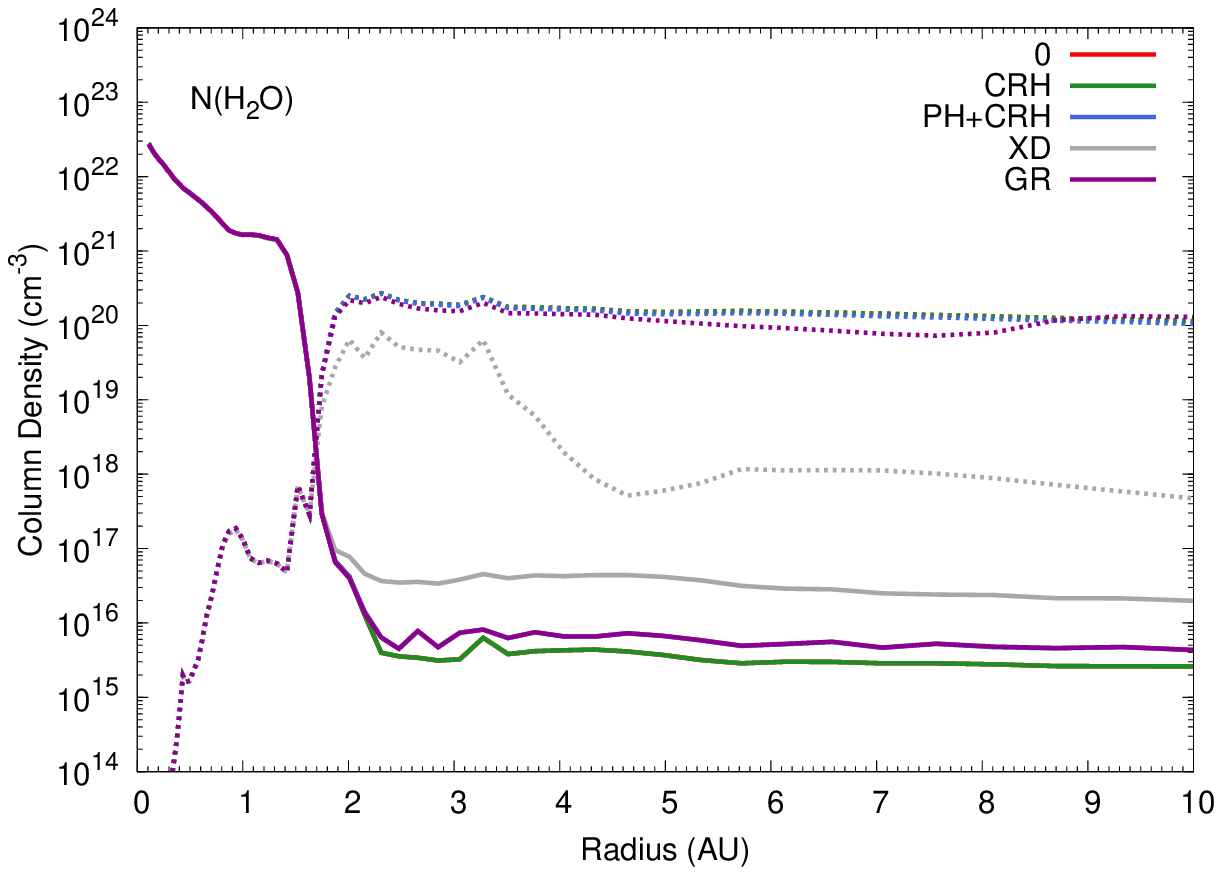}}
\subfigure{\includegraphics[width=0.45\textwidth]{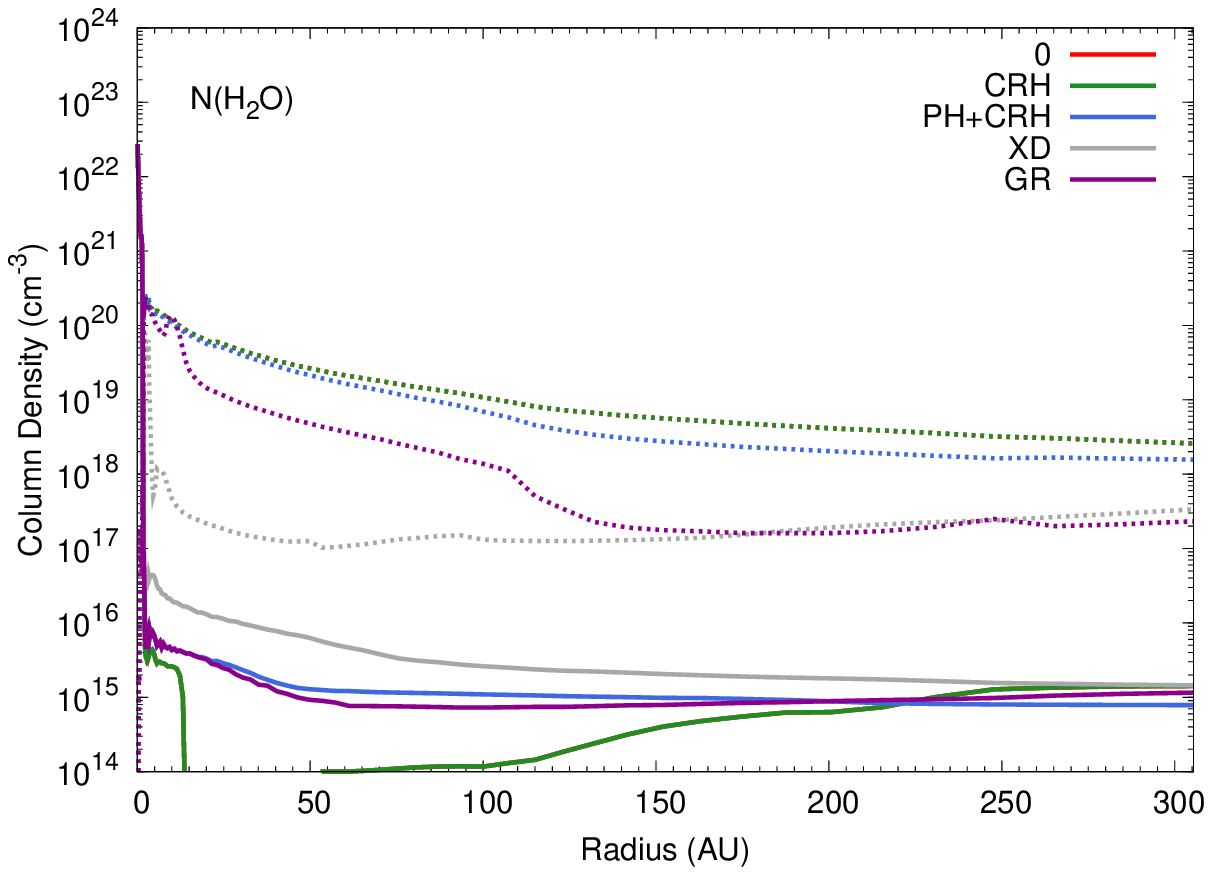}}
\subfigure{\includegraphics[width=0.45\textwidth]{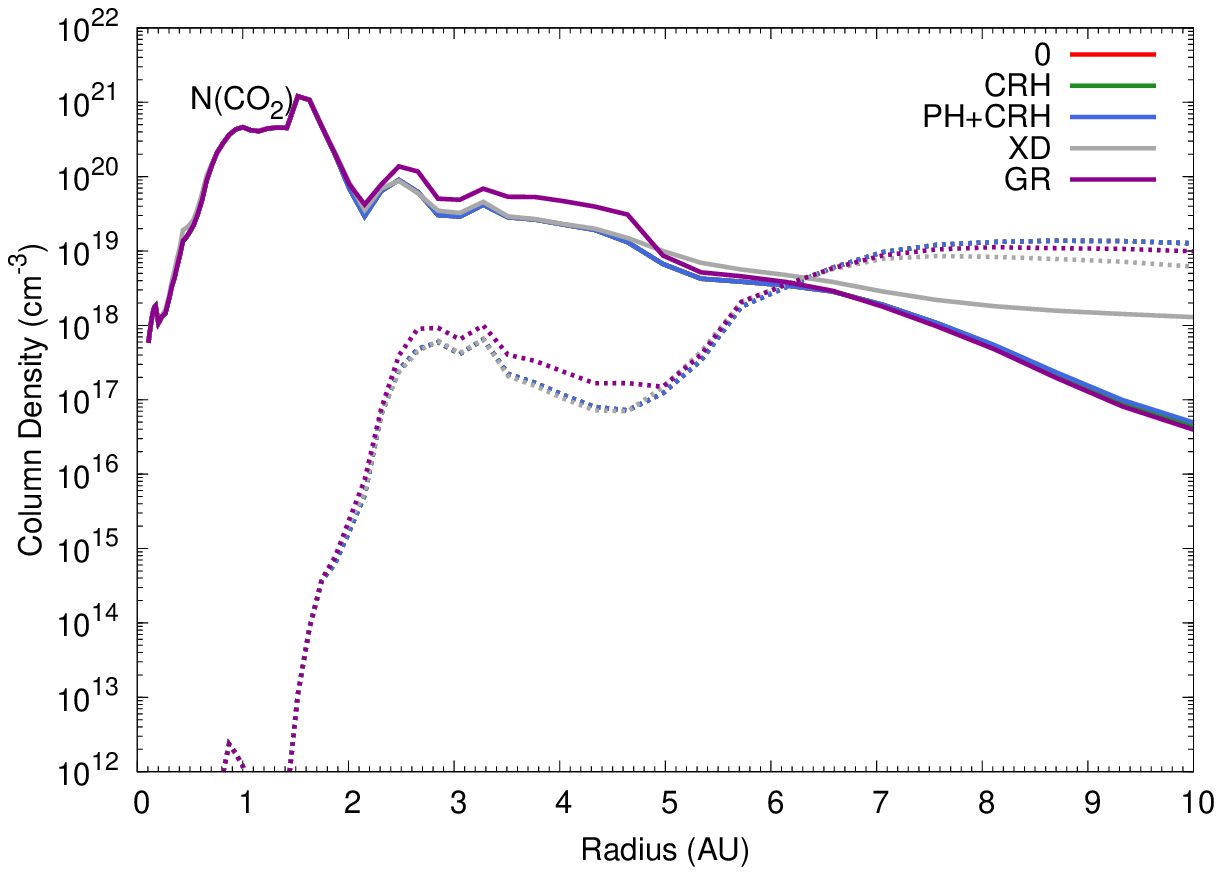}}
\subfigure{\includegraphics[width=0.45\textwidth]{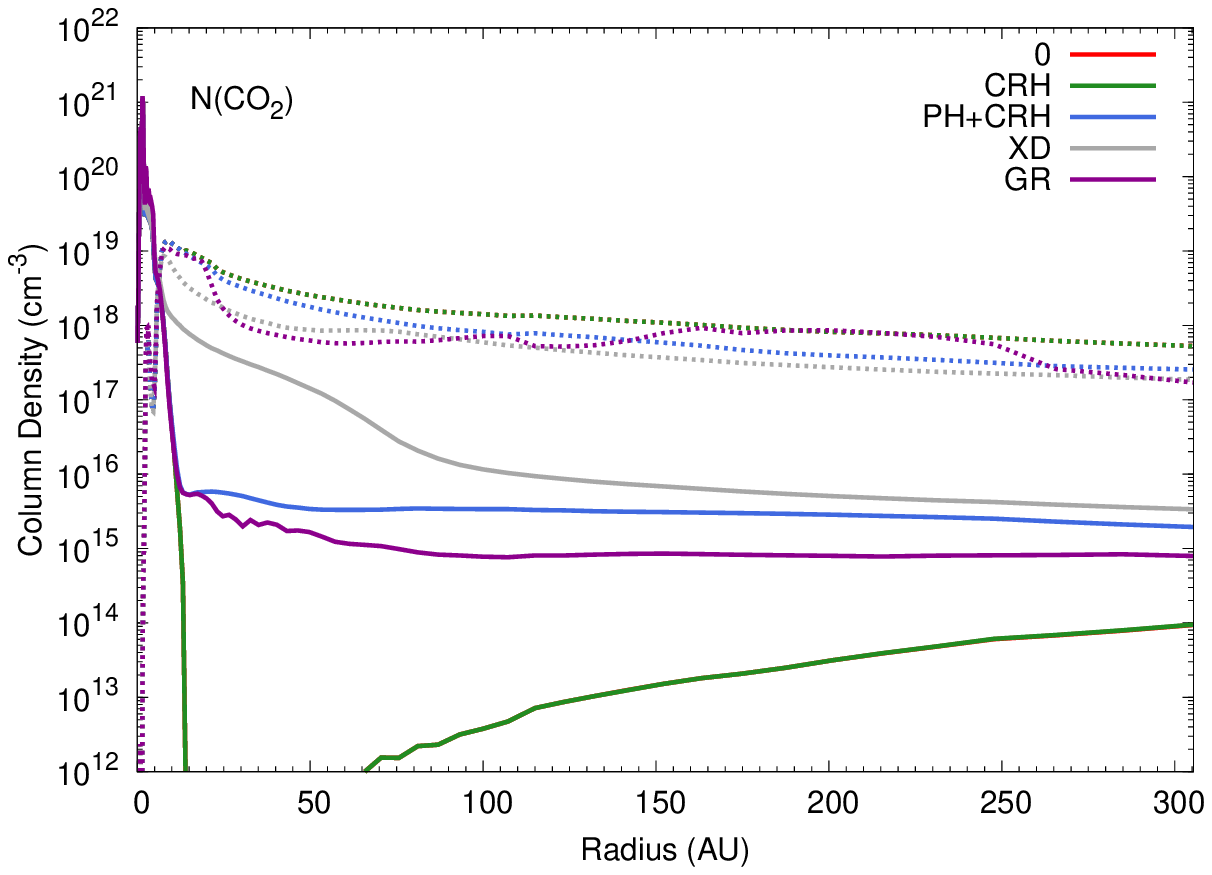}}
\subfigure{\includegraphics[width=0.45\textwidth]{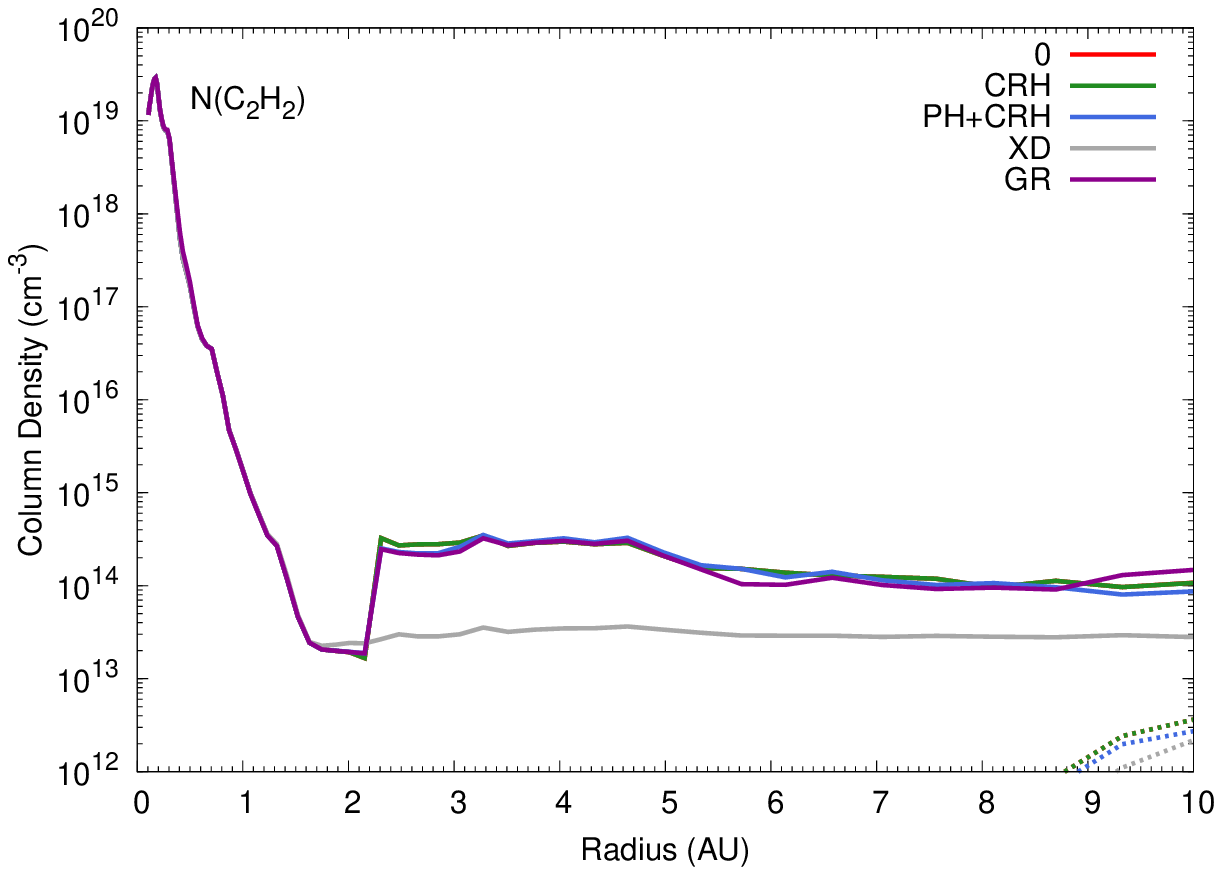}}
\subfigure{\includegraphics[width=0.45\textwidth]{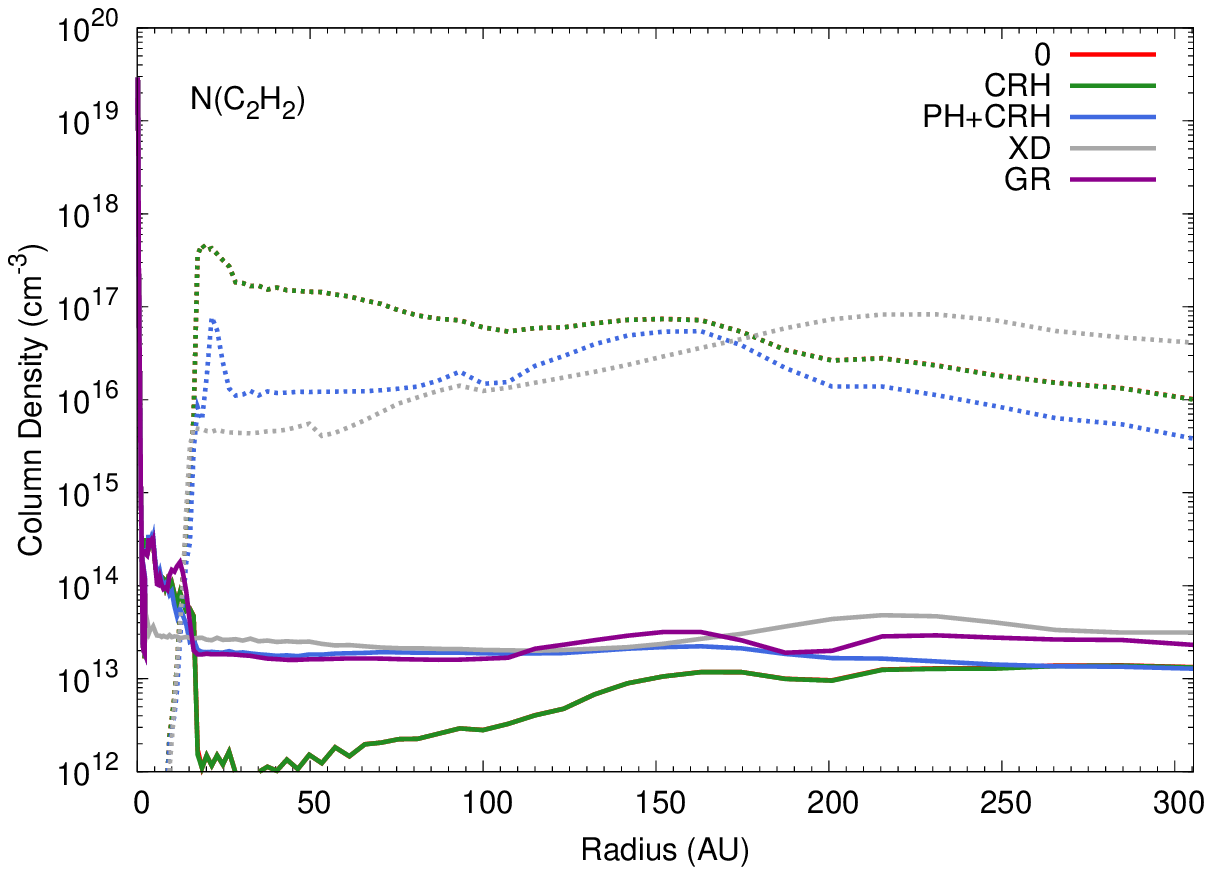}}
\caption{(Continued.)}
\label{figure13}
\end{figure}

\clearpage
\appendix

\section{Disk Physical Structure}
\label{physical appendix}

\subsection{Density}

The number density (cm$^{-3}$), as a function of disk radius and height,
is displayed in the top left and right panels  
of Figure~\ref{figure1} (within a radius of 10~AU and 305~AU, respectively)
and, as a function of $z/r$ (i.e.\ disk height scaled by radius) at a range of radii, 
in the top left panel of Figure~\ref{figure9}.  
The density decreases as a function of increasing disk radius and height with 
the densest region of the disk found in the disk  midplane close to the star
($\sim$~10$^{15}$~cm$^{-3}$) and the most diffuse, in the disk surface at large radii 
($\sim$~10$^{5}$~cm$^{-3}$) so that the density range in our model covers almost 10 
orders of magnitude.  
The density gradient in the vertical direction becomes more extreme with 
decreasing radius.  
At a radius of 0.1~AU, the density ranges from $\sim$~10$^{8}$~cm$^{-3}$ in the disk 
surface 
to $\sim$~10$^{15}$~cm$^{-3}$ in the  midplane, a change of seven orders of magnitude 
over a distance 
of $\approx$~0.03~AU.   
In contrast, at a radius of 305~AU, the density varies by around a factor of 
100 between the  midplane and the surface, over a distance of $\approx$~250~AU.   

\subsection{Gas and Dust Temperature}

Corresponding graphs for the gas temperature are shown in the bottom left and right 
panels of Figure~\ref{figure1} and the bottom left panel of Figure~\ref{figure9}.  
The gas temperature increases as a function of increasing height and 
decreasing radius with the hottest region found in the disk surface close to the star ($\sim$~10$^{4}$~K), 
and the coldest found in the disk  midplane in the outer disk ($\sim$~10~K).  
Within several AU of the star, the temperature increases near the disk  midplane 
due to the influence of viscous heating.    
In the disk surface, the gas and dust temperatures decouple with the dust 
temperature the lower of the two values (represented by dotted lines in Figure~\ref{figure9}).  
At low densities, gas-grain collisions become ineffective so the gas cools 
via radiative line transitions.     
At a radius of 10~AU, the difference in the gas and dust temperatures in the disk surface is 
more than an order of magnitude at $\approx$~2000~K and $\approx$~100~K, respectively.  
The vertical temperature gradients in the disk, although large, are not as extreme as 
those of the density.  

\subsection{UV and X-ray Radiation Fields}

The middle panels of Figure~\ref{figure9} show the UV and X-ray fluxes 
(top and bottom, respectively) as functions of $z/r$ at a range of radii.  
The UV flux is given in units of the average interstellar radiation field (ISRF)
(1.6~$\times$~10$^{-3}$~erg~cm$^{-2}$~s$^{-1}$).  
The extinction coefficient in the disk is very large for UV radiation thus the 
UV flux has an incredibly steep gradient in the vertical direction with the flux in  midplane 
at all radii $\approx$~0.  
In the disk surface, the UV radiation field decreases in strength with distance from the star.  
At a radius of 0.1~AU, the UV flux has a value $\sim$~10$^{9}$ times 
that of the ISRF whereas at 305~AU, this has decreased to $\sim$~10 
times the interstellar value.   

The X-ray flux shows a similar behaviour, however, X-ray photons, with energies $\sim$~keV, 
are more energetic than UV photons and so have greater penetrative power.    
Hence, the X-ray flux has a less steep gradient in the vertical direction than the UV flux.
In the outermost regions of the disk, $r\gtrsim$~10~AU, the X-ray flux 
is attenuated from a value $\sim$~0.1 to $\sim$~10~erg~cm$^{-2}$~s$^{-1}$ in the disk surface 
to $\sim$~10$^{-7}$ to $\sim$~10$^{-6}$~erg~cm$^{-2}$~s$^{-1}$ in the disk  midplane so that there 
remains a small, yet significant, X-ray flux where the UV flux is essentially zero.

\subsection{Cosmic-ray and X-ray Ionisation Rates}

\begin{figure}
\centering
\subfigure{\includegraphics[width=0.32\textwidth]{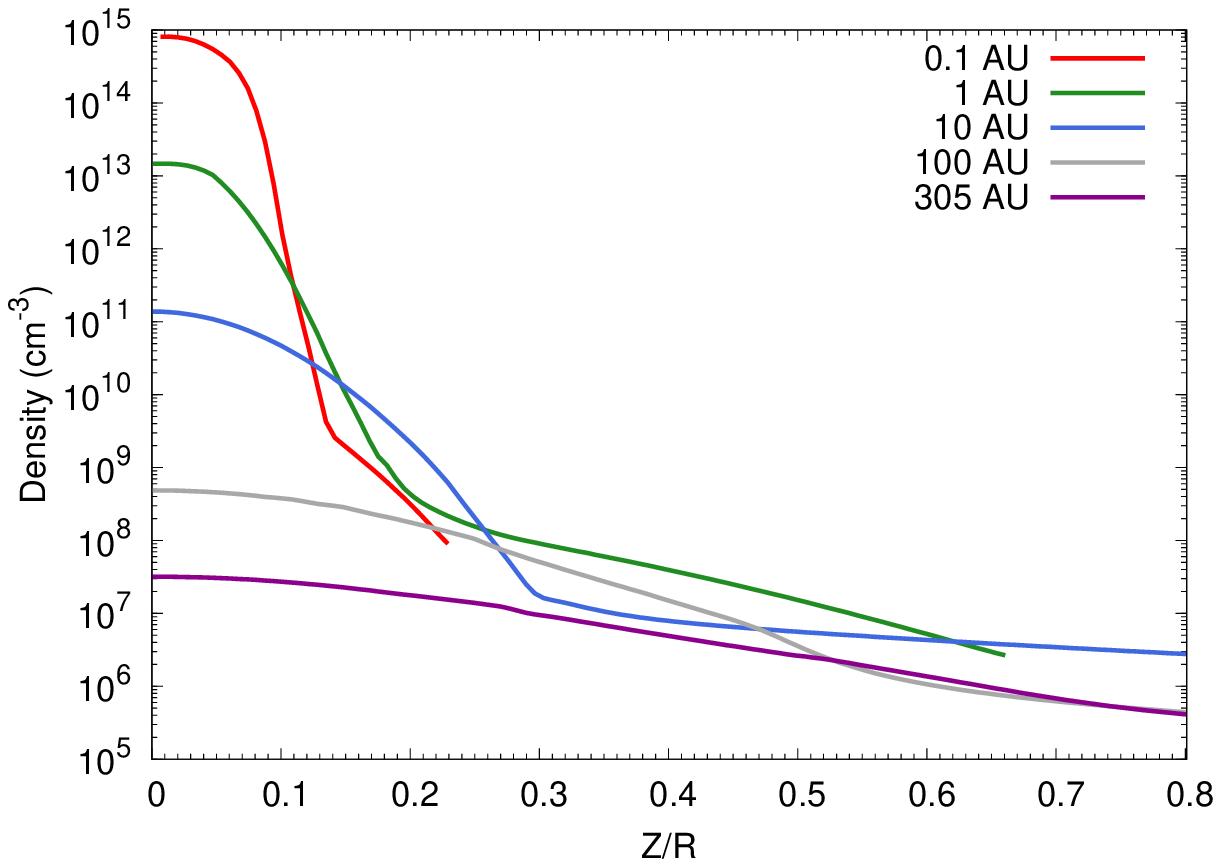}}
\subfigure{\includegraphics[width=0.32\textwidth]{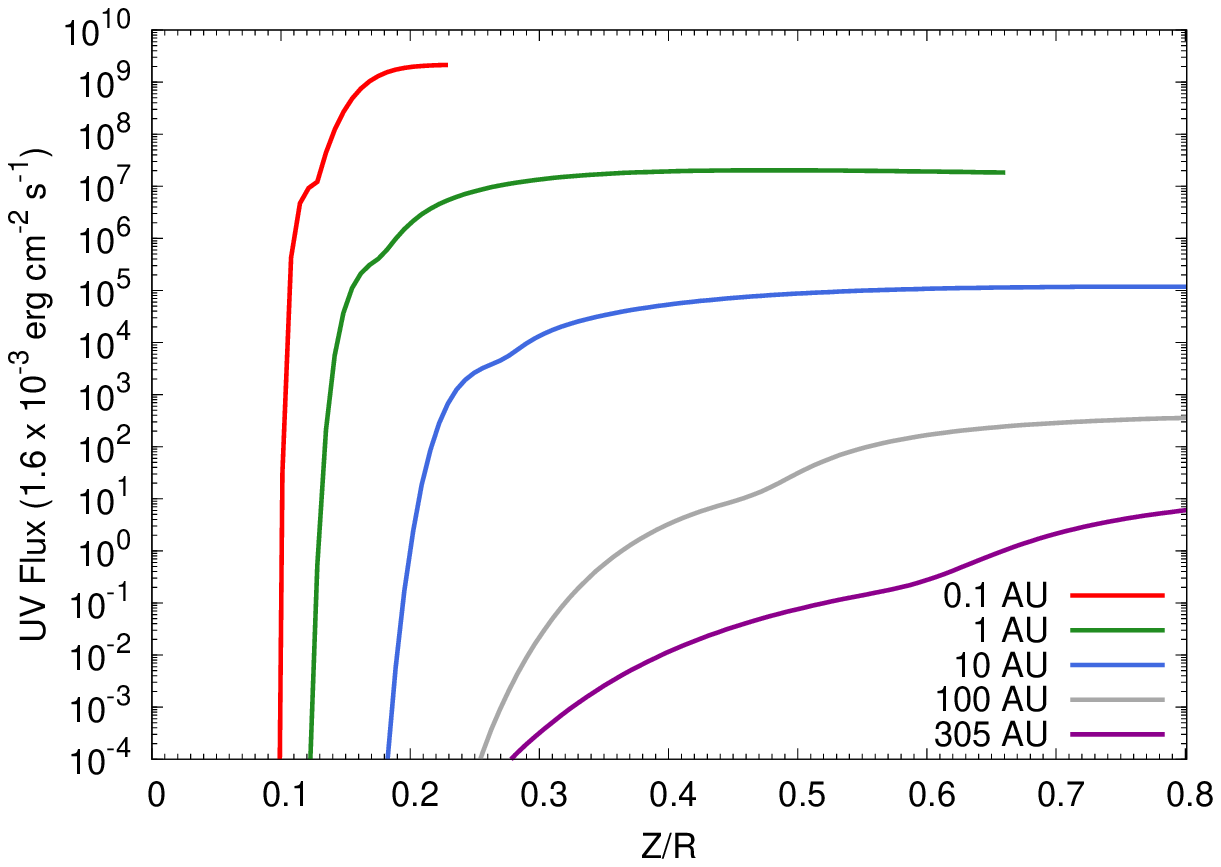}}
\subfigure{\includegraphics[width=0.32\textwidth]{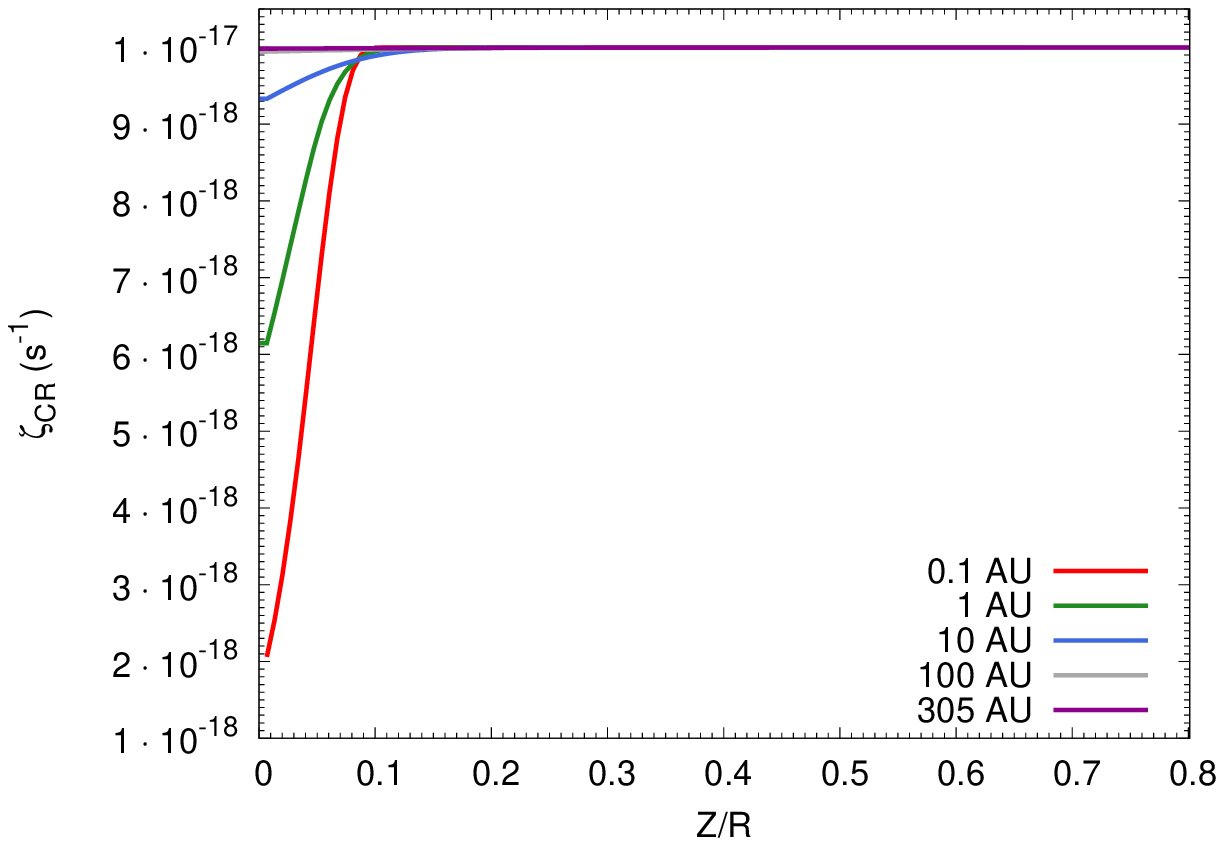}}
\subfigure{\includegraphics[width=0.32\textwidth]{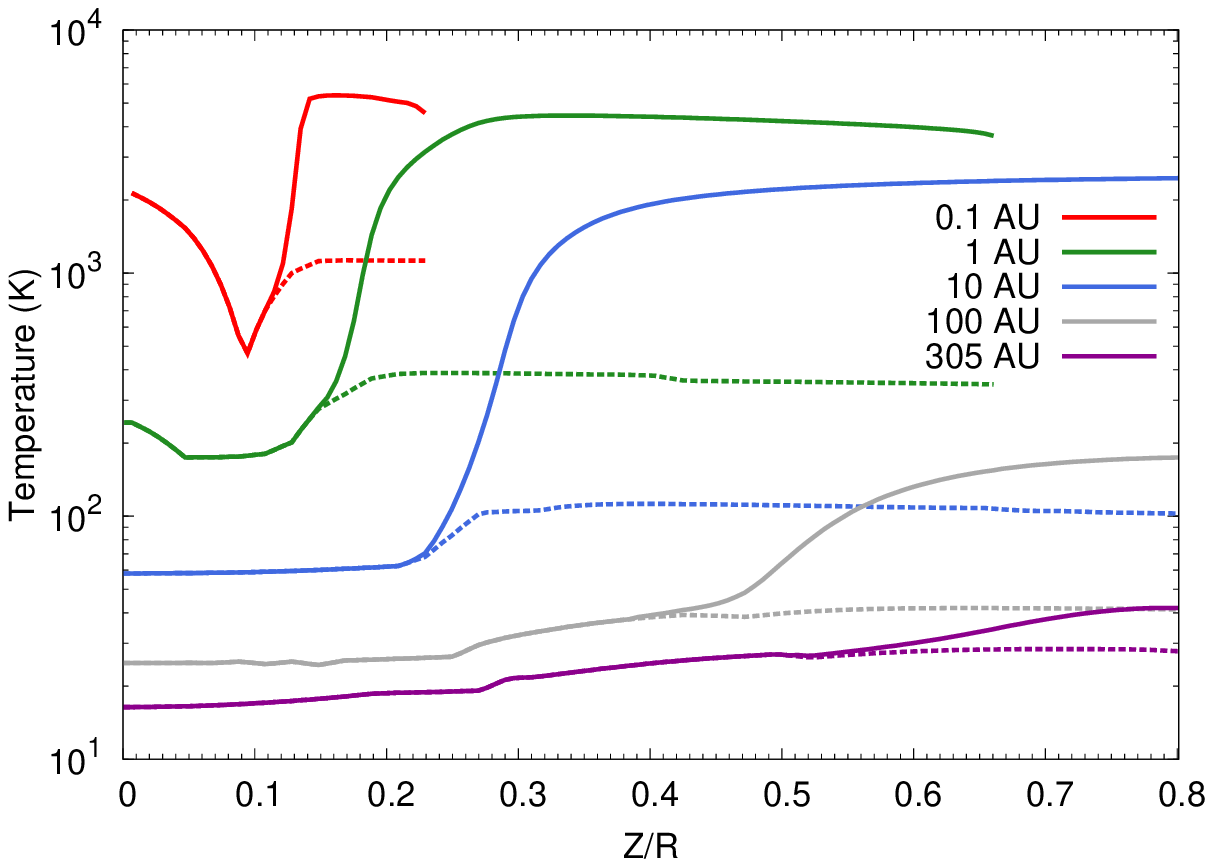}}
\subfigure{\includegraphics[width=0.32\textwidth]{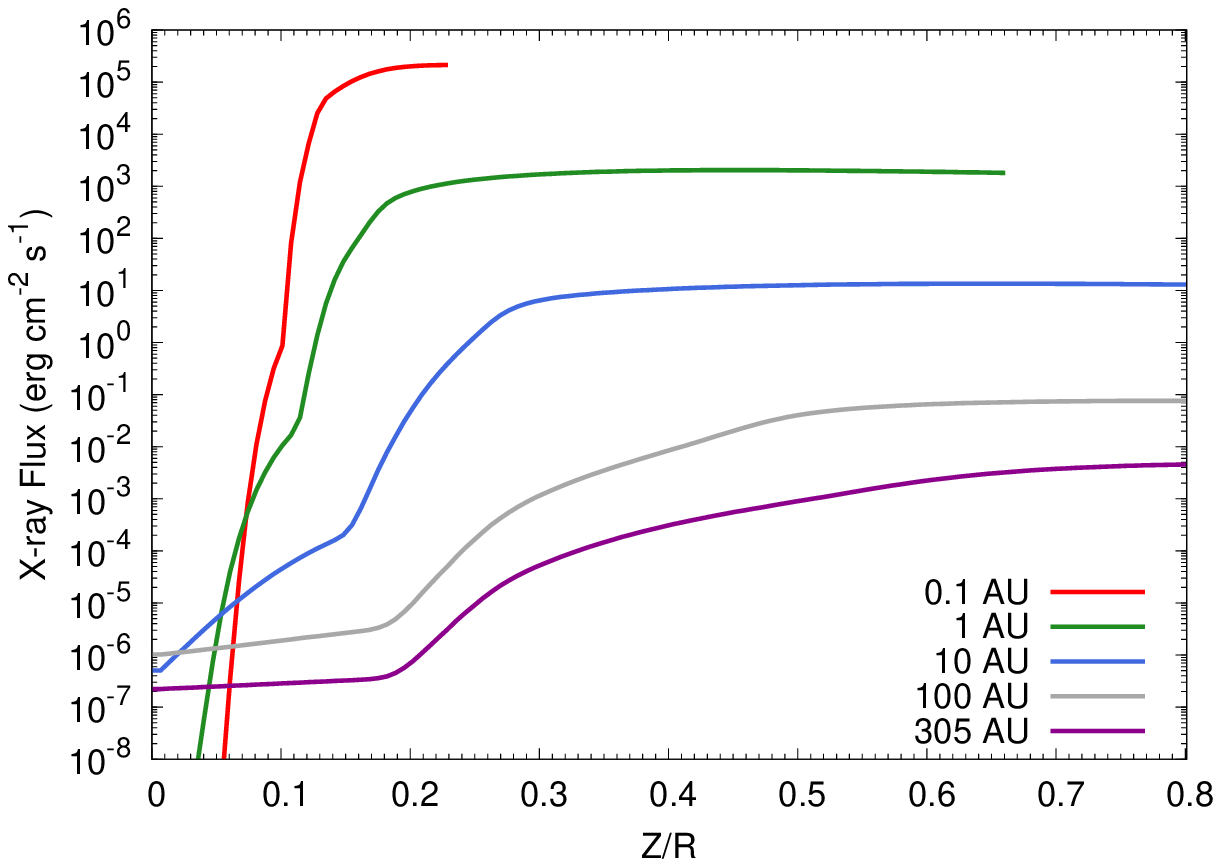}}
\subfigure{\includegraphics[width=0.32\textwidth]{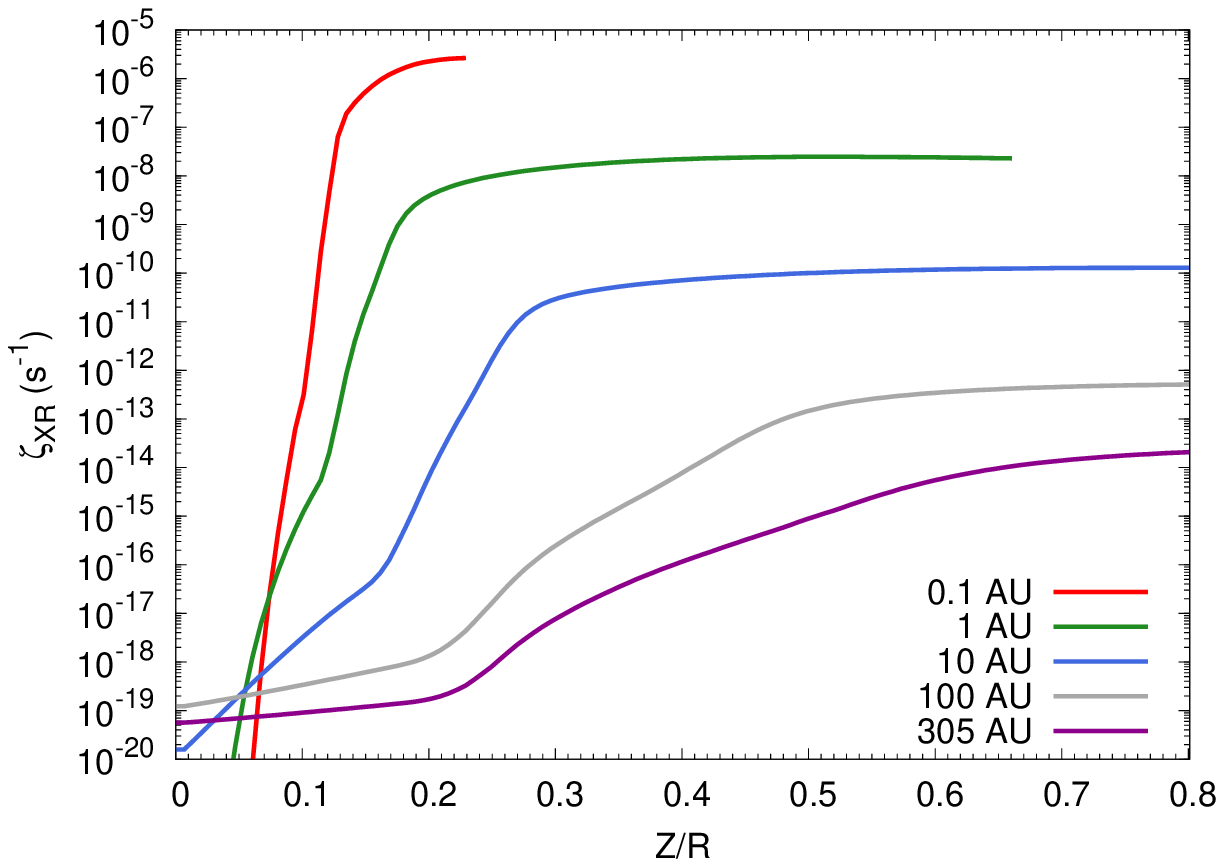}}
\caption{Number density (top left), temperature (bottom left), UV flux (top middle), 
X-ray flux (bottom middle), cosmic-ray ionisation rate (top right) and 
total X-ray ionisation rate (bottom right) as a function of disk height 
at radii, $r$ = 0.1~AU, 1~AU, 10~AU, 100~AU and 305~AU. The gas and dust 
temperatures decouple in the disk surface with the dust temperature 
represented by the dashed lines.  The UV flux 
is given in units of the flux of the interstellar radiation field or ISRF 
(1.6~$\times$~10$^{-3}$~erg~cm$^{-2}$~s$^{-1}$).  
Note that the scale used for the cosmic-ray ionisation rate is linear.}
\label{figure9}
\end{figure}

Finally, the cosmic-ray and X-ray ionisation rates are displayed in Figure~\ref{figure9} 
(top and bottom right panels, respectively) 
 as functions of $z/r$ at a range of radii.  
Note the linear scale used for the cosmic-ray ionisation rate.  
Beyond $r$~$\approx$~10~AU, the cosmic-ray ionisation rate is constant (at the interstellar value 
of $\sim$~10$^{-17}$~s$^{-1}$) throughout the height of the disk as  
cosmic-ray particles, with energies $\sim$~MeV, are highly penetrative.  
In the inner disk, e.g.\ $r$~=~1~AU, the column density of material through to the disk 
 midplane is large enough 
for some attenuation of cosmic-rays, however, the  midplane ionisation rate at this radius 
in our model is still relatively high at $\approx$~60~\% of the interstellar value.  
At even smaller radii, $r \sim$~0.1~AU, the  midplane cosmic-ray ionisation rate is further reduced 
to $\approx$~20~\% the rate in the disk surface.  

The X-ray ionisation rate, unsurprisingly, mirrors the X-ray flux.  
In the outer disk, $r~\geq$~10~AU, the X-ray flux is
significant enough to result in an X-ray ionisation rate $\sim$~10$^{-20}$ to $\sim$~10$^{-19}$
s$^\mathsf{-1}$ in the  midplane, however, here, 
cosmic-ray ionisation ($\sim$~10$^{-17}$~s$^{-1}$) dominates.
The X-ray ionisation rate is significantly higher than the cosmic-ray ionisation rate
in the middle to surface layers of the disk, with values 
ranging from $\sim$~10$^{-13}$~s$^{-1}$ at the maximum radius of our disk model 
to $\sim$~10$^{-6}$~s$^{-1}$ at the minimum.  

\end{document}